# A systematic review of COVID-19 epidemic models with endogenous human behaviour. What's next?

Elena D'Agnese[1,2,3], Alessia Melegaro[3], Vittoria Offeddu[3], Alberto d'Onofrio[4], Nicola Perra[5], Chris T. Bauch[6], Piero Manfredi[1,7]

1. Department of Economics and Management, University of Pisa, Pisa, Italy
2. Department of Statistics, Informatics, and Applications "Giuseppe Parenti", University of Florence, Florence, Italy
3. DONDENA Centre for Research on Social Dynamics and Public Policy, Bocconi University, Milan, Italy
4. Department of Mathematics, Informatics and Geosciences, University of Trieste, Trieste, Italy
5. School of Mathematical Sciences, Queen Mary University of London, London, UK
6. Department of Mathematics, University of Waterloo, Ontario, Canada
7. IRTA-Leonardo, Istituto di Ricerca sul Territorio e l'Ambiente, Pisa, Italy

## Abstract

Human behaviour and epidemic dynamics are intertwined, yet accounting for this feedback remains one of the key challenges of epidemiological modelling. The COVID-19 pandemic was an opportunity to overcome the traditional limitations of the field, raising expectations that data-informed endogenous approaches to behaviour modelling would advance substantially. To quantify the progresses made, we conducted a systematic review of SARS-CoV-2 transmission models endogenously including human behaviour in response to epidemic dynamics.
The COVID-19 pandemic saw great strides in terms of the expanded use of empirical data in epi-behavioural modelling. However, it also showed shortcomings with respect to limited use of behavioural empirical data, lack of innovation in model structure, and limited engagement with other disciplines and decision-makers. Overall, our results suggest that identifying priorities in model design and behavioural data, building an adequate data collection infrastructure, leveraging on AI advancements, and fostering interdisciplinarity are strategies of utmost importance for pandemic preparedness.

Epidemic trajectories are shaped not only by biological processes, but inherently by human behaviour. Decisions such as reducing social contacts, complying with non-pharmaceutical interventions (NPIs), seeking vaccination, or adapting daily routines can alter transmission dynamics as profoundly as pathogen characteristics themselves[1,2]. The COVID-19 pandemic made this interplay extremely visible: infection curves rose and fell in reaction to collective behavioural responses and policy measures, often more rapidly than could be explained by biological factors alone[3]. Infectious diseases are therefore increasingly recognised as behavioural phenomena, and this places human behaviour at the core of epidemic explanations and of public health decision-making[4,5].

Despite this well-recognised central role, integrating human behaviour into epidemic models remains one of the most persistent challenges in communicable disease research. Human behaviour is complex, context-dependent, and shaped by perceptions, preferences, social influence, and institutional constraints[6,7]. There is still no widely accepted framework for representing behavioural responses mechanistically within epidemiological models (e.g., design, calibration, empirical verification) — a challenge that has been described as one of the "hard problems" of epidemiology[8]. Moreover, the chronic shortage of behavioural adaptation data to inform epidemiological models results from the lack of standardised data collection protocols for the quantification of behavioural variables (unlike how public health agencies collect case notification data, for instance) and from limited data collection opportunities[8–10].

The field of Behavioural Epidemiology of Infectious Diseases (BEID) has emerged in response to these challenges[11–15]. Rather than treating behaviour as static, BEID conceptualised epidemics as coupled systems in which individuals adapt their actions in response to information on epidemic trends, and these adaptations in turn reshape transmission dynamics[16]. Within this perspective, over the past two decades, a rich modelling toolbox has been developed, which includes phenomenological formulations, awareness models, evolutionary games, and economic decision frameworks[11,17–24]. Despite this conceptual richness, much of the BEID literature has historically remained theoretical, and a shared, empirically grounded framework for policy applications is still lacking[9,10]. Within this landscape, two broad strategies have emerged for incorporating behaviour into epidemic models, namely endogenous and exogenous approaches (Box 1). While this distinction may appear straightforward, it has important theoretical and practical implications for interpretability and the capacity of models to represent behavioural uncertainty.

*Box 1: Endogenous vs exogenous human behaviour in epidemic modelling:*

Unlike traditional epidemic models[25,26], where human behaviour is either absent or taken as static, BEID focuses on the role of human behaviour in infectious diseases spread and control[27]. This is carried out via **endogenous**, or "coupled", **behaviour-disease models (left panel)** which consider how human behaviour co-evolves in response to the evolving information generated by the epidemic[12,21,27] and how the ensuing effects on epidemic trends eventually feed back on behaviour itself. Coupled models embed behavioural change within the epidemic system itself, preserving the feedback loop between infection dynamics and behavioural adaptation[12,21].

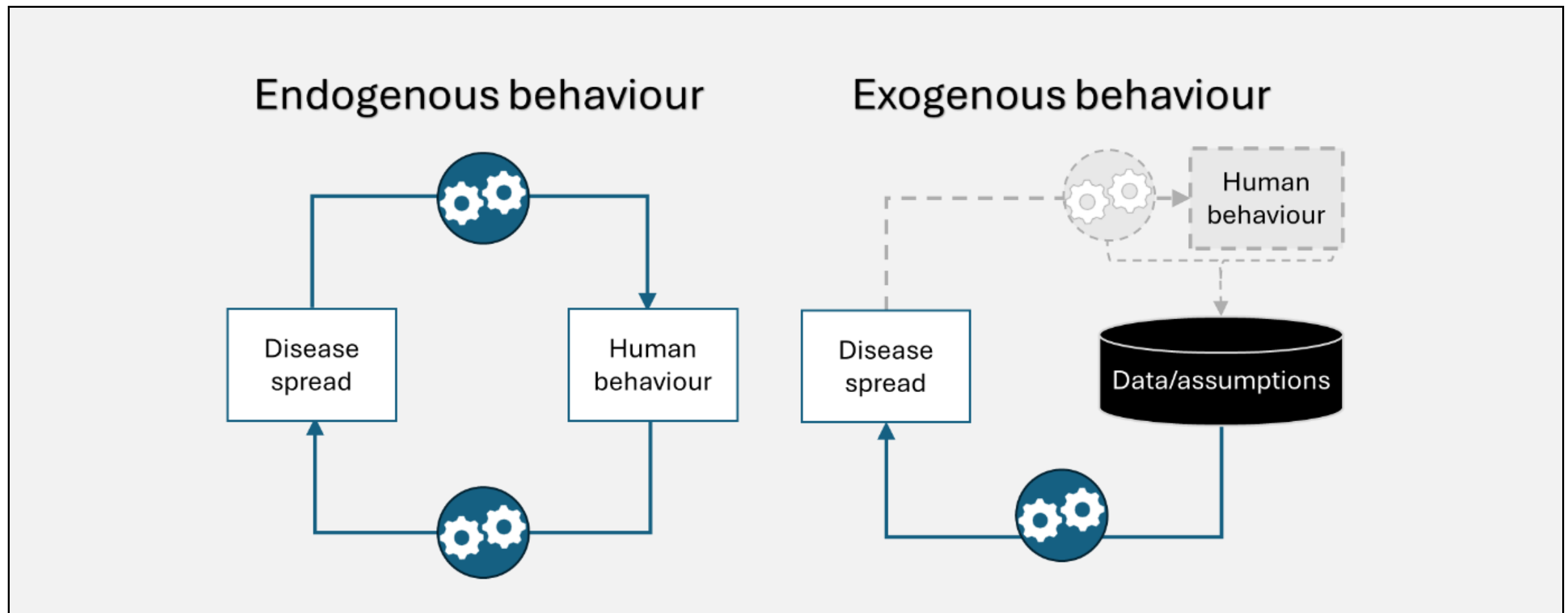

In **exogenous behaviour modelling (right panel)** the feedback loop is artificially removed: the behaviour response is treated as an external input, not resulting from the internal dynamics of the model. This may be done through assumptions (e.g., "We assume that the lockdown reduces the contact rate by 70%"), or directly plugging behavioural data in the model, considered as the outcome of a data-generating "black box" whose description is avoided (e.g., informing the model through e.g., mobility and contact data[28,29], without including a mathematical representation of the mechanisms generating the data itself).

The COVID-19 pandemic, with its exceptional public health interventions and the ensuing unprecedented imposition of behavioural restrictions[30], as well as the dramatically wide spectrum of resulting behavioural responses[8], acted as a large-scale stress test for BEID, while, at the same time, urging institutions to rely on epidemic modelling as never before[31–33]. In parallel, it fuelled data collection efforts worldwide, encompassing not only disease and epidemic surveillance dimensions, but also behavioural ones – for example, mobility data[34], survey-based measures of attitudes and compliance[35,36], longitudinal social contact diaries[37], and digital traces[38], also from social media platforms[39–41]. This caused an extraordinary, heterogeneous surge in publications from multiple fields besides epidemiology, such as behavioural and medical sciences, physics, and economics[42–45]. Yet, although the pandemic appeared to offer a unique opportunity to overcome some of the historical limitations of BEID, human behaviour was mostly represented in models' structure exogenously. This facilitated timely policy evaluation[28,46] but, at the same time, failed to capture the true feedback loops between transmission and behaviour and the related uncertainty[32,47]. Unsurprisingly, many such studies identified the lack of behavioural endogeneity as a major limitation of their models[4,25]. Nevertheless, we still lack a thorough appraisal of whether this acknowledged limitation was addressed during the COVID-19 pandemic through BEID models. In other words, it is still unclear to what extent the emergency situation and the exceptional data availability facilitated the use of endogenous behaviour-epidemic models, fuelling methodological progress, empirical innovations, and a deeper understanding of behavioural mechanisms.

Therefore, in this study we conduct a systematic review of SARS-CoV-2 transmission models endogenously incorporating human behaviour, following PRISMA 2020 guidelines[48]. We aim to: i) identify and classify the theoretical approaches used to model behavioural feedback; ii) assess whether and how empirical data – particularly behavioural data – were employed to inform model calibration; iii) discuss the implications of these findings for the advancement of BEID and epidemic preparedness. By systematically examining the COVID-19 pandemic as a stress test for endogenous behavioural modelling, we seek to clarify what has been learned about integrating human behaviour into epidemic models and what remains to be addressed. In doing so, our contribution is placed within a broader stream of methodological and narrative literature on the topic, both pre- and post-pandemic[12,15,14,49,50,8,7,51,52,10,53–56]. In this context, two post-pandemic reviews[57,58] adopted a systematic approach on the present topic of coupled modelling, but did not focus on the issue of empirical data use.

## Results

We included 128 papers (Supplementary Information S1, Open Science Framework (OSF) Repository[59]), with the largest share published in 2021 (24%) and 2022 (33%). In most studies, behaviour was framed as a purely voluntary response to epidemic dynamics (67%), whereas a smaller proportion explicitly recognised the coexistence of voluntary and mandated behavioural changes (24%). The behaviours modelled were predominantly generic protective actions (50%), defined as broad risk-reduction responses without focusing on a specific behaviour, and social distancing (23%). By contrast, vaccination behaviour—historically one of the most extensively studied topics in BEID—was explicitly analysed in only 9% of the studies. Across the selected papers, behavioural adaptation was most commonly driven by global, prevalence-based information (89%, Box 2), and was mainly incorporated through changes in model parameters (70%), such as the transmission or the vaccination rate. With few exceptions (e.g., agent-based, network, stochastic models), most studies relied on ordinary differential equations (ODEs, 80%). Descriptive statistics from the quality assessment are available in Supplementary Information S2.

*Box 2: Key concepts in BEID framework: the role of information and implicit vs explicit models*

The role of information

Information (denoted by $M$) is the key distinctive feature of behavioural adaptation in contemporary communities compared to the past, where response to infectious threats were mostly based on beliefs (e.g., religious)[60]. Available studies[12] have classified information based on the spatial impact of the source (local *vs* global) and on its type (prevalence- *vs* belief-based). While *global* information denotes information available to everyone (e.g., news), *local* information comes from individuals' closer ties, either physical or metaphorical (e.g., the neighbourhood, or the social sphere). *Prevalence-based* information refers to observable epidemiological indicators describing the state of the epidemic, whereas *belief-based* information incorporates additional layers such as risk perception, personal beliefs, social influence, or network effects.

Building on Funk et al. (2010)[12], we notice that rather than being a 2x2 combination, other possibilities can be envisioned, as in the same model both local and global and/or both prevalence- and belief-based information may be present. Moreover, the term prevalence-based information should be interpreted broadly. Although historically derived from disease prevalence, it often encompasses a wider range of epidemiological indicators, including incidence, mortality, hospitalisations, or other quantities that individuals may observe directly or indirectly and use to guide behavioural decisions.

Implicit vs explicit behavioural models

Within endogenous behaviour modelling, two approaches may be identified: *implicit* and *explicit*.

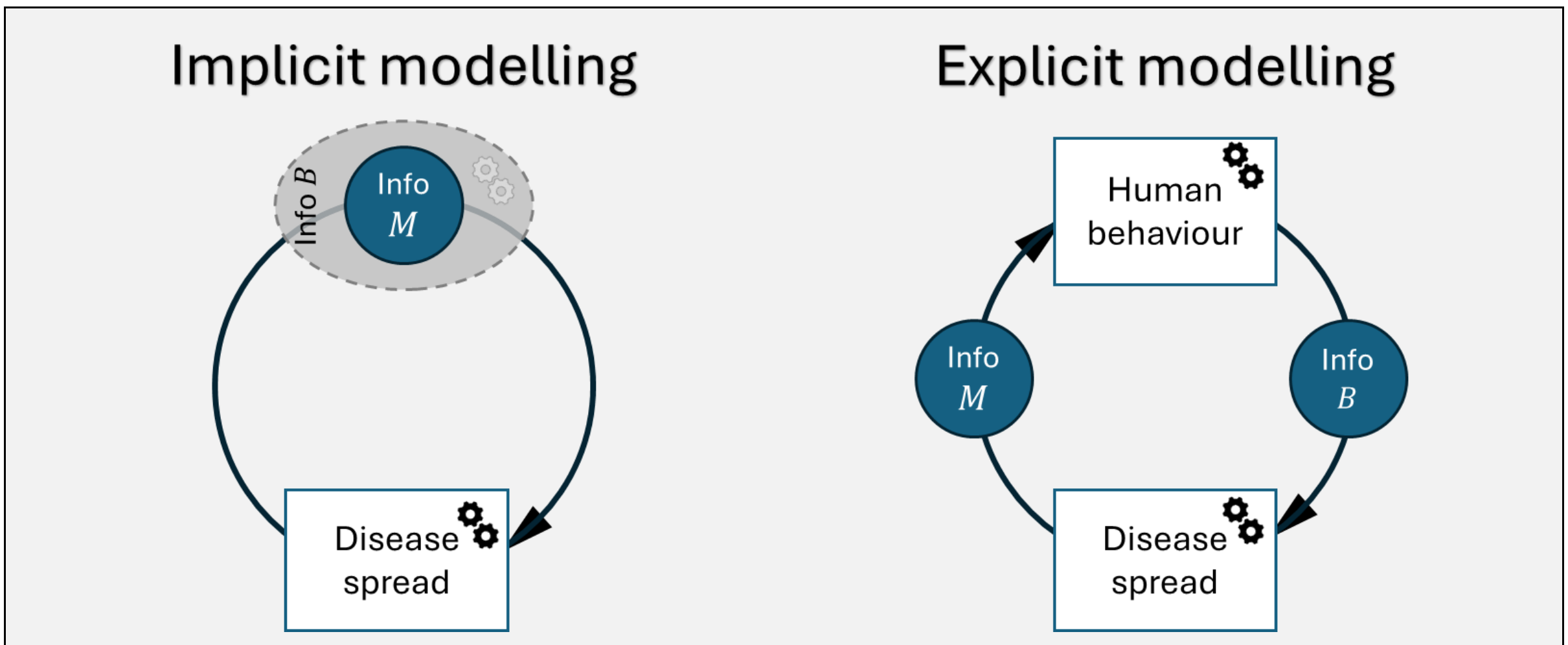

In **implicit behaviour modelling**, behavioural change is represented through a phenomenological link between information on epidemic trends and individual responses, i.e., the behavioural mechanism is not represented *per se,* rather the focus is on the overall effects of behaviour $B$ on epidemiological or policy parameters. Pioneered in Capasso and Serio (1978)[61], this approach was used to model behaviour change during the AIDS pandemic (e.g., [62,63] and references therein) and formalised in later studies[64], landing on the concept of *information index*[21,65,66].

In **explicit behaviour modelling**, behavioural change is explicitly grounded in behavioural theories, whether psychological, economic, sociological. They represent fully coupled systems, in which both disease dynamics and behavioural mechanisms are described via distinct but interacting processes, through the link of information. Foundational modelling solutions of this type include[17,11,20,23,22,19,67,68,24].

Additional information may be found in Supplementary Information S3.

## Theoretical frameworks

To characterise how endogenous behaviour was represented, we developed an *ad hoc* taxonomy, that first distinguishes between implicit (59/128) and explicit (69/128) modelling approaches (Box 2). Within these two broad categories, studies were further classified according to the foundational BEID approach they most closely resembled (Figure 1), namely information index models, awareness-based models, imitation dynamics, and economic epidemiology – for interested readers, we have delineated a short navigation roadmap across BEID approaches in Supplementary Information S3. No entirely new modelling approaches emerged during the COVID-19 pandemic. Supplementary Information S4 provides a brief description of the behaviour modelling solutions adopted by each study within its taxonomy category.

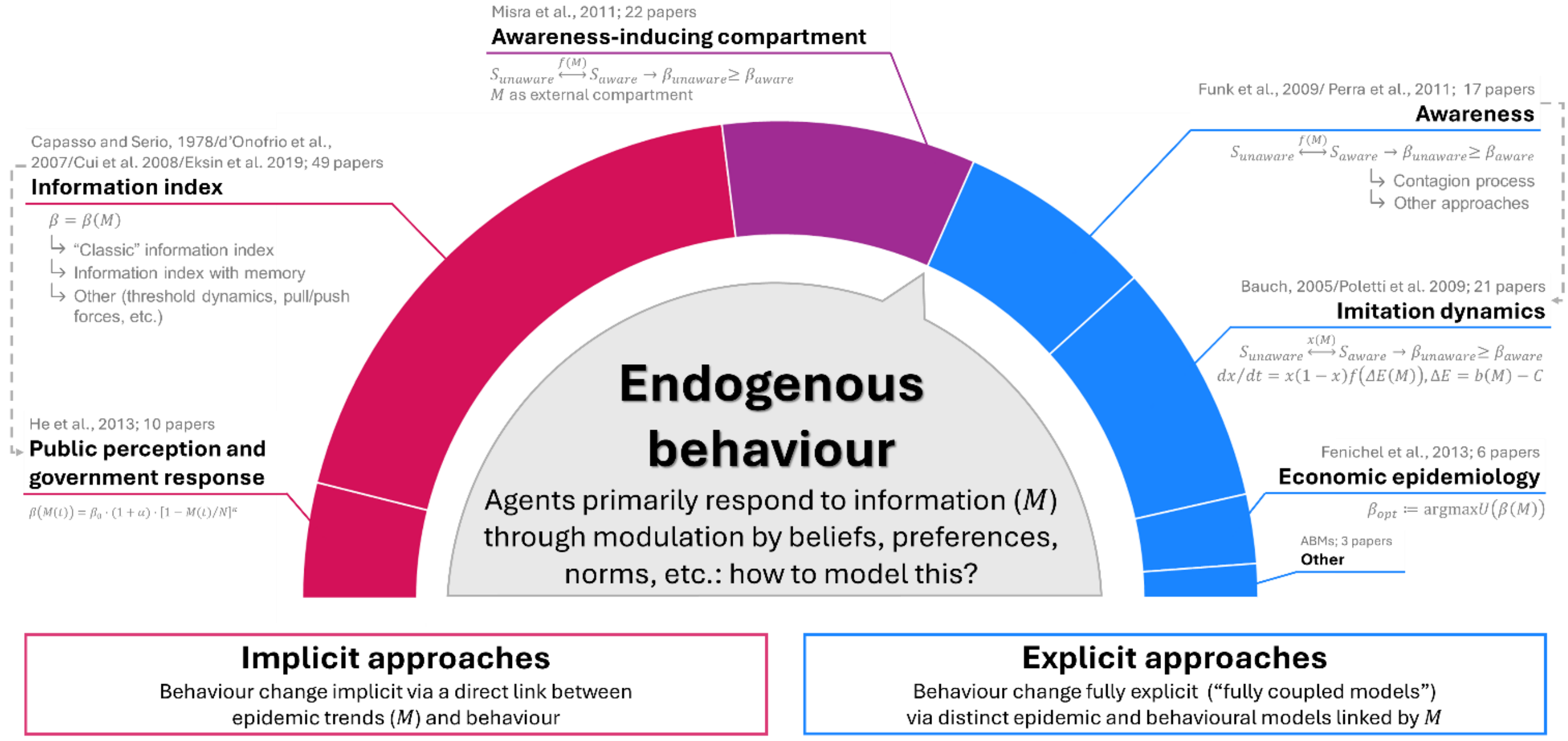

*Figure 1: **Taxonomy of endogenous behaviour modelling approaches identified in the reviewed literature.** The figure provides a methodological "compass" for navigating the behavioural epidemiology modelling toolbox. Sectors in different colours represent the relative share of each category among the 128 included studies. Regardless of their level of complexity or model structure, all approaches can be interpreted as alternative representations of how epidemic-related information M generates behavioural adaptation (e.g., social distancing, vaccination, or other protective actions), which in turn feeds back on disease dynamics. Conceptual links between categories (dotted grey lines) are described in Box 2 and Supplementary Information S3.*

### *Implicit modelling (59 papers)*

**Information index** (n. of studies: 49/59; [52,69–116]). The idea of information as triggering behavioural change relies on the concept of *information index*, namely any model-specific function summarising the epidemic-related information ($M$) available to individuals (e.g., epidemic prevalence, hospitalisations, deaths, etc.). Regardless of its exact formulation, the information index provides a parsimonious representation of how information generated by the epidemic feeds back on behaviour, making it the most widely adopted implicit modelling framework in our dataset. This category includes both applications of the original formulation by Capasso and Serio (1978)[88,99,114] and later adaptations by d'Onofrio et al. (2007), particularly those including both past and current epidemic information[77,78,87]. We also identified later adaptations of Cui et al. (2008)[100] and Eskin et al. (2019)[66,95,116].
Most variations concern the type of information incorporated into the index. Some studies relied exclusively on contemporaneous epidemiological indicators within the index (e.g., [72,83,98]), while others used only past information[81,90]. Several models[73,84,94,110] represent behavioural adaptation as the balance between two opposing forces: a "sense of caution" ("reaction rate") and a "sense of safety" ("restoration rate"). The former increases as infection risk rises, reducing the force of infection through risk-avoidant behaviour - similarly to a classic measure of risk perception -, while the latter increases with vaccination coverage, reflecting behavioural relaxation as perceived risk declines. The information index has also been used to define threshold mechanisms triggering institutional or individual decisions, e.g., enacting and complying with NPIs[82,89]. Information indices have been used altogether to accommodate different epidemic phases[113], or to relate model-based unobserved quantities to information actually available to the population[112]. They have also been adapted to include a forced component of behaviour describing the presence of mandatory NPIs during the pandemic period[93], or to leverage both local and global information about the disease[92].

**Public perception and government responses** (n. of studies: 10/59; [117–126]). Although this category can be viewed as a specific implementation of the information index framework, we discuss it separately because it represents one of the few attempts to model the joint effects of voluntary behavioural adaptation and governmental-imposed interventions. All studies in this group closely follow the formulation originally proposed by He et al. (2013). To account for this interaction, these models introduce a variable representing the strength of governmental measures (replacing the school calendar effect included in the original framework). In most cases, this quantity is assumed constant over time (with the exception of [126]), despite the substantial temporal and geographical variation in policy responses observed during the COVID-19 pandemic. Behavioural adaptation is then driven by a range of information sources, including mortality[118], symptomatic infections[120], or other broader epidemiological indicators[117,123,126]. Overall, the main innovation of this category lies not in the information index itself, but in its explicit attempt to couple individual behavioural responses with institutional action within the same modelling framework.

*Explicit modelling (69 papers)*

**Awareness** (n. of studies: 39/69). Broadly speaking, awareness-based models describe behavioural adaptation as a transition from an "unaware" state associated with normal behaviour to one or more "aware" states associated with altered or risk-reducing behaviour. In these models, awareness of the disease is the driver of behaviour change rather than the disease itself, allowing adaptation to occur even without direct exposure to infection[20,22]. We identified three main ways of operationalising this concept[20,22]:

- Awareness-inducing compartment (n. of studies: 22/39; [127–148]). The largest subgroup focuses on awareness generated through external information channels, such as awareness campaigns, advertisements, traditional media, or social media, following the framework proposed by Misra et al. (2011). Typically, a dedicated awareness compartment - mostly driven by global, prevalence-based information - modifies epidemiological transitions by stratifying individuals into behavioural states characterised by different disease parameters. In some studies, awareness directly modulates the force of infection (e.g., [132,137,139]). Although these models seemingly provide an explicit representation of behaviour *prima facie*, it can be shown that they are equivalent to implicit, information index modelling (Supplementary Information S3). As such, they represent a conceptual bridge between implicit and explicit approaches (Figure 1), and, more generally, between information index formulations and fully explicit awareness dynamics.

- Awareness contagion process (n. of studies: 9/39; [149–157]). This category includes mostly ODE models implementing the three awareness mechanisms described in Perra et al. (2011) (Mod. I, II, and III), briefly summarised in Supplementary Information S3. We identified examples of both the local, prevalence-based information model (Mod. I) [154–156] and the global, prevalence-based information approach (Mod. II)[149,150,153]. These studies use a variety of epidemiological indicators to generate awareness, including infected, isolated, quarantined, and hospitalised individuals. Other studies[151,152,157] combine the original awareness approach described in Funk et al. (2009) with the local information, belief-based model (Mod. III), typically through two-layered networks coupling disease and awareness diffusion.
- Awareness - Other (n. of studies: 8/17; [158–165]). Though not strictly fitting the previous categories, a smaller group of studies extend the awareness framework by introducing more complex transitions across multiple awareness states, often combining explicit awareness dynamics with one or more information-index formulations. As a result, they further highlight the close conceptual links between awareness-based and information-index approaches.

**Evolutionary game theory/Imitation dynamics** (n. of studies: 21/69; [166–186]). In these models, individuals revise their behaviour by comparing the perceived payoffs of alternative strategies during social encounters with other individuals. The approach is flexible enough to capture additional factors as the role of public communication, social norms, and information indices. The main variations with respect to the seminal studies in Bauch (2005) and Poletti et al. (2009) concern the specification of the payoff function, which depends both on the behaviour being modelled and on the complexity of the underlying epidemiological model. For example, the perceived infection risk may depend on the size of the quarantined population (e.g., [173,174]), on broader indicators of infection risk and morbidity (e.g., [166,177]), or on estimates of asymptomatic and undetected infectives (e.g., [170,181]). In other words, the extension of the model structures allows for the inclusions of other disease-dependent dimensions on top of the prevalence, as a simpler model would require. While most studies focus on self-initiated endogenous behaviour, others try to also include a mandatorily-induced component. Some studies examined voluntary compliance with public measures as a protective behaviour[170,173,177]; one incorporated compulsory vaccination for a subset of the population[168]; and others evaluated the support to public policies shaping behaviour[166,182].

**Economic epidemiology** (n. of studies: 6/69; [187–192]). Adopting a micro-economic perspective, this approach assumes that behaviour change is shaped by individuals acting to maximise their (possibly intertemporal) utility, weighing the perceived benefits and costs of adopting specific protective behaviours conditional on their risk perception. The reviewed studies focus on self-initiated behaviours, mainly social distancing[190,191] inducing "reduced" activity levels (e.g., [188,189]). The role of public policy has been seldom considered[191], by maximising the utility of compliance with public recommendations (without directly modelling them). Despite the differences in modelling details (e.g., information sources, utility functional form), the general approach of utility maximisation is very similar across studies. Two studies[190,192] do not directly employ the classic utility maximisation framework, but they show very similar conceptual premises.

**Other** (n. of studies: 3/69; [193–195]). This residual category includes explicit behavioural models that do not closely resemble any of the established BEID frameworks described above and rather use *ad hoc* behavioural recipes/infrastructures. Additional details can be found in Supplementary Information S4.

## Empirical perspective

More than half of the reviewed studies (79/128) incorporated empirical data to inform either disease-related or behaviour-related parameters (Figure 2A). This was achieved through model calibration (disease-related param: 59/128; behaviour-related param.: 52/128), direct parameterisation (8/128 and 5/128, respectively), or a combination of both approaches (5/128 and 2/128, respectively).

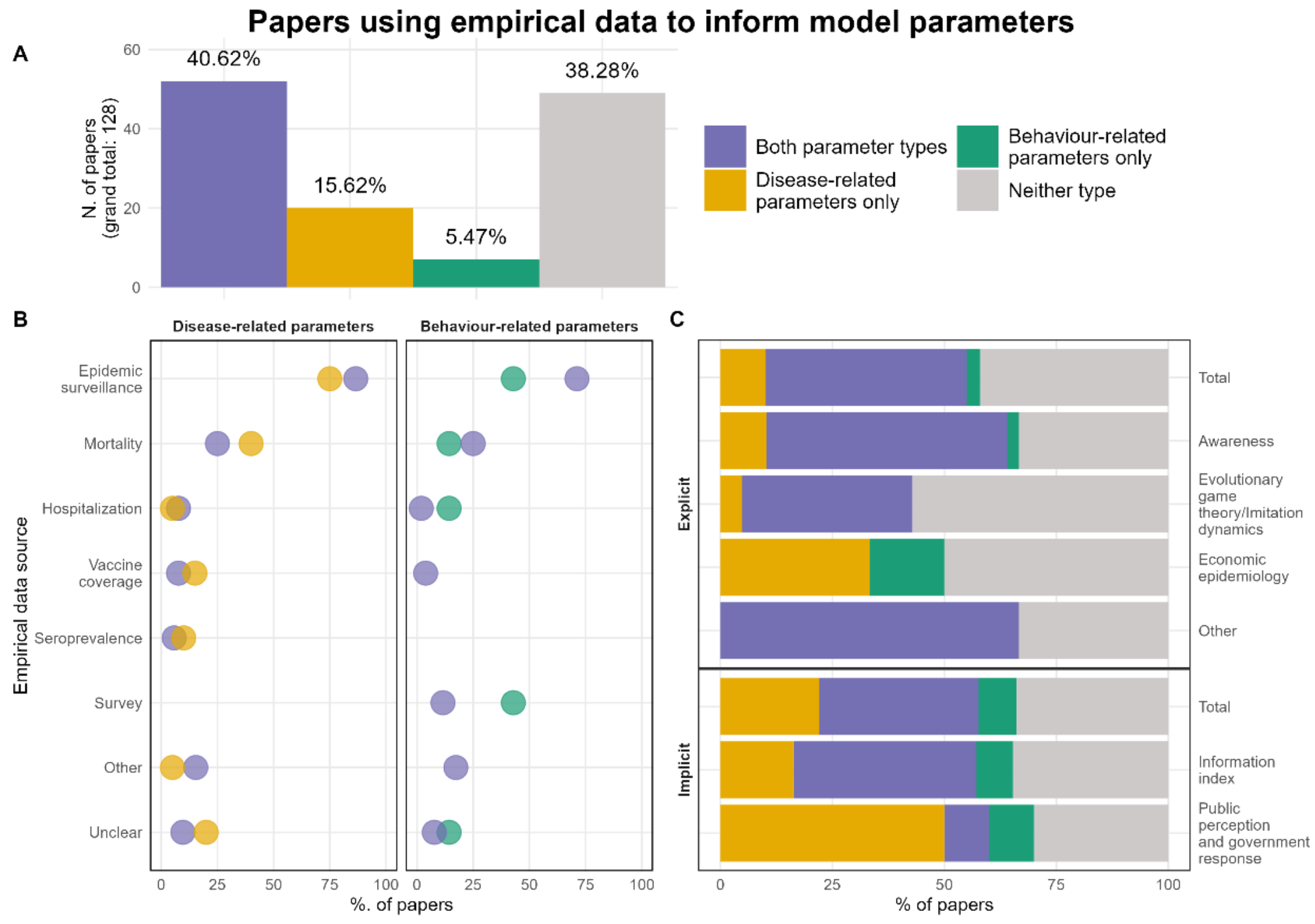

*Figure 2:* ***Use of empirical data to inform disease-related and behaviour-related parameters in models with endogenous behaviour****. Panel A shows the number of studies using empirical data to inform disease-related parameters, behaviour-related parameters, both, or neither. Panel B reports the empirical data sources used for each purpose. Because individual studies may rely on multiple data sources, percentages are calculated relative to the totals shown in Panel A and may exceed 100%. "Other" includes mobility data, testing data, subway ridership data, media data, and social media data. Panel C shows the proportion of studies using empirical data across the theoretical modelling categories identified in Figure 1. Overall, the figure highlights the widespread use of empirical data during the COVID-19 pandemic, while revealing a more limited use of behavioural data to directly inform behavioural mechanisms.*

Examples of behaviour-related quantities informed by data include, among others, the intensity of behavioural responses to epidemic threats, costs of behavioural change to individuals, awareness transition rates, degree of rationality, and risk perception thresholds.

Most studies used empirical data to inform both disease- and behaviour-related parameters (41%), while smaller proportions informed only the former (16%) or the latter type (5%) (Figure 2A). However, the sources of information used for the two purposes were remarkably similar. For both types of parameters epidemic surveillance data - including active, cumulative, and recovered cases, as well as quarantined or symptomatic individuals – represented the dominant empirical source (Figure 2B). COVID-19 mortality data were also frequently used, particularly among studies informing disease-related parameters only. By contrast, studies informing behaviour-related parameters were relatively more likely to incorporate survey data, typically through direct parameterisation, in an attempt to capture attitudes, perceptions, or self-reported behaviours.
A subset of papers did not clearly report the data sources (9/128 for disease-related parameters; 5/128 for behaviour-related parameters; Figure 2B, "Unclear"). Only a minority of papers used hospitalisation, vaccine coverage, seroprevalence, or other epidemiological datasets. A few models leveraged "novel" data streams – such as mobility records or social media analytics - to inform either type of parameters (Figure 2B, "Other").

Studies adopting implicit approaches were somewhat more likely to incorporate empirical data (66%) with respect to those relying on explicit approaches (58%), although differences across modelling categories were generally modest (Figure 2C).

## Discussion and future challenges of coupled infection-behaviour modelling

This work reviewed 128 papers on the endogenous coupling between SARS-CoV-2 spread and human behaviour. By tracing these studies back to the foundational frameworks largely adopted in the BEID literature, we found limited theoretical innovation in how behaviour was represented. This finding likely reflects the need of the field to promptly and usefully engage in the emergency-driven debate using readily-available tools. Indeed, most models could be mapped onto a small number of established approaches - namely information-index formulations, awareness models, imitation dynamics, and economic epidemiology (Figure 1 and Supplementary Information S4). Limited innovation is not necessarily problematic *per se*, as tried-and-true model architectures (e.g., age-structured compartmental models) may still be the optimal tool when parametrised with suitable empirical data and conducted to address questions of interest to decision-makers. Nevertheless, the persistence of a relatively small set of behavioural modelling approaches, even when confronted with an unprecedented public health crisis, suggests a degree of theoretical inertia within BEID.

Most reviewed studies used empirical data to inform their models both for disease- and behaviour-related mechanisms. This surely represents a notable progress compared to pre-COVID literature[15], partly reflecting the large availability of disease data at several aggregation levels collected worldwide. However, behavioural parameters were seldom informed by behavioural data. Instead, they were most often inferred from the epidemiological signal, i.e. by fitting behavioural responses to observed epidemic trends.

From a quality standpoint (Supplementary Information S2), two main issues emerge. First, details of code and data availability were often unclear or unavailable, raising reproducibility concerns. Second, few studies validated their results against independent datasets or provided uncertainty measures for fitted parameters, limiting the robustness and interpretability of their findings. While not unique to behavioural modelling, these issues are particularly relevant here given the central role of behavioural assumptions in shaping model outcomes and their usability for pandemic management.

Looking back at the questions posed by Edmunds et al. (2013) and Funk et al. (2015)[196,9], it is clear that some topics are still unaddressed even in the aftermath of the COVID-19 pandemic – previously defined as an underleveraged opportunity[7]. Our findings suggest that clearer guidance on modelling choices and priorities could improve the efficiency and usability of the BEID modelling toolbox. More fundamentally, the modelling framework appeared insufficiently adaptive to the complexity of behavioural responses observed during COVID-19. Relevant dimensions – such as the interaction between public health policies and voluntary behaviours[30], the interplay between global and local information, or the heterogeneity across different protective behaviours[170,197] – have been only partially explored. Addressing these aspects would require model structure innovation balanced with an appropriate degree of behavioural complexity to avoid too cumbersome or overfitted models. But which details are “relevant” enough to be added to better capture behaviour while preserving a coupled model efficiency and usability for pandemic prediction and management? The answer is not always clear, and such a question is further complicated by the additional requirement of sound behavioural assumptions in BEID models. For example, awareness is often assumed to directly translate into behavioural change, despite this being a strong and not always empirically supported assumption. Indeed, there may be several barriers between individuals’ awareness and actions , e.g., logistical, economic, etc.[198,199]. Additionally, failing to engage with expertise from the behavioural sciences can generate invalid or hardly justifiable assumptions (e.g., triggering behavioural change based on information sources that are not directly observable by individuals), further weakening the behavioural grounding of the models.

Concerning data use, BEID appears to have integrated epidemiological data more readily than behavioural data, despite its stated objective. Over recent decades, BEID highlighted the need to account for the behaviour-epidemic feedback loop, but it seemingly has not (or has been unable to) foster behavioural data collection efforts suitable or strong enough to fully support this goal. The COVID-19 pandemic seems to have only scratched the surface, raising several questions. In light of the large data collection effort witnessed during the pandemic, did we gather behavioural data sufficiently informative for modelling? Were such data accessible (e.g., not restricted by private ownership), transparently described, and usable in practice? And, more fundamentally, are current modelling approaches actually able to incorporate behavioural data meaningfully[197]? Although our review cannot provide definitive answers, it highlights a clear gap between data availability and their integration into coupled models. During the COVID-19 pandemic, novel data streams – such as mobility or digital data – were used in some cases, but their potential appears only partially exploited. More broadly, the infrastructure for collecting, integrating, and standardising behavioural data remains underdeveloped compared with epidemiological surveillance data. Promising future directions include participatory surveillance platforms[200], integration of new and traditional data structures, e.g., tracking behaviour by nesting real-time digital traces of mobility within "classic" social contact matrices[201], and new data collection frameworks explicitly trying to bridge behavioural sciences and epidemiological modelling, e.g., quantitatively capturing the concept of "social influence" as a determinant of behaviour uptake[197].

As previously discussed, exogenous behaviour modelling has often been preferred to endogenous approaches during the COVID-19 pandemic, despite the recognised limitations associated with neglecting behaviour-epidemic feedback loops and the resulting uncertainty[32,47,56]. Therefore, although coupled modelling helped us draw important theoretical conclusions over time, it seemingly has not made a breakthrough in terms of practical applicability during emergency situations, including the COVID-19 pandemic. This raises key questions: has BEID the potential to be fruitfully implemented in pandemic management? Is real-time endogenous behaviour modelling necessary during emergent infectious threats? Is it preferable to exogenous behaviour modelling given emergency response priorities and the compromises both kinds of modelling require? And if so, why was exogenous behaviour modelling more heavily chosen during the COVID-19 pandemic? Answering these questions is crucial as BEID is striving to grow beyond theory, defining its practical role in optimised pandemic management. While endogenous modelling is conceptually relevant in capturing individual decision-making and key behavioural mechanisms, it may be difficult to be operationalised in real time due to the limited parsimony of some of the reviewed approaches and perceived reduced communicability to non-technical stakeholders[33]. Moreover, a gap persists between the BEID theoretical roots and the application-driven needs of decision-makers during crisis times. BEID is still a relatively young field, possibly not yet part of the standard toolbox for emerging infectious disease response in most research groups. Together with the need for emergency-driven promptness and for stronger interdisciplinary collaboration, these factors suggest that BEID was held back by virtue of being still a maturing field.

In this context, and to facilitate BEID maturation towards improved applications, recent advances in Artificial Intelligence (AI) may help address some of the long-standing challenges of the field. First, AI methods – e.g., graph neural networks[202] - may be particularly suitable to capture behavioural processes, often elusive both as inputs and outputs, especially in periods characterised by inherent instability such as emergencies. In principle, they could bridge exogenous and endogenous approaches, moving beyond this dichotomy. Approaches based on physics-informed neural networks[203] or universal differential equations[204] may be very valuable, since they avoid a pure "black box" approach by allowing the modeler to specify known mechanisms and thus meaningfully constrain the solution space. From a behavioural data perspective, AI could support the entire work streamline, enhancing data collection (especially in the identification of the relevant behavioural data to be collected), synthetic data creation[205,206], and data integration[202]. It could also be fruitful to navigate the emerging concept of "epidemic AI foundation models" (AIFM) within BEID. Paraphrasing Lau et al. [207], an even more ambitious and possibly highly rewarding objective would be to create a single AIFM pre-trained on pandemic responses and behavioural traces, that is able to describe known principles of coupled behaviour-epidemic dynamics and uncover other, still unknown, mechanisms. For instance, having a unified system may allow modellers to account for several distinct but interconnected "epi-behavioural" contexts at the same time, and to create a solid baseline for the long-standing issue of public policy evaluation. As epidemics are inherently behavioural, this possibility could greatly shape the future of epidemiology.

In light of our findings, four main priorities emerge for advancing BEID towards future pandemic preparedness and promptness[7]. First, from a theoretical perspective, the field must converge towards a readily-available modelling toolbox, reaching a shared but flexible compromise between modelling priorities – possibly starting from the commonalities we identified in this review. This requires balancing degree of detail/realism, parsimony, and interpretability, while providing clearer guidance on model selection. The toolbox should also be able to predict/reproduce relevant dynamics and to account for key sources of behavioural heterogeneities (e.g., socio-economic features), towards the creation of a benchmark for the evaluation of future modelling efforts. Second, from an empirical data perspective, we need to define a minimal set of desired data and suitable data collection protocols, particularly for behavioural data[197]. This should apply not only during epidemic episodes, but also (and mainly) in-between emergencies, to define a "behavioural baseline" for the practical assessment of risk-driven behaviour changes and to capture the root behavioural drivers of the emergency itself. Third, interdisciplinarity between epi-behavioural modelling and behavioural sciences should be strengthened, towards harmonisation of the BEID objectives and improvement of infectious disease public health practices[198,208]. In particular, future preparedness should prioritise a multi-disciplinary approach to pandemic response through the lenses of all behavioural sciences. Fourth, we must build capacity and familiarity with BEID in infectious disease modelling research groups at both universities and government agencies, so that lack of familiarity with the techniques is not an impediment to using the models when an infectious disease emergency arises. This will also help develop the application side of BEID modelling. Pursuing these objectives first requires an initial step of harmonisation efforts within and across fields, to bridge the distance between research and public health policy makers, e.g., standardising relevant BEID keywords (e.g., information, awareness, fear, distress, adherence) and agreeing on shared, applicable fundamental behavioural principles to be accounted for.

Ultimately, human behaviour remains one of the largest sources of uncertainty in epidemic dynamics. As such, the main challenge of dealing with emerging infectious diseases - including pandemic preparedness and management - could be to foresee it, e.g., detecting and tracking relevant behavioural signals over time[209]. However, notwithstanding the multiple behavioural theories available, we still lack (and may always lack) a unified understanding of human behaviour allowing us to control for all heterogeneity sources possibly affecting predictions and to decide which aspects of behaviour are "relevant". Therefore, towards preparedness, this suggests that the converse path could be pursued instead: rather than acting *ex post*, i.e., treating behaviour as a given unknown to be discovered, we could take an *ex ante* perspective, i.e., acting through education in-between emergency times, striving to shape *a priori* individuals' responses. This would allow to harmonise levels of awareness, risk perceptions, possible compliance, reactions (including fatigue) and needed incentives in the population, tuning our modelling expectations in turn. Moreover, the long-standing issue of policy evaluation could also be tackled through education, by laying the adequate groundwork to anticipate policy resistance or other unintended consequences of human behaviour[210], which are otherwise traditionally difficult to foresee[9]. The COVID-19 pandemic provided an unprecedented opportunity to advance this agenda; whether future epidemics will find the field better prepared depends on our ability to translate behavioural theory, behavioural data, and epidemiological modelling into a coherent scientific framework.

To conclude, we acknowledge some limitations of our work. The string scope and the inclusion of English-language studies only may have excluded relevant works. Given the heterogeneity in the COVID-19 research output, we may have missed papers using field-specific terminology mostly absent from our seed list, e.g., economic epidemiology papers. In parallel, we recognise that the volume of some taxonomy categories may be underestimated, also due to the greater use of preprints and grey literature by different disciplines during the pandemic period. Finally, the absence of a gold-standard quality assessment framework may limit the generalisability of the selected indicators, many of which are also strictly correlated to the articles' journal outlets.

## Methods

Our study followed the PRISMA guidelines for systematic reviews[48]. The PRISMA 2020 checklist is available in Supplementary Information S5. The detailed protocol for this study and the data extraction results are available in the OSF Protocol repository[59].

## Search Strategy

We scoped MEDLINE and Web of Science Core Collection through the Web of Science search platform (available indices[211]: Science Citation Index Expanded, Social Sciences Citation Index, Arts & Humanities Citation Index, Conference Proceedings Citation Index (from 1990)). We retrieved the papers identified by our search string (Table 1) on June 22nd, 2023, and updated the screening pool on June 27th, 2024.

*Table 1: Search string used on Web of Science search platform. Note: TS = title + abstract + authors' keywords + "keywords plus" (from Web of Science)*

| | Component | Search terms |
|---|---|---|
| 1. | **COVID-19** | COVID or SARS-CoV-2 OR nCoV |
| 2. | **Dynamic modelling of infectious disease spread and behavioural epidemiology** | "agent based model*" OR "compartmental model*" OR "computational model*" OR "disease* model*" OR "epidem* model*" OR "game theor* model*" OR "individual based model*" OR "infect* model*" OR "mathematical model*" OR "model-based" OR "network* model*" OR "stochastic model*" OR "transmission model*" OR "SEIR*" OR "SIR*" OR "SEAIR*" OR "transmission dynamic*" OR "dynamic* model*" OR "behavio* epi*" |
| 3. | **Human behavior** | ((protective OR preventive OR mitigat* OR human OR individual* OR *social OR population* OR *seeking OR factor* OR feedback OR chang* OR decision OR voluntary OR health OR respons*) NEAR/0 (behavioral OR behavioural OR behavior* OR behaviour*)) OR (*compliance OR adhere* OR adhesion OR perception* OR adoption OR awareness OR hesitancy OR willingness OR refusal OR ((strateg* OR learning OR feedback OR interaction*) AND game*) OR media)) |
| **Final string: TS = (1) AND TS = (2) AND TS = (3)** | | |

To further strengthen the search strategy, we also inspected the references of the final collected papers.

## Inclusion and exclusion criteria

We selected papers using dynamic transmission models of SARS-CoV-2 endogenously including a human behaviour component (Box 1). We did not impose temporal limitations on the publication time. We included papers written in English only. Exclusion criteria adopted:

- Non-peer-reviewed literature, grey literature, letters, comments, editorials, opinions, reviews, preprints, meta-analyses, proceeding papers, and book chapters;
- Articles with publication dates preceding the end of 2019;

- Papers that did not include a dynamic transmission model of SARS-CoV-2 (e.g., statistical/econometric analyses only, or models focusing on other infectious diseases during the COVID-19 pandemic);
- Papers using dynamic models of SARS-CoV-2 that either did not include human behaviour at all, or that included human behaviour exogenously along previous definitions (Box 1).

## Selection procedure

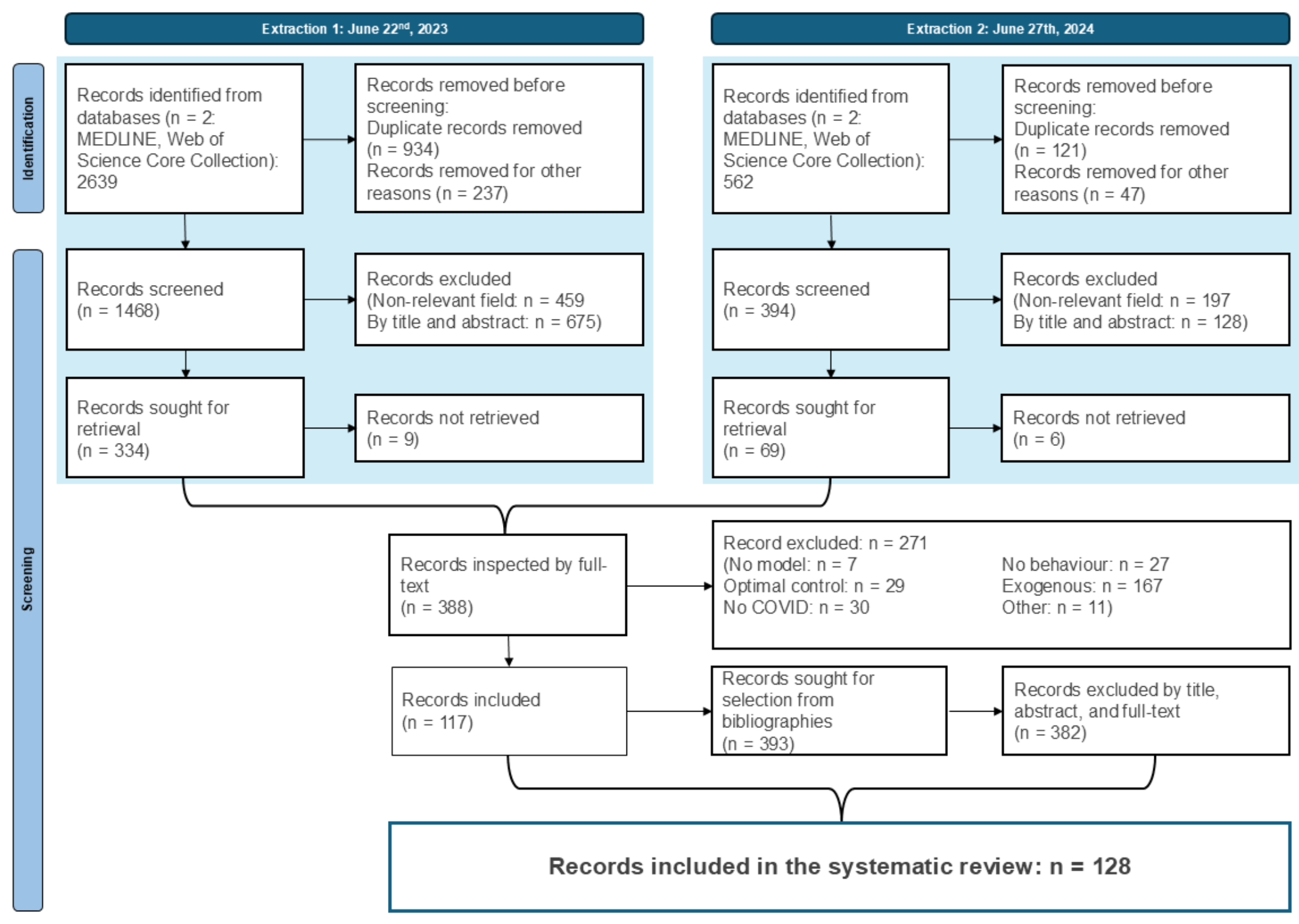

*Figure 3: Paper selection flowchart. The selection procedure has been performed on Excel (Microsoft® Excel® for Microsoft 365 MSO (Version 2503 Build 16.0.18623.20208) 64-bit).*

After retrieval from databases, deduplication and check of technical constraints (EDA), the selection procedure (Figure 3) included 1) selection by title (EDA) 2) selection by title and abstract (EDA + VO) and 3) selection by full-text (AM + EDA). In the first two steps, in doubtful cases, the reviewers adopted a conservative approach, promoting the paper of interest to the next selection step for further inspection. Disagreements were resolved by discussion. The full selection procedure is available in the OSF protocol folder[59].

## Data extraction and quality assessment

The data extraction table includes:

- Information on the paper: publication year, journal, study setting, and others.
- The characteristics of the adopted COVID-19 model: type and structure of the dynamic model

- How behaviour was endogenously incorporated: dichotomous behaviour modelling implicit/explicit, type of behaviour analysed (NPIs, e.g., social distancing, vaccination, response to testing), source of behaviour change (i.e., voluntary, mandatory, both), type of information inducing behaviour change, and others.
- The use of empirical data (e.g., data source, final aim, etc.).

The full data extraction table is available in the OSF protocol folder[59]. Data extraction has been performed by one reviewer (EDA), while the remaining authors checked for correctness.

We decided to proceed with a quality assessment appraisal rather than a risk of bias analysis, as the concept of "risk of bias" is not directly applicable in this context. As, to the best of our knowledge, no gold-standard quality assessment tools are available in the specific field of this review, we adapted the data extraction table used by a scoping review on modelling quality guidelines[212] to create a new tool aiming to capture the further complexity layers introduced by endogenous behaviour modelling (*see* OSF Protocol folder[59]). Quality assessment has been performed by one reviewer (EDA), while the remaining authors checked for correctness.

## Bibliography

1. Bavel, J. J. V. *et al.* Using social and behavioural science to support COVID-19 pandemic response. *Nat Hum Behav* **4**, 460–471 (2020).
2. Michie, S. & West, R. Sustained behavior change is key to preventing and tackling future pandemics. *Nat Med* **27**, 749–752 (2021).
3. Flaxman, S. *et al.* Estimating the effects of non-pharmaceutical interventions on COVID-19 in Europe. *Nature* **584**, 257–261 (2020).
4. Jit, M. *et al.* Reflections On Epidemiological Modeling To Inform Policy During The COVID-19 Pandemic In Western Europe, 2020–23. *Health Affairs* **42**, 1630–1636 (2023).
5. Hadley, L. *et al.* Challenges on the interaction of models and policy for pandemic control. *Epidemics* **37**, 100499 (2021).
6. Davis, R., Campbell, R., Hildon, Z., Hobbs, L. & Michie, S. Theories of behaviour and behaviour change across the social and behavioural sciences: a scoping review. *Health Psychol Rev* **9**, 323–344 (2015).
7. Bedson, J. *et al.* A review and agenda for integrated disease models including social and behavioural factors. *Nat Hum Behav* **5**, 834–846 (2021).
8. Perra, N. Non-pharmaceutical interventions during the COVID-19 pandemic: A review. *Phys Rep* **913**, 1–52 (2021).
9. Funk, S. *et al.* Nine challenges in incorporating the dynamics of behaviour in infectious diseases models. *Epidemics* **10**, 21–25 (2015).
10. Hill, E. M. *et al.* Integrating human behaviour and epidemiological modelling: unlocking the remaining challenges. *Mathematics in Medical and Life Sciences* **1**, 2429479 (2024).
11. Bauch, C. T. Imitation dynamics predict vaccinating behaviour. *Proc Biol Sci* **272**, 1669–1675 (2005).
12. Funk, S., Salathé, M. & Jansen, V. A. A. Modelling the influence of human behaviour on the spread of infectious diseases: a review. *Journal of The Royal Society Interface* **7**, 1247–1256 (2010).
13. Manfredi, P. & d'Onofrio, A. *Modeling the Interplay Between Human Behavior and the Spread of Infectious Diseases*. (Springer, New York, NY, 2013). doi:10.1007/978-1-4614-5474-8.
14. Wang, Z. *et al.* Statistical physics of vaccination. *Physics Reports* **664**, 1–113 (2016).
15. Verelst, F., Willem, L. & Beutels, P. Behavioural change models for infectious disease transmission: a systematic review (2010–2015). *J R Soc Interface* **13**, 20160820 (2016).
16. Bauch, C. T. & Galvani, A. P. Social Factors in Epidemiology. *Science* **342**, 47–49 (2013).
17. Bauch, C. T. & Earn, D. J. D. Vaccination and the theory of games. *Proceedings of the National Academy of Sciences* **101**, 13391–13394 (2004).
18. d'Onofrio, A. & Manfredi, P. Information-related changes in contact patterns may trigger oscillations in the endemic prevalence of infectious diseases. *Journal of Theoretical Biology* **256**, 473–478 (2009).
19. Fenichel, E. P. *et al.* Adaptive human behavior in epidemiological models. *Proceedings of the National Academy of Sciences* **108**, 6306–6311 (2011).
20. Funk, S., Gilad, E., Watkins, C. & Jansen, V. A. A. The spread of awareness and its impact on epidemic outbreaks. *Proceedings of the National Academy of Sciences* **106**, 6872–6877 (2009).

21. d'Onofrio, A., Manfredi, P. & Salinelli, E. Vaccinating behaviour, information, and the dynamics of SIR vaccine preventable diseases. *Theor Popul Biol* **71**, 301–317 (2007).
22. Perra, N., Balcan, D., Gonçalves, B. & Vespignani, A. Towards a Characterization of Behavior-Disease Models. *PLOS ONE* **6**, e23084 (2011).
23. Reluga, T. C. Game Theory of Social Distancing in Response to an Epidemic. *PLOS Computational Biology* **6**, e1000793 (2010).
24. Chen, F. & Toxvaerd, F. The economics of vaccination. *Journal of Theoretical Biology* **363**, 105–117 (2014).
25. Keeling, M. & Rohani, P. *Modeling Infectious Diseases in Humans and Animals*. (2007).
26. Anderson, R. M. & May, R. M. *Infectious Diseases of Humans: Dynamics and Control*. (Oxford University Press, 1991). doi:10.1093/oso/9780198545996.001.0001.
27. Bauch, C., d'Onofrio, A. & Manfredi, P. Behavioral Epidemiology of Infectious Diseases: An Overview. *Modeling the Interplay Between Human Behavior and the Spread of Infectious Diseases* 1–19 (2012) doi:10.1007/978-1-4614-5474-8_1.
28. Colosi, E. *et al.* Screening and vaccination against COVID-19 to minimise school closure: a modelling study. *The Lancet Infectious Diseases* **22**, 977–989 (2022).
29. Di Domenico, L. *et al.* Adherence and sustainability of interventions informing optimal control against the COVID-19 pandemic. *Commun Med* **1**, 1–13 (2021).
30. Hale, T. *et al.* Variation in government responses to COVID-19. http://www.bsg.ox.ac.uk/research/publications/variation-government-responses-covid-19 (2023).
31. Eker, S. Validity and usefulness of COVID-19 models. *Humanit Soc Sci Commun* **7**, 54 (2020).
32. Gozzi, N., Perra, N. & Vespignani, A. Comparative evaluation of behavioral epidemic models using COVID-19 data. *Proc Natl Acad Sci U S A* **122**, e2421993122 (2025).
33. Hadley, L., Rich, C., Tasker, A., Restif, O. & Funk, S. How does policy modelling work in practice? A global analysis on the use of epidemiological modelling in health crises. *PLOS Global Public Health* **5**, e0004675 (2025).
34. Hu, T. *et al.* Human mobility data in the COVID-19 pandemic: characteristics, applications, and challenges. *International Journal of Digital Earth* **14**, 1126–1147 (2021).
35. Johns Hopkins Center for Communication Programs. COVID Behaviors Dashboard. (2021).
36. Atchison, C. *et al.* Early perceptions and behavioural responses during the COVID-19 pandemic: a cross-sectional survey of UK adults. https://doi.org/10.1136/bmjopen-2020-043577 (2021) doi:10.1136/bmjopen-2020-043577.
37. Verelst, F. *et al.* SOCRATES-CoMix: a platform for timely and open-source contact mixing data during and in between COVID-19 surges and interventions in over 20 European countries. *BMC Medicine* **19**, 254 (2021).
38. Ferretti, L. *et al.* Quantifying SARS-CoV-2 transmission suggests epidemic control with digital contact tracing. *Science* **368**, eabb6936 (2020).
39. Dimitrov, D. *et al.* TweetsCOV19 -- A Knowledge Base of Semantically Annotated Tweets about the COVID-19 Pandemic. in *Proceedings of the 29th*

*ACM International Conference on Information & Knowledge Management* 2991–2998 (2020). doi:10.1145/3340531.3412765.
40. Zarei, K., Farahbakhsh, R., Crespi, N. & Tyson, G. A First Instagram Dataset on COVID-19. Preprint at https://doi.org/10.48550/arXiv.2004.12226 (2020).
41. Cinelli, M. *et al.* The COVID-19 social media infodemic. *Sci Rep* **10**, 16598 (2020).
42. Callaway, E., Cyranoski, D., Mallapaty, S., Stoye, E. & Tollefson, J. The coronavirus pandemic in five powerful charts. *Nature* **579**, 482–483 (2020).
43. Teixeira da Silva, J. A., Tsigaris, P. & Erfanmanesh, M. Publishing volumes in major databases related to Covid-19. *Scientometrics* **126**, 831–842 (2021).
44. Whitaker, M. *et al.* How COVID-19 affected academic publishing: a 3-year study of 17 million research papers. *International Journal of Epidemiology* **54**, dyaf058 (2025).
45. Li, G. *et al.* Surging publications on the COVID-19 pandemic. *Clinical Microbiology and Infection* **27**, 484–486 (2021).
46. Franco, N. *et al.* Inferring age-specific differences in susceptibility to and infectiousness upon SARS-CoV-2 infection based on Belgian social contact data. *PLOS Computational Biology* **18**, e1009965 (2022).
47. Keeling, M. J., Dyson, L., Tildesley, M. J., Hill, E. M. & Moore, S. Comparison of the 2021 COVID-19 roadmap projections against public health data in England. *Nat Commun* **13**, 4924 (2022).
48. Page, M. J. *et al.* The PRISMA 2020 statement: an updated guideline for reporting systematic reviews. *BMJ* **372**, n71 (2021).
49. Weston, D., Hauck, K. & Amlôt, R. Infection prevention behaviour and infectious disease modelling: a review of the literature and recommendations for the future. *BMC Public Health* **18**, 336 (2018).
50. Estrada, E. COVID-19 and SARS-CoV-2. Modeling the present, looking at the future. *Physics Reports* **869**, 1–51 (2020).
51. McAdams, D. The Blossoming of Economic Epidemiology. SSRN Scholarly Paper at https://doi.org/10.1146/annurev-economics-082120-122900 (2021).
52. Rahmandad, H., Xu, R. & Ghaffarzadegan, N. Enhancing long-term forecasting: Learning from COVID-19 models. *PLOS Computational Biology* **18**, e1010100 (2022).
53. Reitenbach, A. *et al.* Coupled infectious disease and behavior dynamics. A review of model assumptions. *Rep Prog Phys* https://doi.org/10.1088/1361-6633/ad90ef (2024) doi:10.1088/1361-6633/ad90ef.
54. Li, T. & Xiao, Y. Co-Evolution of Behavior Change and Infectious Disease Transmission Dynamics: A Modelling Review. *CSIAM Transactions on Life Sciences* https://doi.org/10.4208/csiam-ls.SO-2025-0005 (2025) doi:10.4208/csiam-ls.SO-2025-0005.
55. LeJeune, L., Ghaffarzadegan, N., Childs, L. M. & Saucedo, O. Formulating human risk response in epidemic models: Exogenous vs endogenous approaches. *European Journal of Operational Research* **324**, 246–258 (2025).
56. Bagaforo, R. J., Dupas, M.-C., Abrams, S., Dellicour, S. & Hens, N. A scoping review of COVID-19 modelling studies in Belgium 2020-2024: incorporation of behaviour and lessons learned. *Arch Public Health* https://doi.org/10.1186/s13690-026-01959-3 (2026) doi:10.1186/s13690-026-01959-3.

57. Hamilton, A. *et al.* Incorporating endogenous human behavior in models of COVID-19 transmission: A systematic scoping review. *Dialogues in Health* **4**, 100179 (2024).
58. Jo, Y., Sharbayta, S. S. & Buonomo, B. Behavioral change models for infectious disease transmission: a systematic review (2020-2025). Preprint at https://doi.org/10.48550/arXiv.2602.16633 (2026).
59. D'Agnese, E. *et al. OSF Protocol - Advancing Behavioural Epidemiological Modelling: A Systematic Review of Endogenous Human Behaviour in Disease Spread*. https://archive.org/details/osf-registrations-gbn97-v1 (2025) doi:10.17605/OSF.IO/GBN97.
60. Alfani, G. & Melegaro, A. *Pandemie d'Italia - Dalla Peste Nera All'influenza Suina: L'impatto Sulla Società*. (Egea, 2010).
61. Capasso, V. & Serio, G. A generalization of the Kermack-McKendrick deterministic epidemic model. *Mathematical Biosciences* **42**, 43–61 (1978).
62. Kremer, M. Integrating Behavioral Choice into Epidemiological Models of AIDS*. *Q J Econ* **111**, 549–573 (1996).
63. Hadeler, K. P. & Castillo-Chavez, C. A core group model for disease transmission. *Mathematical Biosciences* **128**, 41–55 (1995).
64. Geoffard, P.-Y. & Philipson, T. Disease Eradication: Private versus Public Vaccination. *The American Economic Review* **87**, 222–230 (1997).
65. Cui, J., Sun, Y. & Zhu, H. The Impact of Media on the Control of Infectious Diseases. *J Dyn Diff Equat* **20**, 31–53 (2008).
66. Eksin, C., Paarporn, K. & Weitz, J. S. Systematic biases in disease forecasting – The role of behavior change. *Epidemics* **27**, 96–105 (2019).
67. Misra, A. K., Sharma, A. & Shukla, J. B. Modeling and analysis of effects of awareness programs by media on the spread of infectious diseases. *Mathematical and Computer Modelling* **53**, 1221–1228 (2011).
68. d'Onofrio, A., Manfredi, P. & Poletti, P. The Interplay of Public Intervention and Private Choices in Determining the Outcome of Vaccination Programmes. *PLOS ONE* **7**, e45653 (2012).
69. A Mohsen, A., AL-Husseiny, H. F., Zhou, X. & Hattaf, K. Global stability of COVID-19 model involving the quarantine strategy and media coverage effects. *AIMS Public Health* **7**, 587–605 (2020).
70. Agusto, F. B. *et al.* Impact of public sentiments on the transmission of COVID-19 across a geographical gradient. *PeerJ* **11**, e14736 (2023).
71. Ajbar, A., Alqahtani, R. T. & Boumaza, M. Dynamics of an SIR-Based COVID-19 Model With Linear Incidence Rate, Nonlinear Removal Rate, and Public Awareness. *Front. Phys.* **9**, (2021).
72. Ajbar, A., Alqahtani, R. T. & Boumaza, M. Dynamics of a COVID-19 Model with a Nonlinear Incidence Rate, Quarantine, Media Effects, and Number of Hospital Beds. *Symmetry* **13**, 947 (2021).
73. Al-arydah, M. Mathematical modeling and optimal control for COVID-19 with population behavior. *Mathematical Methods in the Applied Sciences* **46**, 19184–19198 (2023).
74. Barth, J., Li, S., Aprahamian, H. & Gupta, D. Spatiotemporal vaccine allocation policies for epidemics with behavioral feedback dynamics. *Naval Research Logistics (NRL)* **71**, 109–139 (2024).
75. Bozkurt, F., Yousef, A., Abdeljawad, T., Kalinli, A. & Mdallal, Q. A. A fractional-order model of COVID-19 considering the fear effect of the media and social networks on the community. *Chaos Solitons Fractals* **152**, 111403 (2021).

76. Bulai, I. M., Montefusco, F. & Pedersen, M. G. Stability analysis of a model of epidemic dynamics with nonlinear feedback producing recurrent infection waves. *Applied Mathematics Letters* **136**, 108455 (2023).
77. Buonomo, B. Effects of information-dependent vaccination behavior on coronavirus outbreak: insights from a SIRI model. *Ricerche mat* **69**, 483–499 (2020).
78. Buonomo, B., Della Marca, R., d'Onofrio, A. & Groppi, M. A behavioural modelling approach to assess the impact of COVID-19 vaccine hesitancy. *Journal of Theoretical Biology* **534**, 110973 (2022).
79. Buonomo, B., Della Marca, R. & Sharbayta, S. S. A behavioral change model to assess vaccination-induced relaxation of social distancing during an epidemic. *J. Biol. Syst.* **30**, 1–25 (2022).
80. Buonomo, B. & Della Marca, R. Effects of information-induced behavioural changes during the COVID-19 lockdowns: the case of Italy. *Royal Society Open Science* **7**, 201635 (2020).
81. Burridge, J. & Gnacik, M. Public efforts to reduce disease transmission implied from a spatial game. *Physica A: Statistical Mechanics and its Applications* **589**, 126619 (2022).
82. Cantin, G., Silva, C. J. & Banos, A. Mathematical analysis of a hybrid model: Impacts of individual behaviors on the spreading of an epidemic. *NHM* **17**, 333–357 (2022).
83. Chen, T., Li, Z. & Zhang, G. Analysis of a COVID-19 model with media coverage and limited resources. *Math Biosci Eng* **21**, 5283–5307 (2024).
84. Córdova-Lepe, F. & Vogt-Geisse, K. Adding a reaction-restoration type transmission rate dynamic-law to the basic SEIR COVID-19 model. *PLOS ONE* **17**, e0269843 (2022).
85. Deng, J., Tang, S. & Shu, H. Joint impacts of media, vaccination and treatment on an epidemic Filippov model with application to COVID-19. *Journal of Theoretical Biology* **523**, 110698 (2021).
86. Ding, Y., Fu, Y. & Kang, Y. Stochastic analysis of COVID-19 by a SEIR model with Lévy noise. *Chaos: An Interdisciplinary Journal of Nonlinear Science* **31**, 043132 (2021).
87. Dönges, P. *et al.* Interplay Between Risk Perception, Behavior, and COVID-19 Spread. *Front. Phys.* **10**, (2022).
88. Fierro, A., Romano, S. & Liccardo, A. Vaccination and variants: Retrospective model for the evolution of Covid-19 in Italy. *PLOS ONE* **17**, e0265159 (2022).
89. Fosch, A., Aleta, A. & Moreno, Y. Characterizing the role of human behavior in the effectiveness of contact-tracing applications. *Front. Public Health* **11**, (2023).
90. Galanis, G., Di Guilmi, C., Bennett, D. L. & Baskozos, G. The effectiveness of Non-pharmaceutical interventions in reducing the COVID-19 contagion in the UK, an observational and modelling study. *PLoS One* **16**, e0260364 (2021).
91. Gatti, N. & Retali, B. Saving lives during the COVID-19 pandemic: the benefits of the first Swiss lockdown. *Swiss J Econ Stat* **157**, 4 (2021).
92. Gutiérrez-Jara, J. P., Vogt-Geisse, K., Cabrera, M., Córdova-Lepe, F. & Muñoz-Quezada, M. T. Effects of human mobility and behavior on disease transmission in a COVID-19 mathematical model. *Sci Rep* **12**, 10840 (2022).
93. Karatayev, V. A., Anand, M. & Bauch, C. T. Local lockdowns outperform global lockdown on the far side of the COVID-19 epidemic curve. *Proceedings of the National Academy of Sciences* **117**, 24575–24580 (2020).

94. LaJoie, Z., Usherwood, T., Sampath, S. & Srivastava, V. A COVID-19 model incorporating variants, vaccination, waning immunity, and population behavior. *Sci Rep* **12**, 20377 (2022).
95. Li, Z. *et al.* Adaptive behaviors and vaccination on curbing COVID-19 transmission: Modeling simulations in eight countries. *Journal of Theoretical Biology* **559**, 111379 (2023).
96. Liu, P.-Y., He, S., Rong, L.-B. & Tang, S.-Y. The effect of control measures on COVID-19 transmission in Italy: Comparison with Guangdong province in China. *Infectious Diseases of Poverty* **9**, 130 (2020).
97. Lobinska, G., Pauzner, A., Traulsen, A., Pilpel, Y. & Nowak, M. A. Evolution of resistance to COVID-19 vaccination with dynamic social distancing. *Nat Hum Behav* **6**, 193–206 (2022).
98. Mahadhika, C. K. & Aldila, D. A deterministic transmission model for analytics-driven optimization of COVID-19 post-pandemic vaccination and quarantine strategies. *Math Biosci Eng* **21**, 4956–4988 (2024).
99. Maji, C. Impact of Media-Induced Fear on the Control of COVID-19 Outbreak: A Mathematical Study. *International Journal of Differential Equations* **2021**, 2129490 (2021).
100. Maji, C. *et al.* COVID-19 propagation and the usefulness of awareness-based control measures: A mathematical model with delay. *MATH* **7**, 12091–12105 (2022).
101. Mekonen, K. G., Habtemicheal, T. G. & Balcha, S. F. Modeling the effect of contaminated objects for the transmission dynamics of COVID-19 pandemic with self protection behavior changes. *Results in Applied Mathematics* **9**, 100134 (2021).
102. Park, S.-N., Kim, H.-H. & Lee, K. B. Inherently high uncertainty in predicting the time evolution of epidemics. *Epidemiol Health* **43**, e2021014 (2021).
103. Proaño, C. R., Kukacka, J. & Makarewicz, T. Belief-driven dynamics in a behavioral SEIRD macroeconomic model with sceptics. *Journal of Economic Behavior & Organization* **217**, 312–333 (2024).
104. Shastry, V., Reeves, D. C., Willems, N. & Rai, V. Policy and behavioral response to shock events: An agent-based model of the effectiveness and equity of policy design features. *PLOS ONE* **17**, e0262172 (2022).
105. Silal, S. P. *et al.* The National COVID-19 Epi Model (NCEM): Estimating cases, admissions and deaths for the first wave of COVID-19 in South Africa. *PLOS Global Public Health* **3**, e0001070 (2023).
106. Simeonov, O. & Eaton, C. D. Modeling the drivers of oscillations in COVID-19 data on college campuses. *Annals of Epidemiology* **82**, 40–44 (2023).
107. Taki, R., El Fatini, M., El Khalifi, M., Lakhal, M. & Wang, K. Understanding death risks of Covid-19 under media awareness strategy: a stochastic approach. *J Anal* **30**, 79–99 (2022).
108. Thirthar, A. A. *et al.* Mathematical modeling of the COVID-19 epidemic with fear impact. *MATH* **8**, 6447–6465 (2023).
109. Tu, Y. & Meng, X. A reaction–diffusion epidemic model with virus mutation and media coverage: Theoretical analysis and numerical simulation. *Mathematics and Computers in Simulation* **214**, 28–67 (2023).
110. Usherwood, T., LaJoie, Z. & Srivastava, V. A model and predictions for COVID-19 considering population behavior and vaccination. *Sci Rep* **11**, 12051 (2021).
111. Wang, X. Studying social awareness of physical distancing in mitigating COVID-19 transmission. *Math Biosci Eng* **17**, 7428–7441 (2020).

112. Warne, D. J. *et al.* Hindsight is 2020 vision: a characterisation of the global response to the COVID-19 pandemic. *BMC Public Health* **20**, 1868 (2020).
113. Weitz, J. S., Park, S. W., Eksin, C. & Dushoff, J. Awareness-driven behavior changes can shift the shape of epidemics away from peaks and toward plateaus, shoulders, and oscillations. *Proceedings of the National Academy of Sciences* **117**, 32764–32771 (2020).
114. Witbooi, P. J., Vyambwera, S. M. & Nsuami, M. U. Control and elimination in an SEIR model for the disease dynamics of COVID-19 with vaccination. *AIMS Mathematics* **8**, 8144–8161 (2023).
115. Yousef. A fractional-order model of COVID-19 with a strong Allee effect considering the fear effect spread by social networks to the community and the existence of the silent spreaders during the pandemic stage. *AIMS Mathematics* **7**, 10052–10078 (2022).
116. Zhou, Y. *et al.* Transmission trends of the global COVID-19 pandemic with combined effects of adaptive behaviours and vaccination. *Epidemiol Infect* **151**, e39 (2023).
117. Ajbar, A. & Alqahtani, R. T. Bifurcation analysis of a SEIR epidemic system with governmental action and individual reaction. *Advances in Difference Equations* **2020**, 541 (2020).
118. Angulo, W. *et al.* A modified SEIR model to predict the behavior of the early stage in coronavirus and coronavirus-like outbreaks. *Sci Rep* **11**, 16331 (2021).
119. Danane, J., Hammouch, Z., Allali, K., Rashid, S. & Singh, J. A fractional-order model of coronavirus disease 2019 (COVID-19) with governmental action and individual reaction. *Mathematical Methods in the Applied Sciences* **46**, 8275–8288 (2023).
120. Das, M., Samanta, G. & De la Sen, M. A Fractional Order Model to Study the Effectiveness of Government Measures and Public Behaviours in COVID-19 Pandemic. *Mathematics* **10**, 3020 (2022).
121. Furati, K. M., Sarumi, I. O. & Khaliq, A. Q. M. Fractional model for the spread of COVID-19 subject to government intervention and public perception. *Applied Mathematical Modelling* **95**, 89–105 (2021).
122. Kwuimy, C. A. K., Nazari, F., Jiao, X., Rohani, P. & Nataraj, C. Nonlinear dynamic analysis of an epidemiological model for COVID-19 including public behavior and government action. *Nonlinear Dyn* **101**, 1545–1559 (2020).
123. Lin, Q. *et al.* A conceptual model for the coronavirus disease 2019 (COVID-19) outbreak in Wuhan, China with individual reaction and governmental action. *International Journal of Infectious Diseases* **93**, 211–216 (2020).
124. Mushanyu, J., Chazuka, Z., Mudzingwa, F. & Ogbogbo, C. Modelling the impact of detection on COVID-19 transmission dynamics in Ghana. *RMS: Research in Mathematics & Statistics* **8**, 1953722 (2021).
125. Mushayabasa, S., Ngarakana-Gwasira, E. T. & Mushanyu, J. On the role of governmental action and individual reaction on COVID-19 dynamics in South Africa: A mathematical modelling study. *Informatics in Medicine Unlocked* **20**, 100387 (2020).
126. Öz, Y. A Theoretical Model to Investigate the Influence of Temperature, Reactions of the Population and the Government on the COVID-19 Outbreak in Turkey. *Disaster Med Public Health Prep* 1–9 (2022) doi:10.1017/dmp.2020.322.
127. Adeniyi, M. O. *et al.* Dynamic model of COVID-19 disease with exploratory data analysis. *Scientific African* **9**, e00477 (2020).

128. Agaba, G. O. & Soomiyol, M. C. Analysing the Spread of COVID-19 using Delay Epidemic Model with Awareness. *IOSR-JM* **16**, 52–59 (2020).
129. Baba, I. & Baleanu, D. Awareness as the Most Effective Measure to Mitigate the Spread of COVID-19 in Nigeria. *CMC* **65**, 1945–1957 (2020).
130. Chang, X., Wang, J., Liu, M., Jin, Z. & Han, D. Study on an SIHRS Model of COVID-19 Pandemic With Impulse and Time Delay Under Media Coverage. *IEEE Access* **9**, 49387–49397 (2021).
131. Deng, Y., He, D. & Zhao, Y. The impacts of anti-protective awareness and protective awareness programs on COVID-19 outbreaks. *Chaos, Solitons & Fractals* **180**, 114493 (2024).
132. Guo, J. *et al.* Discrete epidemic modelling of COVID-19 transmission in Shaanxi Province with media reporting and imported cases. *MBE* **19**, 1388–1410 (2022).
133. Gutiérrez-Jara, J. P. & Saracini, C. Risk Perception Influence on Vaccination Program on COVID-19 in Chile: A Mathematical Model. *International Journal of Environmental Research and Public Health* **19**, 2022 (2022).
134. Johnston, M. D. & Pell, B. A dynamical framework for modeling fear of infection and frustration with social distancing in COVID-19 spread. *Math Biosci Eng* **17**, 7892–7915 (2020).
135. Lacitignola, D. & Saccomandi, G. Managing awareness can avoid hysteresis in disease spread: an application to coronavirus Covid-19. *Chaos, Solitons & Fractals* **144**, 110739 (2021).
136. Li, T. & Xiao, Y. Linking the disease transmission to information dissemination dynamics: An insight from a multi-scale model study. *Journal of Theoretical Biology* **526**, 110796 (2021).
137. Li, T. & Xiao, Y. Complex dynamics of an epidemic model with saturated media coverage and recovery. *Nonlinear Dyn* **107**, 2995–3023 (2022).
138. Liu, X., Lv, Z. & Ding, Y. Mathematical modeling and stability analysis of the time-delayed SAIM model for COVID-19 vaccination and media coverage. *Math Biosci Eng* **19**, 6296–6316 (2022).
139. Nyabadza, F., Mushanyu, J., Mbogo, R. & Muchatibaya, G. Modelling the Influence of Dynamic Social Processes on COVID-19 Infection Dynamics. *Mathematics* **11**, 963 (2023).
140. Pal, K. K., Sk, N., Rai, R. K. & Tiwari, P. K. Examining the impact of incentives and vaccination on COVID-19 control in India: addressing environmental contamination and seasonal dynamics. *Eur. Phys. J. Plus* **139**, 225 (2024).
141. Rai, R. K., Khajanchi, S., Tiwari, P. K., Venturino, E. & Misra, A. K. Impact of social media advertisements on the transmission dynamics of COVID-19 pandemic in India. *J. Appl. Math. Comput.* **68**, 19–44 (2022).
142. Sardar, T., Nadim, S. S. & Rana, S. Detection of multiple waves for COVID-19 and its optimal control through media awareness and vaccination: study based on some Indian states. *Nonlinear Dyn* **111**, 1903–1920 (2023).
143. Sk, T., Biswas, S. & Sardar, T. The impact of a power law-induced memory effect on the SARS-CoV-2 transmission. *Chaos, Solitons & Fractals* **165**, 112790 (2022).
144. Tiwari, P. K., Rai, R. K., Khajanchi, S., Gupta, R. K. & Misra, A. K. Dynamics of coronavirus pandemic: effects of community awareness and global information campaigns. *Eur. Phys. J. Plus* **136**, 994 (2021).
145. Tripathi, A., Tripathi, R. N. & Sharma, D. A mathematical model to study the COVID-19 pandemic in India. *Model. Earth Syst. Environ.* **8**, 3047–3058 (2022).

146. Wang, Y., Qing, F., Li, H. & Wang, X. Timely and effective media coverage's role in the spread of Corona Virus Disease 2019. *Mathematical Methods in the Applied Sciences* **47**, 3490–3506 (2022).
147. Zhou, W. *et al.* Effects of media reporting on mitigating spread of COVID-19 in the early phase of the outbreak. *MBE* **17**, 2693–2707 (2020).
148. Zuo, C., Zhu, F. & Ling, Y. Analyzing COVID-19 Vaccination Behavior Using an SEIRM/V Epidemic Model With Awareness Decay. *Front. Public Health* **10**, (2022).
149. Abbas, W., A, M. M., Park, A., Parveen, S. & Kim, S. Evolution and consequences of individual responses during the COVID-19 outbreak. *PLOS ONE* **17**, e0273964 (2022).
150. Althobaity, Y., Wu, J. & Tildesley, M. J. Non-pharmaceutical interventions and their relevance in the COVID-19 vaccine rollout in Saudi Arabia and Arab Gulf countries. *Infectious Disease Modelling* **7**, 545–560 (2022).
151. Gao, S., Dai, X., Wang, L., Perra, N. & Wang, Z. Epidemic Spreading in Metapopulation Networks Coupled With Awareness Propagation. *IEEE Transactions on Cybernetics* **53**, 7686–7698 (2023).
152. He, L. & Zhu, L. Modeling the COVID-19 epidemic and awareness diffusion on multiplex networks. *Commun. Theor. Phys.* **73**, 035002 (2021).
153. Kim, S., Ko, Y., Kim, Y.-J. & Jung, E. The impact of social distancing and public behavior changes on COVID-19 transmission dynamics in the Republic of Korea. *PLOS ONE* **15**, e0238684 (2020).
154. Mendoza, V. M. P. *et al.* Managing bed capacity and timing of interventions: a COVID-19 model considering behavior and underreporting. *MATH* **8**, 2201–2225 (2023).
155. Teslya, A. *et al.* Impact of self-imposed prevention measures and short-term government-imposed social distancing on mitigating and delaying a COVID-19 epidemic: A modelling study. *PLOS Medicine* **17**, e1003166 (2020).
156. Teslya, A. *et al.* The importance of sustained compliance with physical distancing during COVID-19 vaccination rollout. *Commun Med (Lond)* **2**, 146 (2022).
157. Zhao, X. *et al.* The impact of awareness diffusion on the spread of COVID-19 based on a two-layer SEIR/V–UA epidemic model. *Journal of Medical Virology* **93**, 4342–4350 (2021).
158. Lux, T. The social dynamics of COVID-19. *Physica A: Statistical Mechanics and its Applications* **567**, 125710 (2021).
159. Molla, J., Farhang-Sardroodi, S., Moyles, I. R. & Heffernan, J. M. Pharmaceutical and non-pharmaceutical interventions for controlling the COVID-19 pandemic. *Royal Society Open Science* **10**, 230621 (2023).
160. Moyles, I. R., Heffernan, J. M. & Kong, J. D. Cost and social distancing dynamics in a mathematical model of COVID-19 with application to Ontario, Canada. *Royal Society Open Science* **8**, 201770 (2021).
161. Pant, B., Safdar, S., Santillana, M. & Gumel, A. B. Mathematical Assessment of the Role of Human Behavior Changes on SARS-CoV-2 Transmission Dynamics in the United States. *Bull Math Biol* **86**, 92 (2024).
162. Retzlaff, C. O. *et al.* Fear, Behaviour, and the COVID-19 Pandemic: A City-Scale Agent-Based Model Using Socio-Demographic and Spatial Map Data. *Journal of Artificial Societies and Social Simulation* **25**, 1–3 (2022).

163. Worby, C. J. & Chang, H.-H. Face mask use in the general population and optimal resource allocation during the COVID-19 pandemic. *Nat Commun* **11**, 4049 (2020).
164. Yao, S., Fan, N. & Hu, J. Modeling the spread of infectious diseases through influence maximization. *Optim Lett* **16**, 1563–1586 (2022).
165. Yedomonhan, E. *et al.* Modeling the effects of Prophylactic behaviors on the spread of SARS-CoV-2 in West Africa. *MBE* **20**, 12955–12989 (2023).
166. Agusto, F. B. *et al.* To isolate or not to isolate: the impact of changing behavior on COVID-19 transmission. *BMC Public Health* **22**, 138 (2022).
167. Amaral, M. A., Oliveira, M. M. de & Javarone, M. A. An epidemiological model with voluntary quarantine strategies governed by evolutionary game dynamics. *Chaos, Solitons & Fractals* **143**, 110616 (2021).
168. Ge, J. & Wang, W. Vaccination games in prevention of infectious diseases with application to COVID-19. *Chaos, Solitons & Fractals* **161**, 112294 (2022).
169. He, S., Zhou, W., Wang, X. & Tang, S. Modelling and analysis of the cross-impact of age heterogeneity and behavioural changes on the evolution of disease transmission. *Comp. Appl. Math.* **43**, 142 (2024).
170. Jentsch, P. C., Anand, M. & Bauch, C. T. Prioritising COVID-19 vaccination in changing social and epidemiological landscapes: a mathematical modelling study. *The Lancet Infectious Diseases* **21**, 1097–1106 (2021).
171. Jia, S., Meng, X. & Zhang, T. The effectiveness of human interventions against COVID-19 based on evolutionary game theory. *Journal of Applied Analysis and Computation* **12**, 1748–1762 (2022).
172. Jing, S., Milne, R., Wang, H. & Xue, L. Vaccine hesitancy promotes emergence of new SARS-CoV-2 variants. *J Theor Biol* **570**, 111522 (2023).
173. Kabir, K. A., Chowdhury, A. & Tanimoto, J. An evolutionary game modeling to assess the effect of border enforcement measures and socio-economic cost: Export-importation epidemic dynamics. *Chaos, Solitons & Fractals* **146**, 110918 (2021).
174. Kabir, K. M. A. & Tanimoto, J. Evolutionary game theory modelling to represent the behavioural dynamics of economic shutdowns and shield immunity in the COVID-19 pandemic. *Royal Society Open Science* **7**, 201095 (2020).
175. Li, T., Xiao, Y. & Heffernan, J. Linking Spontaneous Behavioral Changes to Disease Transmission Dynamics: Behavior Change Includes Periodic Oscillation. *Bull Math Biol* **86**, 73 (2024).
176. Liu, S., Zhao, Y. & Zhu, Q. Herd Behaviors in Epidemics: A Dynamics-Coupled Evolutionary Games Approach. *Dyn Games Appl* **12**, 183–213 (2022).
177. Madeo, D. & Mocenni, C. Identification and Control of Game-Based Epidemic Models. *Games* **13**, 10 (2022).
178. Martcheva, M., Tuncer ,Necibe & and Ngonghala, C. N. Effects of social-distancing on infectious disease dynamics: an evolutionary game theory and economic perspective. *Journal of Biological Dynamics* **15**, 342–366 (2021).
179. Meng, X. *et al.* Coupled disease-vaccination behavior dynamic analysis and its application in COVID-19 pandemic. *Chaos, Solitons & Fractals* **169**, 113294 (2023).
180. Ngonghala, C. N., Goel, P., Kutor, D. & Bhattacharyya, S. Human choice to self-isolate in the face of the COVID-19 pandemic: A game dynamic modelling approach. *J Theor Biol* **521**, 110692 (2021).

181. Ovi, M. A., Nabi, K. N. & Kabir, K. M. A. Social distancing as a public-good dilemma for socio-economic cost: An evolutionary game approach. *Heliyon* **8**, e11497 (2022).
182. Pedro, S. A. *et al.* Conditions for a Second Wave of COVID-19 Due to Interactions Between Disease Dynamics and Social Processes. *Front. Phys.* **8**, (2020).
183. Steinegger, B., Arola-Fernández, L., Granell, C., Gómez-Gardeñes, J. & Arenas, A. Behavioural response to heterogeneous severity of COVID-19 explains temporal variation of cases among different age groups. *Philosophical Transactions of the Royal Society A: Mathematical, Physical and Engineering Sciences* **380**, 20210119 (2021).
184. Tang, B., Zhou, W., Wang, X., Wu, H. & Xiao, Y. Controlling Multiple COVID-19 Epidemic Waves: An Insight from a Multi-scale Model Linking the Behaviour Change Dynamics to the Disease Transmission Dynamics. *Bull Math Biol* **84**, 106 (2022).
185. Vivekanandhan, G. *et al.* Investigation of vaccination game approach in spreading covid-19 epidemic model with considering the birth and death rates. *Chaos, Solitons & Fractals* **163**, 112565 (2022).
186. Zhao, S. *et al.* Imitation dynamics in the mitigation of the novel coronavirus disease (COVID-19) outbreak in Wuhan, China from 2019 to 2020. *Ann Transl Med* **8**, 448 (2020).
187. Baril-Tremblay, D., Marlats, C. & Ménager, L. Self-isolation. *J Math Econ* **93**, 102483 (2021).
188. Chan, T.-L., Yuan, H.-Y. & Lo, W.-C. Modeling COVID-19 Transmission Dynamics With Self-Learning Population Behavioral Change. *Front. Public Health* **9**, (2021).
189. Dasaratha, K. Virus dynamics with behavioral responses. *Journal of Economic Theory* **214**, 105739 (2023).
190. Di Guilmi, C., Galanis, G. & Baskozos, G. A Behavioural SIR Model: Implications for Physical Distancing Decisions. *RBE* **9**, 45–63 (2022).
191. Espinoza, B., Marathe, M., Swarup, S. & Thakur, M. Asymptomatic individuals can increase the final epidemic size under adaptive human behavior. *Sci Rep* **11**, 19744 (2021).
192. Gostoli, U. & Silverman, E. Self-Isolation and Testing Behaviour During the COVID-19 Pandemic: An Agent-Based Model. *Artif Life* **29**, 94–117 (2023).
193. de Mooij, J. *et al.* A framework for modeling human behavior in large-scale agent-based epidemic simulations. *SIMULATION* **99**, 1183–1211 (2023).
194. Silk, M. J., Carrignon, S., Bentley, R. A. & Fefferman, N. H. Improving pandemic mitigation policies across communities through coupled dynamics of risk perception and infection. *Proceedings of the Royal Society B: Biological Sciences* **288**, 20210834 (2021).
195. Zhai, S., Zhang, F., Tang, X., Liu, P. & Qu, H. Optimizing the impact of virus transmission on society: decreasing infection rates and enhancing public awareness. *Nonlinear Dyn* **112**, 10689–10701 (2024).
196. Edmunds, W. J., Eames, K. & Keogh-Brown, M. Capturing Human Behaviour: Is It Possible to Bridge the Gap Between Data and Models? in *Modeling the Interplay Between Human Behavior and the Spread of Infectious Diseases* (eds Manfredi, P. & D'Onofrio, A.) 311–321 (Springer, New York, NY, 2013). doi:10.1007/978-1-4614-5474-8_19.

197. Offeddu, V. *et al.* Advancing coupled behavioural-epidemic models: An interdisciplinary framework for the collection of empirical data. 2025.10.07.25337410 Preprint at https://doi.org/10.1101/2025.10.07.25337410 (2025).
198. European Centre for Disease Prevention and Control. ECDC Lighthouse. https://www.ecdc.europa.eu/en/social-and-behavioural-sciences/ecdc-lighthouse (2025).
199. Michie, S., van Stralen, M. M. & West, R. The behaviour change wheel: A new method for characterising and designing behaviour change interventions. *Implementation Sci* **6**, 42 (2011).
200. Casigliani, V., Diadema, E., Gesualdo, F., Fava, V. & Rizzo, C. Engaging the public in surveillance: building a platform to monitor preventive behaviours. *Eur J Public Health* **35**, ckaf161.685 (2025).
201. Di Domenico, L., Bosetti, P., Sabbatini, C. E., Opatowski, L. & Colizza, V. Mobility-driven synthetic contact matrices as a scalable solution for real-time pandemic response modeling. *Nat Commun* **17**, 1845 (2026).
202. Kraemer, M. U. G. *et al.* Artificial intelligence for modelling infectious disease epidemics. *Nature* **638**, 623–635 (2025).
203. Millevoi, C., Pasetto, D. & Ferronato, M. A Physics-Informed Neural Network approach for compartmental epidemiological models. *PLOS Computational Biology* **20**, e1012387 (2024).
204. Kuwahara, B. & Bauch, C. T. Predicting Covid-19 pandemic waves with biologically and behaviorally informed universal differential equations. *Heliyon* **10**, (2024).
205. Binz, M. *et al.* A foundation model to predict and capture human cognition. *Nature* **644**, 1002–1009 (2025).
206. Meng, J. AI emerges as the frontier in behavioral science. *Proceedings of the National Academy of Sciences* **121**, e2401336121 (2024).
207. Lau, M. S. Y., Metcalf, C. J. E., Liu, Z., Grenfell, B. T. & Jin, W. Toward AI foundation models for epidemics: Promise, challenges, and paths forward. *Proceedings of the National Academy of Sciences* **123**, e2526192123 (2026).
208. Kretzschmar, M. E. *et al.* Challenges for modelling interventions for future pandemics. *Epidemics* **38**, 100546 (2022).
209. Pananos, A. D. *et al.* Critical dynamics in population vaccinating behavior. *Proceedings of the National Academy of Sciences* **114**, 13762–13767 (2017).
210. Bhattacharyya, S. & Bauch, C. T. Emergent Dynamical Features in Behaviour-Incidence Models of Vaccinating Decisions. in *Modeling the Interplay Between Human Behavior and the Spread of Infectious Diseases* (eds Manfredi, P. & D'Onofrio, A.) 243–254 (Springer, New York, NY, 2013). doi:10.1007/978-1-4614-5474-8_15.
211. Liu, W. The data source of this study is Web of Science Core Collection? Not enough. *Scientometrics* **121**, 1815–1824 (2019).
212. Chaturvedi, M. *et al.* A scoping review of guidelines on reporting and assessing dynamic mathematical models of infectious diseases. 2024.11.27.24318060 Preprint at https://doi.org/10.1101/2024.11.27.24318060 (2024).

# Supplementary material

# A systematic review of COVID-19 epidemic models with endogenous human behaviour. What's next?

Elena D'Agnese[1,2,3], Alessia Melegaro[3], Vittoria Offeddu[3],
Alberto d'Onofrio[4], Nicola Perra[5], Chris T. Bauch[6], Piero Manfredi[1,7]

1. Department of Economics and Management, University of Pisa, Pisa, Italy
2. Department of Statistics, Informatics, and Applications "Giuseppe Parenti", University of Florence, Florence, Italy
3. DONDENA Centre for Research on Social Dynamics and Public Policy, Bocconi University, Milan, Italy
4. Department of Mathematics, Informatics and Geosciences, University of Trieste, Trieste, Italy
5. School of Mathematical Sciences, Queen Mary University of London, London, UK
6. Department of Mathematics, University of Waterloo, Ontario, Canada
7. IRTA-Leonardo, Istituto di Ricerca sul Territorio e l'Ambiente, Pisa, Italy

**Supplementary material list:**

- S1: Descriptive of included papers
- S2: Quality assessment results
- S3: Models for behaviour change: a "compass for navigators"
- S4: Modelling solutions adopted by papers
- S5: PRISMA 2020 Checklist

# Supplementary Material S1: Descriptives of included papers

For more detailed information, e.g. characteristics by included paper, we refer to the reader to the data extraction table available in the OSF Repository[1].

*Table 1: Descriptive characteristics of the 128 included studies, overall and by modelling approach (implicit versus explicit behaviour modelling). Percentages may exceed 100% for "Behaviour inspected", "Effects of behaviour change on disease model", and "Type of dynamic model", as studies could belong to more than one category*

| Var. | Item | Total | | Implicit | | Explicit | |
|---|---|---|---|---|---|---|---|
| | | # | % | # | % | # | % |
| - | Modelling dichotomy | 128 | - | 59 | - | 69 | - |
| Source of behaviour change | Voluntary | 86 | 67.19 | 29 | 49.15 | 57 | 82.61 |
| | Voluntary + Mandated | 31 | 24.22 | 24 | 40.68 | 7 | 10.14 |
| | Unclear/Unspecified | 11 | 8.59 | 6 | 10.17 | 5 | 7.25 |
| Behaviour inspected | Enhanced hygiene | 1 | 0.78 | 0 | 0.00 | 1 | 1.45 |
| | Unspecified | 14 | 10.94 | 12 | 20.34 | 2 | 2.90 |
| | Self-isolation/quarantine | 15 | 11.72 | 2 | 3.39 | 13 | 18.84 |
| | Social distancing | 30 | 23.44 | 13 | 22.03 | 17 | 24.64 |
| | Vaccination | 12 | 9.38 | 4 | 6.78 | 8 | 11.59 |
| | Contact tracing | 1 | 0.78 | 1 | 1.69 | 0 | 0.00 |
| | Generic protective behaviour[2] | 64 | 50.00 | 30 | 50.85 | 34 | 49.28 |
| | Face mask wearing | 1 | 0.78 | 0 | 0.00 | 1 | 1.45 |
| | Testing | 1 | 0.78 | 0 | 0.00 | 1 | 1.45 |
| | Mobility | 3 | 2.34 | 3 | 5.08 | 0 | 0.00 |
| Information type triggering behaviour change (*see* Funk et al. 2010)[3] | Global info, prevalence-based | 114 | 89.06 | 57 | 96.61 | 57 | 82.61 |
| | Mixed source, prevalence-based | 1 | 0.78 | 0 | 0.00 | 1 | 1.45 |
| | Local info, prevalence-based | 7 | 5.47 | 2 | 3.39 | 5 | 7.25 |
| | Global info, mixed type | 1 | 0.78 | 0 | 0.00 | 1 | 1.45 |
| | Local info, belief-based | 2 | 1.56 | 0 | 0.00 | 2 | 2.90 |
| | Mixed info, mixed type | 1 | 0.78 | 0 | 0.00 | 1 | 1.45 |
| | Global info, belief-based | 1 | 0.78 | 0 | 0.00 | 1 | 1.45 |
| | Mixed info, prevalence-based | 1 | 0.78 | 0 | 0.00 | 1 | 1.45 |
| Effect of behaviour change on disease model | Change in model structure | 31 | 24.22 | 1 | 1.69 | 30 | 43.48 |
| | Change in model parameters | 90 | 70.31 | 57 | 96.61 | 33 | 47.83 |

[1] Elena D'Agnese et al., *OSF Protocol - Advancing Behavioural Epidemiological Modelling: A Systematic Review of Endogenous Human Behaviour in Disease Spread*, Protocol (Open Science Framework, 2025), https://doi.org/10.17605/OSF.IO/GBN97.

[2] Here, we reference papers that describe the adoption of protective behaviour without explicitly tailoring the analysis to a specific kind of protective behaviour.

[3] We extended the original categorization to allow cross-contamination of information sources and types ("mixed information" – global + local - and "mixed type" – prevalence- and belief-based -, respectively)

| Var. | Item | Total | | Implicit | | Explicit | |
|---|---|---|---|---|---|---|---|
| | | # | % | # | % | # | % |
| | Change in disease state | 7 | 5.47 | 1 | 1.69 | 6 | 8.70 |
| | Change in social contact structure | 4 | 3.12 | 1 | 1.69 | 3 | 4.35 |
| Type of dynamic disease model | Ordinary Differential Equations | 102 | 79.69 | 46 | 77.97 | 56 | 81.16 |
| | Agent based model | 5 | 3.91 | 2 | 3.39 | 3 | 4.35 |
| | Network model | 9 | 7.03 | 1 | 1.69 | 8 | 11.59 |
| | Stochastic disease model | 8 | 6.25 | 4 | 6.78 | 4 | 5.80 |
| | Other (PDEs, Fractional order modelling, Filippov system) | 9 | 7.03 | 8 | 13.56 | 1 | 1.45 |

Considering that some studies analysed more than one country, the United States (23/128), China (21/128), and Italy (18/128) were the most frequently modelled contexts, while 34 studies did not specify a geographical focus. Fourteen studies examined at least one African country, and ten included at least one country from the Latin America and Caribbean region (United Nations Statistics Division, 2025). When considering last authors' affiliations, China emerged as the most represented country (28/128), followed by the United States (19/128). In comparison, Italy accounted for 6 studies, African countries for 11, and the Latin America and Caribbean region for 4.

## Supplementary Material S2: Quality assessment results

The quality assessment table has been adapted from Chaturvedi et al. (2024). Each paper was assigned a score of either 0, 0.5, or 1 for each dimension specified in Figure 1. Additional details can be found on the study protocol on OSF (D'Agnese et al., 2025).

We can see that the highest scores have been given in the "Publication Details A" dimension, which was expected given the procedure that we followed to include the papers in the pool. Other subdimensions dealing with clarity in the paper objective and rationale have relatively high mean scores as well ("Applicability" dimension). However, this generally does not translate in a clear description of the methods used to achieve those objectives. Indeed, subdimensions as "Description of Methods", "Calibration", and "Limitations" show relatively lower scores. Lowest scores have been attributed to the "External validity" and "Parameter uncertainty" subdimensions, which means that in general the included papers do not compare their results with additional, independent data, and they do not provide uncertainty measures for their calibrated parameters (such as distributions and/or credible intervals). Few papers publish their code ("Code Availability A"), hindering reproducibility, while on the contrary only a minority of papers has unavailable data ("Code Availability B"), which however is expected as most of these papers use publicly available data.

We do not see major differences across taxonomy categories in terms of mean subdimension scores. Papers with explicit approaches tend to perform relatively better in terms of "Structural assumptions" (both behavioural and epidemiological), "Description of Methods", "Interpretation", and "Calibration", possibly mirroring the greater complexity of some of the model mechanisms described and thus the need for enhanced clarity.

| Dimension | Explanation | Mean score |
|---|---|---|
| **Decision problem** | Is there a clear statement of the aim/decision problem? | |
| **Study design** | Is the study design clearly stated and/or applicable to the decision problem? | |
| **Structural assumptions - A** | Are structural and methodological assumptions clearly stated and reasonable? (Epidemiological model) | |
| **Structural assumptions - B** | Are structural and methodological assumptions clearly stated and reasonable? (Behavioural component) | |
| **Time** | Is the time horizon and/or time step clearly stated and appropriate? | |
| **Parameters** | Are parameter values transparent and reasonable? Are data sources and their translation into parameter values transparent and justified? | |
| **Calibration** | If applicable, were methods and data sources used for calibration described in sufficient detail? Were the subsequent results presented clearly? | |
| **External validity** | Has the model been validated against independent data sources? | |
| **Parameter uncertainty** | Have the methods used to assess uncertainty surrounding parameters been clearly described? | |
| **Interpretation** | Is the presented interpretation of the results reasonable, fair, and/or balanced? | |
| **Code availability - A** | Is the code available? | |
| **Code availability - B** | Are the data available? | |
| **Description of methods** | Are methods described transparently and in enough detail that they can be reproduced? | |
| **Implementation** | Has the choice of software been stated? | |
| **Limitations** | Are limitations clearly described? | |
| **Publication details - A** | Is the type of study clearly identified in the title or abstract? | |
| **Publication details - B** | Has the model developer been stated? | |
| **Funding sources** | Are the funding sources and the role of the funder in the study clearly stated? | |
| **Conflicts of interest** | Are there any potential conflicts of interest? | |

*Figure 1: Mean quality assessment scores across evaluation dimensions. Grey bars represent overall mean scores, blue circles represent mean scores for explicit approaches, and red squares represent mean scores for implicit approaches.*

# Supplementary Material S3: Models for behaviour change: a "compass for navigators"

In what follows, we provide some details on the foundational ideas and works of the Behavioural Epidemiology of Infectious Diseases (BEID) that directed the construction of our taxonomy (*see* Fig. 1 in the Main text). This allows us to outline a history of the field, highlighting the main ideas of these studies and the relationships among them, thus providing a "compass" for readers approaching the subject for the first time. We also take advantage of this compass to clarify potentially ambiguous terminologies e.g., "global" vs "local" information, by linking them to classical ideas from information diffusion theory.

For the sake of simplicity, and given the focus of this review on COVID-19, we develop this guide departing from a few key tools and concepts to swiftly illustrate the frameworks:

- **(a) The (deterministic) epidemic SIR model**. This is the most widely known model of mathematical epidemiology and it has been the "training field" of all key models of BEID. As argued in the Discussion of the main text, our perspective is that universal and simple models (rather than complex) should always be considered as baseline models for behavioural change.
- **(b) Information on epidemic trends is the primary driver of behaviour change**. This does not rule out a role for other factors, even purely exogenous (intrinsic or "belief-based" motivations, social norms, ideology etc., think e.g., to "ideological" hesitancy to vaccination). It rather emphasises the main trait of modern "post-truth" societies, dominated by fast-spreading, multiple sourced, often "disordered", information (Iyengar and Massey 2019).
- **(c) Social distancing as the major instance of behavioural response.** The entire theory of behavioural epidemiology was developed around two main types of behaviour, namely vaccine refusal and risk avoidance during severe epidemics. Though all infections are "behavioural" and all policy actions can trigger nontrivial behavioural responses – as widely shown by COVID-19 - in our subsequent illustrations we focus on social distancing which represents the key non-pharmaceutical (NP) response. In the basic SIR model, this is primarily reflected into endogenous changes over time of the transmission rate $\beta(t)$, which straightforwardly reflects into changes of the controlled reproduction number $R_E(t) = \beta(t)S(t)/\gamma$.

**(d) Focus on individual responses** to the epidemic threat (with or without a public response action). This strengthens the idea that, even in the presence of a public response, no mandate becomes automatically effective in terms of epidemic control, rather it requires to be implemented by individuals through (spontaneous) adherence.

**Mathematical epidemiology: from classical to behavioural models**.

The SIR model, which is the main cornerstone of classical mathematical epidemiology, was built by the fathers of the discipline (Kermack and McKendrick 1927) applying the model of statistical physics to social processes and contagion. In particular, social contacts between individuals (the first ingredient of transmission) were taken as the purely random collisions of particles of a perfect gas, whereas contagion (the second key ingredient) of a susceptible individual ($S$) during an "adequate" social encounter with an infective one ($I$), was taken as a chemical reaction transforming an $(S, I)$ pair into a new $(I, I)$ pair with a constant probability. Consequently, both underlying parameters, namely the average social contact rate per unit time ($C$) and the transmission probability per single contact ($q$), were taken as "physical constant" of the socio-epidemiological environment, summarised by the single constant $\beta = qC,$ universally known as the contact rate. This is often written as $\beta = -C \cdot \log(1-q)$ (for example in Keeling and Rohani 2007).

Capasso and Serio 1978 were first in challenging the statistical physics approach. They suggested that in most nontrivial cases transmission is prevalence-dependent i.e., $\beta = \beta(I),$ as a psychological response of individuals to the perceived "magnitude" of the infective alert. Capasso and Serio's approach was used extensively since the onset of the HIV/AIDS pandemic which triggered the first cohort of behavioural change models. However, systematisation only started with the BEID revolution. In what follows, we report the key features of main BEID models with special focus on their relationships.

**The unifying principle: Information index and the SIR(M) model** The information index concept was fully developed in d'Onofrio et al. 2007 (vaccinating behaviour) and d'Onofrio and Manfredi 2009 (social distancing for recurrent infections). The idea that in modern societies individuals primarily respond to infective threats based on the *available information* on current and past epidemic trends (e.g., infection incidence and prevalence, hospitalisations, deaths) is compactly represented by the concept of *information-dependent* transmission rate i.e., $\beta = \beta(M)$ where $\beta$ is a decreasing function of the *information index* $M$. $M$ summarises the relevant news about epidemic risks (e.g., epidemic prevalence $I$ or incidence, hospitalisations $H$, deaths $D$) people can learn from available sources ($M$ = "media"). A general mathematical form of $M$ is the following:

$$M(t) = \int_{-\infty}^{t} f(I(\tau), H(\tau), D(\tau)) \cdot K(t-\tau)\, d\tau, \qquad (1)$$

where $K$ denotes the information-memory kernel. For example, if the information index $M$ only includes current information about epidemic prevalence, then $M(t) = f(I(t))$, the "classical" prevalence-dependent case.

The ensuing SIR(M) model is the most typical example of a behaviour-implicit formulation i.e., one treating the link between information and individual action as a black-box. Its main merits are that it is by far the simplest for analytical purposes and, especially, the most parsimonious for empirical validation.

**Awareness-inducing compartment: information as a separate variable.** As a first step towards more explicit approaches (i.e., opening the black box), several contributions during COVID-19 times analysed SIRM-type models considering $M$ as a further dynamic variable (Misra et al. 2011) with its equation accounting for the creation and dilution of information:

$$M'(t) = f(I(t), S(t), H(t), D(t)) - \sigma M \quad (2)$$

Though models as (2) are often described as awareness-type, they are immediately reducible to (1) by a straightforward time-integration. Moreover, the flexible forms of kernel $K$ makes (1) more general than (2).

**Awareness based approaches (ASD): truly opening the black box.** Rather than focusing on the disease itself, this approach takes disease awareness (or fear) - defined as the amount of personal information that allows the individual to respond to the infection threat - as the true driver of behaviour change. The approach, first proposed in a network setting (Funk et al. 2009 *- first endogenous BEID model explicitly coupling an infection and a behaviour network)* accounts for local spreading of both information and infection under a countable distribution of awareness levels which individuals can reach (or lose) through *word-of-mouth contacts* with individuals of higher (lower) awareness. The higher the awareness level, the lower the infection hazard for susceptible individuals. Though this approach had the merit to uncover many of the complex facets of the infection-behaviour interplay (Funk et al. 2010) a great deal of intuition of such facets can be achieved by simpler formulations (Epstein et al. 2008; Perra et al. 2011; Misra et al. 2011; Perra and Vespignani 2013) where fully susceptible ($S$) individuals who acquire awareness of risk thanks to information circulation, switch from “normal” to “reduced” ($S_A$, “altered”, or “aware”) risky behaviour. The switch from $S$ to $S_A$ can be modelled in many different forms each having meaningful – yet not necessarily unique - behavioural interpretations:

**(i) “Global”, “public”, information-driven, awareness:** the $S \rightarrow S_A$ flow has the form $\alpha_A(M)S$ driven by at an information-dependent rate $\alpha_A(M)$. In the language of information diffusion (e.g., Bass 1969) this reflects “media” or “public” information, often termed “global” in the BEID literature to contrast with information acquired through social contacts or imitation of others which **are local processes in space** (Funk et al. 2010; Perra et al. 2011). Note that under the typical prevalence-dependent form $\alpha_A(I) = e^{-\beta_A I}$ , for small $\beta_A$ one gets $M \cong \beta_A I$ (Perra et al. 2011). This is equivalent to say that the flow $S \rightarrow S_A$ has the form $\beta_A SI$, which is equivalent to assume that awareness is generated by inter-human social contacts (real or virtual) of $S$-type individuals with severely symptomatic ($I$) individuals at some intrinsic rate $\beta_A$.

**(ii) Local, "word-of-mouth", awareness.** In this case, the $S \to S_A$ flow occurs via explicit contacts representing "word-of-mouth" or imitation (Bass 1969) contagion, often termed "local" in the BEID literature (Funk et al. 2009; Perra et al. 2011; Misra et al. 2011) between either $S$ or and $I$ or $S$ and $S_A$ individuals. In early studies of awareness (Funk et al. 2009; Epstein et al. 2008; Perra et al. 2011) these opinion transitions are modelled by constant rates.

**(iii) Local, "word-of-mouth", "information-driven", awareness.** A straightforward extension could assume that the intensity of such flows is information dependent e.g., of the form $e^{-\beta_L I} S \cdot S_A$ or $e^{-\beta_L I} S \cdot I$. The first expression amounts to consider behavioral changes promoted by *local* contacts between unaware ($S$) and aware ($S_A$) individuals at an intensity modulated by epidemic severity represented through *global* prevalence. For small $\beta_L$, the first expression yields the 3-linear incidence $\beta_L S \cdot S_A \cdot I,$ while the second yields $\beta_A S \cdot S_A \cdot I^2$. This recalls that the presence of nonlinearities in data (e.g., contact or mobility) might indeed be a hint of behavioural change phenomena.

**(iv) "belief-based" awareness**, if the flow from $S$ to $S_A$ occurs at a purely constant rate $\alpha_A$, reflecting an intrinsic (Funk et al. 2010) tendency to seek protection.

**Evolutionary game (EG), the SIRx model.** First proposed in the biosciences (Taylor and Jonker 1978) and first applied in BEID in studies of vaccination choices (Bauch 2005; d'Onofrio et al. 2011) or of social distancing (Poletti et al. 2009, 2012), this approach has become a dominant one in applied sciences thanks to its simplicity in plugging ideas of game-theory in a dynamical system setting. In its basic form, individuals randomly mixing over the "behavioural" network (i.e., mixing "locally" according to the previous terminology) can switch between either of two alternative strategies as is the case of e.g., normal ($N$) vs altered ($A$) behaviour (or, adhering or not adhering to governmental mandates) after a comparison of the Delta payoff they could gain from playing the other strategy instead of the current one. This model is typically specified by a single equation for the proportion $x$ adopting strategy $A$ (e.g., Poletti et al. 2012):

$$x'(t) = k\, x(1-x)\, \Delta E(I,H,D,x), \quad \Delta E = f(I,H,D) - \delta x \tag{3}$$

where $k$ represents the speed of transmission of opinions about the two strategies and $\Delta E$ denotes the expected gain in payoff. If $\Delta E > 0$ this will generate a positive flow towards strategy $A$. The payoff gain is often specified as $\Delta E = f(I,H,D) - \delta x$ i.e., as the balance between the relative benefit of playing $A$ – given by a reduction in the risk of infection - and the relative cost of playing $A$. This approach elegantly summarises and integrates previous approaches since: a) it is an awareness-based approach where individuals switch between two states of different awareness, and indeed (3) can be derived in a more traditional form (e.g., Cabriel et al. 2026) by interpreting the payoff components as intensities driving individuals between the two awareness levels; (b) it typically relies on appropriate information-indices to specify the net payoff of infection; (c) it can be specified in a pure “information index” form when the speed $k$ is high keeping $x(t)$ close to its quasi equilibrium $\Delta E = f(I, H, D) - \delta x \cong 0$, yielding $x \cong \delta^{-1} \cdot f(I, H, D)$ (d’Onofrio et al. 2011). In this manner the SIRx model straightforwardly reduces to an SIR(M) model; (iv) it easily allows to include global (i.e., public) information (d’Onofrio et al. 2012) as well as beliefs and social norms (Oraby et al. 2014), (v) it has been extended to model truly local spatial information in a (partial) differential equations setting by the use of spatio-temporal information indices.

**Economic epidemiology/epidemiology of infectious diseases.** Foundational studies of this area date back to the 1990s and were mainly motivated by the application of microeconomic approaches to behavioral change and risk avoidance related to HIV/AIDS (Kremer and Morcom 1998) and rational vaccine refusal (Philipson 2000). In the most typical microeconomic approach (e.g., Gersovitz 2013; Auld et al. 2025), a representative selfish agent chooses rationally - i.e., by optimising some objective function (e.g., an intertemporal utility function) - the desired levels of risky behavior, of prevention (i.e., risk avoidance), of therapy, etc., given incentives to do so. Choices are bounded by the range of options that are accessible to the agent. Such options identify the related tradeoffs and the constraints of the optimisation. In particular, a selfish agent is one who, whilst making her choices, acts purely following her self-interest. In a communicable diseases setting, this means that the individual will not take into account the possible risks/costs that she is potentially causing to other people if their social/sexual contacts are uninfected. This creates the potential for an *externality*. The aggregation of individual-level (sometimes, household-level) behavior determines the social outcome in terms of risky behavior and prevention.

Pairwise, at the level of public policy, a social planner chooses, along the same lines, the best levels of intervention, prevention and therapy at the societal level. The emerging discrepancy between the costs and benefits of optimal actions from the viewpoint of individuals and the totality of these costs and benefits for the society as a whole (this is the aforementioned concept of externality) creates a rationale for policy to intervene in order align private and public choices. A major example is that of vaccination.

The COVID-19 pandemic represented a take-off point in the interest of economic science for infectious diseases. An indisputable merit of economic research starting with early work on COVID-19 control (e.g., Alvarez et al. 2021; Eichenbaum et al. 2021) has been that of combining in unified frameworks the entire set of impacts of an epidemic event, including the outbreak dynamics, private and public responses, as well their microeconomic impact and macro-level consequences for the society as a whole.

## Bibliography

Alvarez, Fernando, David Argente, and Francesco Lippi. 2021. “A Simple Planning Problem for COVID-19 Lock-down, Testing, and Tracing.” *American Economic Review: Insights* 3 (3): 367–82. https://doi.org/10.1257/aeri.20200201.

Auld, M. Christopher, Eli P. Fenichel, and Flavio Toxvaerd. 2025. “The Economics of Infectious Diseases.” *Journal of Economic Literature* 63 (4): 1281–330. https://doi.org/10.1257/jel.20241649.

Bass, Frank M. 1969. “A New Product Growth for Model Consumer Durables.” *Management Science* 15 (5): 215–27.

Bauch, Chris T. 2005. “Imitation Dynamics Predict Vaccinating Behaviour.” *Proceedings of the Royal Society B: Biological Sciences* 272 (1573): 1669–75. https://doi.org/10.1098/rspb.2005.3153.

Cabriel, Lorenzo, Chris T. Bauch, Fabio Anselmi, Piero Manfredi, and Alberto d’Onofrio. 2026. “Irrational Behaviour-Induced Suppression, Enhancement and Oscillations of Voluntary Vaccination: A Stochastic Model with Delays.” *Physica D: Nonlinear Phenomena* 489 (May): 135143. https://doi.org/10.1016/j.physd.2026.135143.

Capasso, Vincenzo, and Gabriella Serio. 1978. “A Generalization of the Kermack-McKendrick Deterministic Epidemic Model.” *Mathematical Biosciences* 42 (1): 43–61. https://doi.org/10.1016/0025-5564(78)90006-8.

Eichenbaum, Martin S., Sergio Rebelo, and Mathias Trabandt. 2021. “The Macroeconomics of Epidemics.” *The Review of Financial Studies* 34 (11): 5149–87. https://doi.org/10.1093/rfs/hhab040.

Epstein, Joshua M., Jon Parker, Derek Cummings, and Ross A. Hammond. 2008. “Coupled Contagion Dynamics of Fear and Disease: Mathematical and Computational Explorations.” *PLOS ONE* 3 (12): e3955. https://doi.org/10.1371/journal.pone.0003955.

Funk, Sebastian, Erez Gilad, Chris Watkins, and Vincent A. A. Jansen. 2009. “The Spread of Awareness and Its Impact on Epidemic Outbreaks.” *Proceedings of the National Academy of Sciences* 106 (16): 6872–77. https://doi.org/10.1073/pnas.0810762106.

Funk, Sebastian, Marcel Salathé, and Vincent A. A. Jansen. 2010. “Modelling the Influence of Human Behaviour on the Spread of Infectious Diseases: A Review.” *Journal of The Royal Society Interface* 7 (50): 1247–56. https://doi.org/10.1098/rsif.2010.0142.

Gersovitz, Mark. 2013. “Mathematical Epidemiology and Welfare Economics.” In *Modeling the Interplay Between Human Behavior and the Spread of Infectious*

*Diseases*, edited by Piero Manfredi and Alberto D'Onofrio. Springer. https://doi.org/10.1007/978-1-4614-5474-8_12.

Iyengar, Shanto, and Douglas S. Massey. 2019. "Scientific Communication in a Post-Truth Society." *Proceedings of the National Academy of Sciences* 116 (16): 7656–61. https://doi.org/10.1073/pnas.1805868115.

Keeling, Matt, and Pejman Rohani. 2007. *Modeling Infectious Diseases in Humans and Animals*. Princeton University Press.

Kermack, W. O., and A. G. McKendrick. 1927. "A Contribution to the Mathematical Theory of Epidemics." *Proceedings of the Royal Society of London Series A, Containing Papers of a Mathematical and Physical Character* 115 (772): 700–721.

Kremer, Michael, and Charles Morcom. 1998. "The Effect of Changing Sexual Activity on HIV Prevalence." *Mathematical Biosciences* 151 (1): 99–122. https://doi.org/10.1016/S0025-5564(98)10010-X.

Misra, A. K., Anupama Sharma, and J. B. Shukla. 2011. "Modeling and Analysis of Effects of Awareness Programs by Media on the Spread of Infectious Diseases." *Mathematical and Computer Modelling* 53 (5): 1221–28. https://doi.org/10.1016/j.mcm.2010.12.005.

Onofrio, Alberto d', and Piero Manfredi. 2009. "Information-Related Changes in Contact Patterns May Trigger Oscillations in the Endemic Prevalence of Infectious Diseases." *Journal of Theoretical Biology* 256 (3): 473–78. https://doi.org/10.1016/j.jtbi.2008.10.005.

Onofrio, Alberto d', Piero Manfredi, and Piero Poletti. 2011. "The Impact of Vaccine Side Effects on the Natural History of Immunization Programmes: An Imitation-Game Approach." *Journal of Theoretical Biology* 273 (1): 63–71. https://doi.org/10.1016/j.jtbi.2010.12.029.

Onofrio, Alberto d', Piero Manfredi, and Piero Poletti. 2012. "The Interplay of Public Intervention and Private Choices in Determining the Outcome of Vaccination Programmes." *PLOS ONE* 7 (10): e45653. https://doi.org/10.1371/journal.pone.0045653.

Onofrio, Alberto d', Piero Manfredi, and Ernesto Salinelli. 2007. "Vaccinating Behaviour, Information, and the Dynamics of SIR Vaccine Preventable Diseases." *Theoretical Population Biology* 71 (3): 301–17. https://doi.org/10.1016/j.tpb.2007.01.001.

Oraby, Tamer, Vivek Thampi, and Chris T. Bauch. 2014. "The Influence of Social Norms on the Dynamics of Vaccinating Behaviour for Paediatric Infectious Diseases." *Proceedings of the Royal Society B: Biological Sciences* 281 (1780): 20133172. https://doi.org/10.1098/rspb.2013.3172.

Perra, Nicola, Duygu Balcan, Bruno Gonçalves, and Alessandro Vespignani. 2011. “Towards a Characterization of Behavior-Disease Models.” *PLOS ONE* 6 (8): e23084. https://doi.org/10.1371/journal.pone.0023084.

Perra, Nicola, and Alessandro Vespignani. 2013. “Modeling Contact and Mobility Based Social Response to the Spreading of Infectious Diseases.” In *Modeling the Interplay Between Human Behavior and the Spread of Infectious Diseases*, edited by Piero Manfredi and Alberto D’Onofrio. Springer. https://doi.org/10.1007/978-1-4614-5474-8_7.

Philipson, Tomas. 2000. “Economic Epidemiology and Infectious Diseases.” In *Handbook of Health Economics*, vol. 1. Elsevier. https://doi.org/10.1016/S1574-0064(00)80046-3.

Poletti, Piero, Marco Ajelli, and Stefano Merler. 2012. “Risk Perception and Effectiveness of Uncoordinated Behavioral Responses in an Emerging Epidemic.” *Mathematical Biosciences* 238 (2): 80–89. https://doi.org/10.1016/j.mbs.2012.04.003.

Poletti, Piero, Bruno Caprile, Marco Ajelli, Andrea Pugliese, and Stefano Merler. 2009. “Spontaneous Behavioural Changes in Response to Epidemics.” *Journal of Theoretical Biology* 260 (1): 31–40. https://doi.org/10.1016/j.jtbi.2009.04.029.

Taylor, Peter D., and Leo B. Jonker. 1978. “Evolutionary Stable Strategies and Game Dynamics.” *Mathematical Biosciences* 40 (1): 145–56. https://doi.org/10.1016/0025-5564(78)90077-9.

## Supplementary Material S4.I: Modelling solutions adopted by papers in the "**Information Index**" category

| Paper | Title | Behavioural modelling solution |
|---|---|---|
| Agusto et al. 2023 | Impact of public sentiments on the transmission of COVID-19 across a geographical gradient | $\beta_M = \beta_1 + (\beta_0 - \beta_1)e^{-mC_I(t)}$<br>$C_I(t) = (1-q)\sigma \int_0^t E(\tau)d\tau$<br>$m = \frac{1}{\varepsilon}(y_p + y_n)$<br>$y_p = a_p t + b_p; y_n = a_n(t) + b_n$<br><br>$\beta_0, \beta_1$ := before and after lockdown infection rates, respectively<br>$\beta_M$ := sentiment-related infection rate<br>$C_I(t)$ := cumulative number of symptomatic infectious individuals in the community<br>$m$ := sentiment parameter, dependent on $y_p, y_n$, i.e. fitted straight lines representing positive and negative sentiments<br>$q$ := proportion developing asymptomatic infections<br>$E$ := Exposed individuals<br><br>→<br>$\frac{dS}{dt} = \frac{\beta_M[I(t)+\eta_A A(t)+\eta_Q Q(t)+\eta_H H(t)]S(t)}{N(t)}$<br>where $I(t), A(t), Q(t), H(t)$ denote infective, asymptomatic, quarantined, and hospitalized individuals, respectively |
| Ajbar et al. 2021 | Dynamics of an SIR-Based COVID-19 Model With Linear Incidence Rate, Nonlinear Removal Rate, and Public Awareness | $\frac{dS}{dt} = \mu N - \frac{\beta_1 SI}{N} - \frac{\beta_2 SI^2}{(M+I)N}$<br><br>$\frac{\beta_2 I}{(M+I)}$ := effect of media<br>$M$ := half-saturation constant<br>$S, I, N$ := susceptible, infected, total number of individuals, respectively |
| Ajbar et al. 2021 | Dynamics of a COVID-19 Model with a Nonlinear Incidence Rate, | $\frac{dS}{dt} = \mu N - \frac{\beta_1 SI}{(1+\alpha_1 I)N} - \frac{\beta_2 SI^2}{(\alpha_2 + I)N} - \mu S$ |

| Paper | Title | Behavioural modelling solution |
| --- | --- | --- |
| | Quarantine, Media Effects, and Number of Hospital Beds | $\frac{1}{(1+\alpha_1 I)}$ ≔ inhibition caused by change in attitude by susceptible individuals<br>$\frac{\beta_2 I}{(\alpha_2+I)}$ ≔ effect of media, dependent on the awareness rate $\beta_2$<br>$\alpha_1, \alpha_2$ ≔ half-saturation constants<br>$S, I, N$ ≔ susceptible, infected, total number of individuals, respectively |
| Al-arydah 2023 | Mathematical modeling and optimal control for COVID-19 with population behavior | $f_m(I) = \frac{f_M I^m}{I^m + H_i^m}$<br>$g_n(V) = \frac{g_M V^n}{V^n + H_v^n}$<br><br>$I, V$ ≔ infected and vaccinated individuals, respectively<br>$f_m(I)$ ≔ sense of caution ($f_M \leq 1$: saturation level of caution)<br>$g_n(V)$ ≔ sense of safety ($g_M \leq 1$: saturation level of safety)<br>$H_i, H_v$ ≔ half-saturation constants for the sense of caution and sense of safety functions<br><br>→<br>$\beta(I, V) = \beta_0 \left(1 - f_m(I)\left(1 - g_n(V)\right)\right)$<br>$\beta$ ≔ transmission rate, where $\beta_0$ is the background transmission rate |
| Barth et al. 2023 | Spatiotemporal vaccine allocation policies for epidemics with behavioral feedback dynamics | $\beta(t) = \bar{\beta}\left(1 - M(t)\right)$<br>$M(t) = \frac{1}{1 + e^{-u(t)}}$<br>$u(t) = \eta^I \left( \sum_{t'=1}^{T} \Delta I(t - t') - \sum_{t'=T+1}^{2T} \Delta I(t - t') \right) + \eta^D \left( \sum_{t'=1}^{T} \Delta D(t - t') - \sum_{t'=T+1}^{2T} \Delta D(t - t') \right)$<br>$M(t)$ ≔ fraction of the population that follows strict social distancing protocols<br>$u(t)$ ≔ strength and directionality of individuals' response to epidemic developments within their attention span<br>$\beta(t), \bar{\beta}$ ≔ effective and baseline contact rate, respectively<br>$I, D$ ≔ infective and dead individuals, respectively<br>$\eta^I, \eta^D$ ≔ strength of individuals' reactions<br><br>→ |

| Paper | Title | Behavioural modelling solution |
|---|---|---|
| | | $\dot{S}(t) = -\beta(t)S(t)\left(\epsilon\left(P(t)+A(t)\right)+I(t)\right)$ |
| Bozkurt et al. 2021 | A fractional-order model of COVID-19 considering the fear effect of the media and social networks on the community | $f_1(\alpha_1, I) = \frac{1}{1+\alpha_1 I}$<br>$f_2(\alpha_2, D) = \frac{1}{1+\alpha_2 D}$<br>$f_1, f_2 \coloneqq$ fear to be infected and to die from COVID-19<br>$I, D \coloneqq$ infected and dead individuals, respectively<br>$\alpha_1, \alpha_2 \coloneqq$ level of the fear to be infected and to die from COVID-19, respectively<br><br>→<br>$D^\alpha S(t) = \Lambda + S(t)r\left(1-\frac{S(t)}{K_1}\right)\frac{1}{1+\alpha_1 I(t)} - \beta_1 E(t)S(t) - \gamma_1 I(t)S(t) - \eta S(t)$<br>$D^\alpha Q(t) = Q(t)\left(1-\frac{Q(t)}{K_4}\right)\frac{1}{1+\alpha_2 D(t)} + \beta_2 I(t) - \eta Q(t) - \mu Q(t) - \gamma_2 Q(t)$<br>Where $S$ denotes the susceptible population and $Q$ the quarantined population. |
| Bulai et al. 2022 | Stability analysis of a model of epidemic dynamics with nonlinear feedback producing recurrent infection waves | $\frac{dM}{dt} = H - \frac{M}{\tau}$<br>$f(M) = \frac{1}{1+\frac{M}{K_p}}$<br>$M \coloneqq$memory variable, that enters in the model through the feedback function $f(M)$<br>$H \coloneqq$ hospitalized individuals<br>$K_p \coloneqq$ value of $M$ needed for a 50% transmission rate reduction, i.e. $f(K_p) = 1/2$<br><br>→<br>$\frac{dS}{dt} = -f(M)\frac{\beta SI}{N}$ |
| Buonomo 2020 | Effects of information-dependent vaccination behavior on coronavirus outbreak: insights from a SIRI model | $\varphi(M) = \varphi_0 + \varphi_1(M)$<br>$M(t) = \int_{-\infty}^{t} \tilde{g}\left(S(\tau), I(\tau), V(\tau)\right)K(t-\tau)d\tau$<br>$\tilde{g} = \begin{cases} 0, & \text{if } t < 0 \\ g(I), & \text{if } t \geq 0 \end{cases}, \quad g(I) = kI$<br>$\varphi_1(M) = (1-\varphi_0-\varepsilon)\frac{DM}{1+DM}$ |

| Paper | Title | Behavioural modelling solution |
|---|---|---|
| | | $M(t)$ ≔ information index<br>$\varphi_0, \varphi_1$ ≔ fraction of the population that gets vaccinated regardless of rumours and that gets vaccinated influenced by information, respectively<br>$D$ ≔ positive constant tuning reactivity factor of voluntary vaccination<br>$\tilde{g}$ ≔ function describing information relevant to the agents<br>$k$ ≔ information coverage parameter<br>$K(t)$ ≔ delay kernel s.t. $\int_0^{+\infty} K(t)dt = 1$<br><br>→<br>$\dot{S} = -\beta SI - \varphi(M)S$<br>$\dot{V} = -\psi\beta VI + \varphi(M)S$ |
| Buonomo et al. 2020 | Effects of information-induced behavioural changes during the COVID-19 lockdowns: the case of Italy | $M(t) = \int_{-\infty}^{t} k\big(Q(\tau) + I_s(\tau)\big)Erl_{1,a}(t-\tau)d\tau$<br>$FoI = \beta(M)\frac{\varepsilon_p I_p + \varepsilon_m I_m + \varepsilon_s I_s}{N - Q}$<br>$\beta(M) = \frac{\beta_b - \beta_0}{1 + \alpha M}$<br>$\gamma(M) = \gamma_0 + \gamma_1(M), \qquad \gamma_1(M) = (1 - \gamma_0 - \zeta)\frac{DM}{1 + DM}$<br>$M(t)$ ≔ information index<br>$\beta_b, \beta_0$ ≔ baseline transmission rate and mandatory social distancing transmission rate, respectively<br>$\gamma(M)$ ≔ information-dependent quarantine rate, where $\gamma_0$ is the fraction of detected population forced into home isolation and $\gamma_1(M)$ is the fraction of undetected population that voluntarily undergoes home isolation based on available information<br>$D$ ≔ positive constant tuning reactivity factor of voluntary quarantine<br>$\zeta$ ≔ constant s.t. $1 - \zeta$ is the ceiling overall quarantine rate<br>$k, a$ ≔ information coverage and inverse of the average information delay, respectively<br><br>→<br>$\dot{S} = \Lambda - \beta(M)\frac{S}{N - Q}\big(\varepsilon_p I_p + \varepsilon_m I_m + \varepsilon_s I_s\big) - \mu S$<br>$\dot{I}_m = p\eta I_p - \gamma(M)I_{,} - \sigma_m I_m - \nu_m I_m - \mu I_m$<br>$\dot{Q} = \gamma(M)I_m - \sigma_q Q - \nu_q Q - \mu Q$ |

| Paper | Title | Behavioural modelling solution |
|---|---|---|
| | | where $S, I_m, I_p, I_s, Q$ denote susceptible, asymptomatic/mildly symptomatic, post-latent, severely symptomatic/hospitalized, and quarantined individuals, respectively. |
| Buonomo et al. 2022 | A behavioural modelling approach to assess the impact of COVID-19 vaccine hesitancy | $M(t) = \int_{-\infty}^{t} kaI_s(\tau)e^{-a(t-\tau)}d\tau$<br>$\varphi_0 + \varphi_1(M)$ ≔ vaccination rate<br>$\varphi_0, \varphi_1$ ≔ fraction of the population that gets vaccinated regardless of rumours and that gets vaccinated influenced by information, respectively<br>$I_s$ ≔ symptomatic infectious<br>$k, a$ ≔ information overage and inverse of average time delay in the collection of the information of the disease, respectively<br><br>→<br>$\dot{S} = \Lambda - (\varphi_0 + \varphi_1(M))S - \beta S(\varepsilon_a I_a + \varepsilon_s I_s) - \mu S$<br>$\dot{V} = (\varphi_0 + \varphi_1(M))S - \sigma\beta V(\varepsilon_a I_a + \varepsilon_s I_s) - \mu V$ |
| Buonomo et al. 2022 | A BEHAVIORAL CHANGE MODEL TO ASSESS VACCINATION-INDUCED RELAXATION OF SOCIAL DISTANCING DURING AN EPIDEMIC | $\mathcal{M}(t) = \int_{\infty}^{t} g(I(\tau))Erl_{1,a}(t-\tau)d\tau$<br>$\mathcal{V}(t) = \int_{\infty}^{t} h(V(\tau))Erl_{1,a}(t-\tau)d\tau$<br>$\varphi(\mathcal{M}) = \varphi_0 + \varphi_1(\mathcal{M})$<br>$\beta(\mathcal{V}) = \beta_{\max} - \beta_0 + \beta_1(\mathcal{V})$<br>$\mathcal{M}(t), \mathcal{V}(t)$ ≔ information indices<br>$\beta_{\max}, \beta_0, \beta_1(\mathcal{V})$ ≔ baseline transmission rate, mandatory social distancing transmission rate, and social distancing relaxation rate induced by the vaccination rollout, respectively<br>$g(\cdot), h(\cdot)$ ≔ message functions describing information that individuals consider relevant<br>$a$ ≔ inverse of the average information time delay<br>$\varphi_0, \varphi_1(\mathcal{M})$ ≔ information-independent constant and information-dependent vaccination rate, respectively<br><br>→<br>$\dot{S} = \mu - \beta(\mathcal{V})SI - \varphi(\mathcal{M})S + \gamma I + \vartheta V - \mu S$<br>$\dot{I} = \beta(\mathcal{V})SI + \sigma\beta(\mathcal{V})VI - \gamma I - \mu I$<br>$\dot{V} = \varphi(\mathcal{M})S - \sigma\beta(\mathcal{V}) - \vartheta V - \mu V$<br>Where $S, I, V$ denote susceptible, infective, and vaccinated individuals, respectively |

| Paper | Title | Behavioural modelling solution |
|---|---|---|
| Burridge et al. 2022 | Public efforts to reduce disease transmission implied from a spatial game[a] | $F(u) = logit^{-1}(-\beta_0 + \beta_1 u)$<br>$\beta_0 = c_0 - c_1 I(t-T)$<br>$F(u)$ ≔ response function depending on the cooperation rate $u$ (i.e., whether people act to reduce the change of catching/transmission the disease or not)<br>$\beta_0, \beta_1$ ≔ time-varying and time-constant coefficients driving behavioural changes<br>$T$ ≔ information time lag<br>→<br>$\dot{u} = \frac{1}{\tau}(F(u) - u)$<br>$\dot{S} = \xi R - \epsilon(1-u)^2 IS$<br>$\dot{I} = \epsilon(1-u)^2 IS - pI$<br>Where $S, I$ are susceptible and infective populations, respectively |
| Cantin et al. 2022 | Mathematical analysis of a hybrid model: impacts of individual behaviours on the spreading of an epidemic | $\rho_I(D_i, t_{s+1}) = \frac{I_i(t_{s+1})}{N_i(t_{s+1})}$.<br>$\rho_{\mathcal{N}}(D_i, t_{s+1}) = \frac{1}{N_i}\sum_{a\in\mathcal{U}} \mathcal{N}(I, a, t_{s+1})$.<br>$\rho_I(D_i, t_{s+1})$ ≔ action threshold for decision-maker; if greater than $\mathcal{T}_1$, imposes increase in fraction of protected individuals (according to $p_{i+1}(t_{s+1}) = p_i(t_s) \times (1 + d_1)$); if greater than $\mathcal{T}_2$ ($\mathcal{T}_2 > \mathcal{T}_1$), imposes quarantine in region $D_i$<br>$\rho_{\mathcal{N}}(D_i, t_{s+1})$ ≔ action threshold for individuals; if greater than $\mathcal{T}_3$, they will oppose to the decision-maker's strategies (according to $u_i(t_{s+1}) = u_i(t_s) \times (1 + d_2)$), otherwise they will decrease opposition by a fraction $u$ (according to $u_i(t_{s+1}) = u_i(t_s) \times (1 - d_2)$),<br>$\sum_{a\in\mathcal{U}} \mathcal{N}(I, a, t_{s+1})$ ≔ number of infected neighbors evaluated by agent $a$ of the set of agents $\mathcal{U}$ in region $D_i$.<br><br>→<br>$\frac{dS}{dt} = \Lambda - \beta\left(1 - p(1-u)\right)\frac{\left(\theta A(t) + I(t)\right)}{N}S(t) - \phi p(1-u)S(t) + \omega P(t) - \mu S(t)$<br>$\frac{dP}{dt} = \phi p(1-u)S(t) - \omega P(t) - \mu P(t)$<br>Where $S, P, A, I$ denote susceptible, protected, asymptomatic, and infected individuals, respectively. |

[a] We refer here to the model in Sec. 4; the first model proposed in the paper (i.e., the spatial game model), does not show a feedback loop between disease and behaviour, but focuses on the effect of behaviour on disease only.

| Paper | Title | Behavioural modelling solution |
|---|---|---|
| Chen et al. 2024 | Analysis of a COVID-19 model with media coverage and limited resources. | $dS(t) = \left[\Lambda - (\mu + \nu)S(t) - \left(\beta_1 - \frac{\beta_2 I(t)}{\alpha + I(t)}\right)S(t)I(t)\right]dt$<br>$\frac{\beta_2 I(t)}{\alpha + I(t)}$ :=effect of media coverage on the transmission rate, where $\beta_2$ is the maximum contact rate that the media coverage can reduce<br>$S, I$ := susceptible and infected individuals, respectively |
| Cordova-Lepe et al. 2022 | Adding a reaction-restoration type transmission rate dynamic-law to the basic SEIR COVID-19 model | $g[t, I] = \mu(t)\frac{I}{I + I_m}$<br>$f[t, \beta] = \nu(t)\left(\frac{\lvert\beta(t) - \beta_0\rvert}{\beta_0}\right)^{\alpha}$<br>$g[t, I]$ := reaction rate<br>$f[t, \beta]$ := restoration rate<br>$I$ := number of active cases<br>$I_m$ := half-saturation constant<br>$\beta_0$ := baseline transmission rate (observed at the beginning of the pandemic)<br>$\mu(t), \nu(t)$ := reaction and restoration coefficients, respectively<br><br>→<br>$\beta' = -g[t, I]\beta + f[t, \beta]\beta, \qquad \beta(0) = \beta_0$<br>$S' = -\beta SI/N$ |
| Deng et al. 2021 | Joint impacts of media, vaccination and treatment on an epidemic Filippov model with application to COVID-19 | $\epsilon = \begin{cases} 0, & M(t) < 0 \\ 1, & M(t) > 0 \end{cases}$<br>$M(t) = a_1 I + a_2 \dot{I}\rvert_{S_2}$<br>$M(t)$ := threshold value involving media coverage<br>$I, \dot{I}\rvert_{S_2}$ := infected population and changing rate of infected population for subsystem $S_2$, respectively<br>$a_1, a_2$ := dependence of media impact on the size of infected population and its changing rate, respectively<br><br>→<br>$\frac{dS}{dt} = \Lambda - e^{-\epsilon M(t)}\beta SI - \mu S - \frac{\gamma_1 S}{1+(1-\epsilon)\omega_1 S}$<br>$\frac{dI}{dt} = e^{-\epsilon M(t)}\beta SI - \mu I - \delta I - \frac{\gamma_2 I}{1+(1-\epsilon)\omega_2 I}$ |

| Paper | Title | Behavioural modelling solution |
|---|---|---|
| Ding et al. 2021 | Stochastic analysis of COVID-19 by a SEIR model with Levy noise | $dS = \left(A - \left(\beta_1 - \frac{\beta_2 I}{\alpha + I}\right)\frac{SI}{N} - \mu S\right)dt + \sigma_1 S dB_1(t) + \int_Y q_1(y)S(t-)\tilde{N}(dt,dy)$<br>$\frac{\beta_2 I}{\alpha+I}$ := reduced contact rate influenced by the mass media alert<br>$S, I$ := susceptible and infected individuals, respectively<br>$B_1(t)$ := Brownian motion with independent components<br>$\tilde{N}(dt,dy)$ := Poisson counting measure with compensator |
| Doenges et al. 2022 | Interplay Between Risk Perception, Behavior, and COVID-19 Spread | $H_R(t) := ICU^{tot} \cdot \mathcal{G}_{p_R b_R} = \int_{-\infty}^{t} dt' ICU^{tot}(t')\mathcal{G}_{p_R b_R}(-t'+t)$<br>$H_{u,w}(t) := ICU^{tot} \cdot \mathcal{G}_{p_{vac}, b_{vac}} = \int_{-\infty}^{t} dt' ICU^{tot}(t')\mathcal{G}_{p_{vac}, b_{vac}}(-t' - \tau_{u,w} + t)$<br>$H_R(t), H_{u,w}(t)$ := health risk in the recent past, dependent on the sum of all patients in ICU treatment $ICU^{tot}$, for decisions on adoption of protective behaviours and vaccination willingness (for first $u$ and booster $w$ doses), respectively<br>$\mathcal{G}_{p_R b_R}, \mathcal{G}_{p_{vac}, b_{vac}}$ := Gamma distributions modulating time memory for decision on adoption of protective behaviours and vaccination willingness, respectively<br><br>→<br>$\sum_j C_{ji}\mathcal{I}_j$ := Infection term, where $C_{ji}$ is the overall contact matrix and $\mathcal{I}_j$ describes the infectiousness of age group i. Each layer $\nu$ of the contact matrix $C_{ji}^{\nu}$ (representing settings) is weighted by reduction factors $k_{\mathrm{NPI,self}}^{\nu}(H_R)$<br>$u_i^{\mathrm{willing}}$ := willingness to be vaccinated (i.e. to receive the first dose), described by $u_i^{willing} = u_i^{base} + (u_i^{\max} - u_i^{base})(1 - \exp(-\alpha_u H_u))$<br>$w_i^{\mathrm{willing}}$ := willingness to get booster doses, described by<br>$w_i^{willing} = w_i^{base} + (w_i^{\max} - w_i^{base})(1 - \exp(-\alpha_w H_w))$<br>$\frac{dS_i}{dt} = -S_i \sum_j C_{ji}\mathcal{I}_j - M_i \phi_i(H_u)\left(\frac{S_i}{S_i + W_i^n}\right)$<br>$\frac{dW_i^n}{dt} = -W_i^n \sum_j C_{ji}\mathcal{I}_j - M_i \phi_i(H_u)\left(\frac{W_i^n}{S_i + W_i^n}\right) + \Omega R_i$<br>$\frac{dW_i^v}{dt} = -W_i^v \sum_j C_{ji}\mathcal{I}_j - M_i u_i^{current}\varphi_i(H_w) + \Omega V_i + \Omega R_i^v$ |

| Paper | Title | Behavioural modelling solution |
| --- | --- | --- |
| | | $\frac{dV_i}{dt} = M_i(\phi_i(H_u) + u_i^{current}\varphi_i(H_w)) - \Omega V_i$<br>$\frac{du_i^{current}}{dt} = \phi_i(H_u)$<br>$\frac{dw_i^{current}}{dt} = \varphi_i(H_w)$<br>Where $S, W^n, W^v, R, V$ represent susceptible, susceptible with waned natural immunity, susceptible with waned vaccine immunity, recovered, and vaccinated individuals respectively. Vaccinated individuals can't get infected. The transmission rate is influenced by the changes in the contact matrix due to self-regulation of contacts. |
| Fierro et al. 2022 | Vaccination and variants: Retrospective model for the evolution of Covid-19 in Italy | $\alpha^i(t) = \alpha_c(t) \cdot \sigma^i(t), \quad i \in \{u, v\}$<br>$\sigma(t) = \sigma_{aw}(t) \cdot \sigma_{term}(t) \cdot \sigma_{var}(t) \cdot \sigma_0^i$<br>$\sigma_{aw}(t) = \frac{1}{I_D(t)}$<br>$\alpha^i(t) :=$ transmission rate from unvaccinated and vaccinated undiagnosed individuals, dependent on the pure contact rate $\alpha_c(t)$ and the susceptibility term $\sigma_t$<br>$\sigma_{aw}(t) :=$ awareness mechanism, dependent on diagnosed infected $I_D(t)$<br>$\sigma_{term}(t), \sigma_{var}(t), \sigma_0^i :=$ modification of transmissibility due to seasonal variability, emergency of new variants, and the vaccination status, respectively<br><br>→<br>$\dot{S}(t) = -S(t)\sum_{i=v,u}\{\alpha^i[I_A^i(t) + I_S^i(t)] + \gamma I_D^i(t)\} - \chi_1 S(t) + \phi R(t)$<br>$\dot{V}_1(t) = \chi_1 S(t) - f_1 V_1(t)\sum_{i=v,u}\{\alpha^i[I_A^i(t) + I_S^i(t)] + \gamma I_D^i(t)\} - \chi_2 V_1(t)$<br>$\dot{V}_2(t) = \chi_2 V_1(t) - f_2 V_2(t)\sum_{i=v,u}\{\alpha^i[I_A^i(t) + I_S^i(t)] + \gamma I_D^i(t)\} - \chi_3 V_2(t)$<br>$\dot{V}_3(t) = \chi_3 V_2(t) - f_3 V_3(t)\sum_{i=v,u}\{\alpha^i[I_A^i(t) + I_S^i(t)] + \gamma I_D^i(t)\}$<br>$\dot{E}^u(t) = S(t)\sum_i\{\alpha^i[I_A^i(t) + I_S^i(t)] + \gamma I_D^i(t)\} - \delta E^u(t)$<br>$\dot{E}^v(t) = \sum_{j=1}^3 f_j V_j(t)\sum_{i=v,u}\{\alpha^i[I_A^i(t)+I_S^i(t)] + \gamma I_D^i(t)\} - \delta E^v(t)$<br><br>Where $S, V_1, V_2, V_3, E^u, E^v, I_A, I_S, I_D, R$ denote susceptible, vaccinated with first, second, and third dose, infective asymptomatic, symptomatic, diagnosed, and recovered compartments, respectively. $i \in \{u, v\}$ stratifies compartments into unvaccinated and vaccinated status, respectively. |

| Paper | Title | Behavioural modelling solution |
|---|---|---|
| Fosch et al. 2023 | Characterizing the role of human behavior in the effectiveness of contact-tracing applications | $I_{thr}$ =7-day incidence/100,000 inh.<br>$I_{thr}$ ≔ reluctancy threshold, i.e. minimal level of infection incidence triggering the adoption of the contact-tracing app in the population (aside the existence of both fixed non-adopters and early adopters)<br><br>→<br>compliant app users who test positive for SARS-CoV-2 will report their infection in the app, activating the warning system, and causing their Bluetooth contacts who are also active users to enter a 10-day preventive quarantine (thus influencing the transmission process) |
| Galanis et al. 2021 | The effectiveness of non-pharmaceutical interventions in reducing the outcomes of the covid-19 epidemic in the UK, an observational and modelling study | $m_t = \lambda_0 c_{t-1} + \lambda_1$<br>$\beta_t = P m_t = \zeta c_{t-1} + \varepsilon, \qquad \varepsilon = P\lambda_0, \zeta = P\lambda_1$<br>$\xrightarrow{c_t = \rho\alpha E_{t-1}} \beta_t = \zeta\rho\alpha E_{t-2} + \varepsilon$<br>$m_t$ ≔mobility at time $t$<br>$c_t$ ≔ confirmed cases at time $t$<br>$E_t$ ≔ exposed individuals at time $t$<br>$P$ ≔ probability of getting infected if susceptible<br>$\zeta, \varepsilon$ ≔ capture the relative importance of the behavioural component related to observed infections and NPIs, respectively<br>$\lambda_0, \lambda_1, \rho, \alpha$ ≔ effect of daily confirmed cases, effect of other influences including the effects of NPIs, fraction of symptomatic infected out of the total infected, and inverse of the average incubation period in days, respectively<br><br>→<br>$S_{t+1} = S_t - \beta_t S_t I_t / N$<br>$E_{t+1} = (1-\alpha)E_t + \beta_t S_t I_t / N$ |
| Gatti et al. 2021 | Saving lives during the COVID-19 pandemic: the benefits of the first Swiss lockdown. | $\frac{d\beta_a}{dt} = \frac{\beta_0}{\exp\left(\alpha_D\left(\frac{dD_a}{dt} + 0.27\left(\sum_{i=1, i\neq a}^{8} \phi_{i,3}\frac{dD_i}{dt}\right)\right) - \psi\right)} - \beta_a$ for $a \in \{1,2,4,5,6,7,8\}$<br>$\frac{d\beta_a}{dt} = \frac{\beta_0}{\exp\left(\alpha_D\left(\frac{dD_a}{dt} + 0.27\left(\sum_{i=1, i\neq 3}^{8} \phi_{i,4}\frac{dD_i}{dt}\right)\right) - \psi\right)} - \beta_a$ for $a \in \{3\}$<br><br>$\beta_a, \beta_0$ ≔age-stratified behavioural and baseline transmission rates, respectively<br>$\frac{dD_a}{dt}$ ≔ death rate of individuals in age group $a$ |

| Paper | Title | Behavioural modelling solution |
|---|---|---|
| | | $\phi_{i,3}, \phi_{i,4}$ ≔normalization coefficients to take into account how much importance it is given to death rates of age group $a$ having as reference group age group 20-29 and 30-39, respectively, using the Value of Statistical Life survey values<br>$\psi$ ≔ parameter capturing seasonal patterns<br>$\alpha_D$ ≔ weighting parameter<br><br>→<br>$\frac{dS_a}{dt} = -\frac{\beta_a \sum_{a=1}^{8} I_a}{\sum_{a=1}^{8} N_a} * S_a$<br>$\frac{dI_a}{dt} = \frac{\beta_a \sum_{a=1}^{8} I_a}{\sum_{a=1}^{8} N_a} * S_a - \gamma I_a$<br>$\frac{dR_a}{dt} = \gamma I_a - \theta R_a$<br>$\frac{dD_a}{dt} = \delta a \theta R_a$<br>Where $S_a, I_a, R_a, D_a$ denote susceptible, infected, resolving and dead individuals of age group $a$, respectively |
| Gutierrez-Jara et al. 2022 | Effects of human mobility and behavior on disease transmission in a COVID-19 mathematical model | $\dot{D}_y = -\lambda_1(D_y - D_y^*) + \lambda_2^y \sum_y (I_y + dQ_i)/N$<br>$\Lambda_y(t) = \beta_a^y(D_y, \Theta_y)\frac{A_y(t)}{N_y(t)} + \beta_i^y(D_y, \Theta_y)\frac{I_y(t)}{N_y(t)}, \quad y \in \{h.c.w,o\}$<br>$D_y(t)$ ≔ interaction distance in the setting $y \in \{h, w, c, o\}$, dependent on the rate $\lambda_1$ of resistance to change with respect to the natural distance $D_y^*$, and the per capita rate $\lambda_2^y$ of change of the distance as a reaction to the number of infectious individuals. The formula for $\Theta_y$ is setting-dependent.<br>$\Lambda_y(t)$ ≔ force of infection, dependent on both infectious and asymptomatic individuals<br>$\beta_a^y(D_y, \Theta_y), \beta_i^y(D_y, \Theta_y)$ ≔ transmission rates upon a contact of a susceptible with an asymptomatic and symptomatic infectious individual, respectively<br><br>→<br>$\dot{S}_h = -\delta(\sum_x \alpha_x)S_h + \sum_x \frac{1}{\sigma_x} S_x - S_h\Lambda_h - \frac{\psi_h S_h I_h}{N_h} + \nu Q_s + \epsilon R$<br>$\dot{S}_x = \alpha_x \delta S_h - \frac{1}{\sigma_x} S_x - S_x\Lambda_x - \frac{\psi_k S_x I_x}{N_x}$<br>Where $h$ , $x \in \{w, c, o\}$ denote settings, and $S, I, Q, R, N$ denote susceptible, infectious, quarantined, removed, and total population, respectively |

| Paper | Title | Behavioural modelling solution |
|---|---|---|
| Karatayev et al. 2020 | Local lockdowns outperform global lockdown on the far side of the COVID-19 epidemic curve | $F_j(t) = w\left(1 - \epsilon C_j(t)\right) + (1-w)\left(1 - \epsilon\left(1 - \exp\left(-\frac{\omega P_j^+}{P_j}\right)\right)\right), \quad P_j^+ = P_j^{AK} + P_j^{IK}$<br>$F_j(t)$ ≔ fraction of contacts remaining after physical distancing in county $j$ at time $t$, dependent on confirmed cases reported in their county $P_j^{AK}, P_k^{IK}$ (tested asymptomatic and infected cases respectively) and on public closures $C_j(t)$ (school and workplaces)<br>→<br>$\lambda_j(t) = 1 - \Pi_{D,T}\left[1 - F_j(t) f_t \beta_{D,j}\left(\frac{\xi}{1 + cP_{j,t}^*} + \frac{1-\xi}{P_{j,t}^*}\right)\right]^{P_{j,t}^{*DT}}$<br>Where $\lambda_j(t)$ denotes the daily probability that a susceptible person in county $j$ is infected by an infectious person. |
| LaJoie et al. 2022 | A COVID-19 model incorporating variants, vaccination, waning immunity, and population behavior. | $f_I = e^{-d_I I}$<br>$f_V = \frac{1}{f_I} + \left(1 - \frac{1}{f_I}\right) e^{-d_V V}$<br>$f_I$ ≔ sense of caution<br>$f_V$ ≔ sense of safety<br>$I, V$ ≔ Infected and vaccinated population, respectively<br>$d_I, d_V$ ≔ population's level of caution and safety factors, respectively<br>→<br>$\beta_\nu = \beta_{\nu 0} f_I f_V$<br>$\frac{dS^{(i)}}{dt} = -S^{(i)} \sum_j^{immunity} \sum_\nu^{variants}\left(\beta_\nu I_\nu^{(j)}\left(1 - \eta_\nu^{(i)}\right)\right)$<br>$\frac{dI_\nu^{(i)}}{dt} = I_\nu^{(i)}\left(\beta_\nu S^{(i)}\left(1 - \eta_\nu^{(i)}\right) - \mu_\nu \gamma_\nu - \gamma_\nu\right)$<br>Where $S^{(i)}, I_\nu^{(i)}$ represent susceptible and infected individuals respectively, with immunity level $i$ and infected by variant $\nu$. $\beta_{\nu 0}$ is the baseline transmission rate for variant $\nu$. |
| Li et al. 2023 | Adaptive behaviors and vaccination on curbing COVID-19 transmission: Modeling simulations in eight countries | $\frac{dS(t)}{dt} = -\lambda f \frac{S(t)}{N}\left[I_p(t) + \alpha I_s(t) + \beta I_c(t)\right] - \vartheta\theta S(t-\tau)$<br>$\frac{dE(t)}{dt} = \lambda f \frac{S(t)}{N}\left[I_p(t) + \alpha I_s(t) + \beta I_c(t)\right] - \eta E(t)$<br>$f_1 = e^{-kI_c}, f_2 = e^{-k(I_c + \phi R)}, f_3 = \left(1 - \frac{I_c}{N}\right)^k, f_4 = \left(1 - \frac{I_c + \phi R}{N}\right)^k$<br>$f_1, f_3$ ≔ short-term awareness, dependent on the number of clinical cases $I_c$ |

| Paper | Title | Behavioural modelling solution |
| --- | --- | --- |
| | | $f_2, f_4 :=$ long-term awareness, dependent on the number of clinical cases $I_c$ and removed individuals $R$<br>$k, \phi :=$ parameters (not described)<br><br>→<br>Short- and long-term awareness (i.e. information on either current or total infection) reduce the transmission rate |
| Liu et al. 2020 | The effect of control measures on COVID-19 transmission in Italy: Comparison with Guangdong province in China | $\beta(t) = \begin{cases} \beta_0, & if \ \frac{1}{klog(H(t))} > 1 \\ \beta_0 \frac{1}{klog(H(t))} & \end{cases}$<br>$H(t) :=$ reported confirmed cases<br>$\beta_0 :=$ baseline transmission rate<br><br>→<br>$\frac{dS}{dt} = -(\beta c + cq(1-\beta))S\frac{I+\theta A}{N} + \lambda S_q$<br>$\frac{dS_q}{dt} = (1-\beta)cqS\frac{(I+\theta A)}{N} - \lambda S_q$<br>$\frac{dH}{dt} = \delta_I I + \delta_q E_q - (\alpha + \gamma_H)H$<br>Where $S, S_q, E_q, I, A,$ denote susceptible, quarantined susceptible, suspected quarantined, symptomatic infected, and asymptomatic infected, respectively. $q$ is the quarantine rate due to contact tracing, and $c$ is the contact rate. |
| Lobinska et al. 2022 | Evolution of resistance to COVID-19 vaccination with dynamic social distancing. | Each day, if the number of new infections exceeds (resp., is less than) $L$, then $s$ is decreased (resp., increased) by a random, uniformly distributed number between 0 and $s_0$.<br>$s :=$ social distancing parameter, dependent on the threshold number of new daily infections $L$; $s = 1$ is unconstrained social interaction, $s = 0$ is complete lockdown<br>$c :=$ daily rate of vaccination<br><br>→<br>$\dot{x} = -\beta sxy - \frac{cx}{x+z}$<br>$\dot{y} = \beta sxy - ay$<br>Where $x, y, z$ denote susceptible, infected, and recovered individuals, respectively. |

| Paper | Title | Behavioural modelling solution |
|---|---|---|
| Mahadhika et al. 2024 | A deterministic transmission model for analytics-driven optimization of COVID-19 post-pandemic vaccination and quarantine strategies | $\frac{dS}{dt} = \Lambda_h - u_{1S} - \beta_1 SI - \mu S$<br>$\frac{dV}{dt} = u_{1S} - \frac{\beta_2}{1 + b_1 I} VI - \mu V$<br>$\frac{dI}{dt} = \beta_1 SI + \frac{\beta_2}{1 + b_1 I} VI - u_{2I} - \gamma_0 I - \mu I$<br>$\frac{\beta_2}{1+b_1 I}$ ≔ media effect<br>$\beta_1, \beta_2$ ≔ contact rate between susceptible-infected and vaccinated-infected, respectively<br>$b_1$ ≔saturation coefficient<br>$S, V, I$ ≔ susceptible, vaccinated, and infected individuals, respectively<br><br>→<br>Vaccinated individuals are more propense to heightened awareness (assumption of the paper), and thus their infection term is modulated by media reports |
| Maji et al. 2021 | Impact of Media-Induced Fear on the Control of COVID-19 Outbreak: A Mathematical Study | $\frac{dS}{dt} = r - \frac{\beta SI}{1 + \alpha I} - \mu S$<br>$\frac{dE}{dt} = \frac{\beta SI}{1 + \alpha I} - (c + \mu)E$<br>$\frac{1}{1+\alpha I}$ ≔ fear effect<br>$S, E, I$ ≔ susceptible, exposed, infected individuals, respectively<br>$\alpha$ ≔ saturation constant[b] |
| Maji et al. 2022 | COVID-19 propagation and the usefulness of awareness-based control measures: A mathematical model with delay | $\frac{dS}{dt} = r - \overbrace{\beta e^{-mI(t-\tau)}}^{\eta(I):=} - \mu S$<br>$\frac{dE}{dt} = \beta e^{-mI(t-\tau)} - (c + \mu)E$<br>$\eta(I)$ ≔ infection rate dependent on public awareness<br>$S, E, I$ ≔ susceptible, infected but not infectious, and infected individuals, respectively |
| Mekonen et al. 2021 | Modeling the effect of contaminated objects for the transmission dynamics of COVID- | $\lambda = \frac{\beta_1(E + \nu I)}{N} + \frac{\beta_2 M}{M + K}$<br>$e(\lambda) = \frac{\lambda^n}{\lambda_0^n + \lambda^n}$ |

[b] In the paper, it is described as "the effect of fear through media coverage" in Table 1

| Paper | Title | Behavioural modelling solution |
|---|---|---|
| | 19 pandemic with self protection behavior changes | $\lambda$ ≔ force of infection<br>$e(\lambda)$ ≔ behaviour-change function, where $n$ is the Hill coefficient that portrays the rate of reaction by the population<br>$E, I, M$ ≔ latently infected, infected individuals, and contaminated materials/surfaces, respectively<br>$\beta_1, \beta_2$ ≔ transmission rate from humans and from environment, respectively<br>$K, \nu$ ≔ pathogen concentration in the environment and modification parameter, respectively<br><br>→<br>$\frac{dS}{dt} = \Lambda - (\lambda + \alpha e + \mu)S$<br>$\frac{dE}{dt} = \alpha e S - (\theta + \mu + (1 - \eta)\lambda)E$<br>Where $S$ denotes the susceptible individuals. |
| Mohsen et al. 2020 | Global stability of COVID-19 model involving the quarantine strategy and media coverage effects. | $\frac{dS(t)}{dt} = A - \left(\beta_1 - \beta_2 \frac{I}{m + I}\right) SI - dS$<br>$\beta_2 \frac{I}{m+I}$ ≔ media effect<br>$\beta_1, \beta_2$ ≔ baseline transmission and reduction term, respectively<br>$S, I$ ≔ susceptible and infected individuals, respectively<br><br>→<br>Awareness through media reports reduces the transmission rate (modulated by a constant $\beta_2$) |
| Park et al. 2021 | Inherently high uncertainty in predicting the time evolution of epidemics | $\beta_j = \beta_{j-1} \left\{1 - \delta \frac{I(t_j) - I(t_{j-1})}{I(t_j)}\right\}, \quad j = 1,2,3, \ldots$<br>$\delta \frac{I(t_j) - I(t_{j-1})}{I(t_j)}$ ≔ media/fear effect<br>$I(t_j)$ ≔ infectious individuals at time $t_j$<br>$\delta$ ≔ multiplication parameter<br><br>→<br>$\dot{S} = -\beta_j IS/N$<br>$\dot{E}(t) = \beta_j IS/N - \kappa E$ |
| Proano et al. 2024 | Belief-driven dynamics in a behavioral SEIRD macroeconomic model with sceptics | $\beta_t = \max\{0, \beta_0 - \phi_g(1 - \omega_t)G_t\}$ |

| Paper | Title | Behavioural modelling solution |
|---|---|---|
| | | $\omega_t = \frac{\exp(\mu_y U_t^S)}{3 + \exp(\mu_y U_t^S) + \exp(\mu_i U_t^C)}, S \in \{N, Y\}, C \in \{I, D\}$<br>$U_t^I = \Delta I_t, U_t^D = \Delta D_t, U_t^N = 0, U_t^Y = -y_t$<br>$\beta_0$ ≔ baseline transmission rate<br>$\omega_t$ ≔ share of sceptics: $C$ represent pull forces and $S$ represent push forces from/towards scepticism<br>$\phi G_t$ ≔ social distancing measures<br>$I_t, D_t$ ≔ infectious and dead individuals, respectively<br>$y_t$ ≔ economic consequences of social distancing measures<br><br>→<br>$\Delta S_{t+1} = \gamma_S R_t - \beta_t S_t I_t / N_t$<br>$\Delta E_{t+1} = \beta_t S_t I_t / N_t - \sigma E_t$<br>Where $S, E, R$ denote susceptible, exposed, and recovered individuals, respectively |
| Rahmandad et al. 2022 | Enhancing long-term forecasting: Learning from COVID-19 models | $\beta = \beta_0 w \frac{1}{(1 + af')^\gamma}$<br>$f'$ ≔ lagged daily deaths deaths<br>$w$ ≔ weather impact<br><br>→<br>$\frac{dS}{dt} = -\frac{\beta SI}{N}$<br>$\frac{dE}{dt} = \frac{\beta SI}{N} - \frac{E}{\tau_1}$<br>Where $S, E, I$ are susceptible, exposed, infectious individuals, respectively, and $\tau_1$ is the exposure period. |
| Shastry et al. 2022 | Policy and behavioral response to shock events: an agent-based model of the effectiveness and equity of policy design features | $baseline = \frac{transmissibility}{1 - transmissibility}$<br>$own = \frac{riskTolerance}{1 - riskTolerance}$<br>$local = \frac{mean(localRisk)}{1 - mean(localRisk)}$<br>$\Pr(event) = \frac{baseline + own + local}{1 + (baseline + own + local)}$<br>$riskTolerance_i = \frac{\exp(\beta_0 + \beta_1 Age_i + \beta_2 SickFrac_i)}{1 + \exp(\beta_0 + \beta_1 Age_i + \beta_2 SickFrac_i)}$ |

| Paper | Title | Behavioural modelling solution |
|---|---|---|
| | | $\Pr(event)$ ≔ probability of infected agent of exposing susceptible agents involved in an interaction with them, dependent on their own risk tolerance $riskTolerance$, the risk tolerance of all other agents $localRisk$, and a baseline transmissibility level $transmissibility$ (beta coefficients: free parameters)<br>$Age_i$ ≔ age of the agent $i$<br>$SickFrac_i$ ≔ proportion of agent $i$ social network that are either symptomatic or hospitalized<br><br>→<br>Agents will determine whether they will comply with the shelter-in-place order by ceasing social interactions, s.t. those with the highest tolerance will not comply. |
| Silal et al. 2023 | The National COVID-19 Epi Model (NCEM): Estimating cases, admissions and deaths for the first wave of COVID-19 in South Africa. | $behResp1$ = 110 deaths/day<br>$behResp2$ = 30 deaths/day<br><br>→<br>when district death rates reach (resp., go below) the threshold, people reduce (resp., increase) their interactions |
| Simeonov et al. 2023 | Modeling the drivers of oscillations in COVID-19 data on college campuses. | $f(I;r;C_{\max}) = \begin{cases} \left(1-\frac{I}{C_{\max}}\right)^r, & I < C_{\max} \\ 0, & I \geq C_{\max} \end{cases}$<br>$C_{\max}$ ≔ number of cases that prompt institutions to impose a total lockdown<br>$r$ ≔ individuals' perception of risk<br>$I$ ≔ infected individuals<br><br>→<br>$\frac{dS}{dt} = -\beta f(I,r,C_{\max})SI + qR$<br>Where $S, R$ denote susceptible and recovered individuals, respectively |
| Taki et al. 2022 | Understanding death risks of Covid-19 under media awareness strategy: a stochastic approach | $g(S,U) = \left(\alpha - \alpha_1 f(U)\right)\frac{SU}{N}$<br>$S, U$ ≔ susceptible and unquarantined individuals, respectively<br>$\alpha_1 f(U)$ ≔ media effect, where $\alpha_1$ is the baseline contact rate, and $f(U)$ is s.t. $f(0) = 0, f' > 0, \lim_{U\to\infty} f(U) = 1$[c] |

[c] Authors inspect some functional forms for $f_1(u)$ in the simulations. We do not report them here for conciseness.

| Paper | Title | Behavioural modelling solution |
|---|---|---|
| | | → <br> $\dot{S} = \Lambda N - g(S,U) - \mu_1 S$ <br> $\dot{U} = g(S,U) - \gamma_1 U - \mu_2 U - (1-\gamma_1)\delta U - \lambda_1 U$ |
| Thirthar et al. 2023 | Mathematical modeling of the COVID-19 epidemic with fear impact | $\frac{dS(t)}{dt} = \frac{A}{1+\alpha I(t)} - \beta(1-\gamma)(1-\delta)S(t)I(t) - \pi S(t)B(t) - (\mu_1 + \gamma + \delta)S(t) + \frac{aI(t)}{b+I(t)}$ <br> $\frac{1}{1+\alpha I(t)}$ ≔ fear effect, where $\alpha$ is defined as the rate of fear <br> $S, I, B$ ≔ susceptible, infected, and pathogen population, respectively <br><br> → <br> Awareness through media reports reduces the recruitment rate in the susceptible compartment and affects the recovery rate as well. |
| Tu et al. 2023 | A reaction-diffusion epidemic model with virus mutation and media coverage: Theoretical analysis and numerical simulation | $\frac{\partial S(x,t)}{\partial t} - \mathbb{D}_S \Delta S = \Lambda(x) - \frac{\beta_1(x)SI_1}{1+a_1 I_1} - \frac{\beta_2(x)SI_2}{1+a_2 I_2} - d(x)S + \sigma(x)R,\ x \in \Omega, t > 0$ <br> $\frac{\partial I_1(x,t)}{\partial t} - \mathbb{D}_{I_1}\Delta I_1 = \frac{\beta_1(x)SI_1}{1+a_1 I_1} - \gamma_1(x)I_1 - d(x)I_1 - \mu_1(x)I_1$ <br> $\frac{\partial I_2(x,t)}{\partial t} - \mathbb{D}_{I_2}\Delta I_2 = \frac{\beta_2(x)SI_2}{1+a_2 I_2} - \gamma_2(x)I_2 - d(x)I_2 - \mu_2(x)I_2$ <br> $S(x,t)$ ≔ proportion of susceptible individuals at space $x$ and time $t$ <br> $I_1, I_2$ ≔ infected with ordinary and mutated COVID-19 virus, respectively <br> $\frac{1}{1+a_i I_i}$ ≔ media effect, $i \in \{1,2\}$ denoting the COVID-19 strain of interest and $\alpha_1, \alpha_2$ denoting the corresponding media frequency <br><br> → <br> Media effect reduces the transmission rate |
| Usherwood et al. 2021 | A model and predictions for COVID-19 considering population behavior and vaccination | $f_I = e^{-d_I I}$ <br> $f_V = \frac{1}{f_I} + \left(1 - \frac{1}{f_I}\right) e^{-d_V V}$ <br> $f_I$ ≔ sense of caution <br> $f_V$ ≔ sense of safety <br> $I, V$ ≔ infectious and vaccinated individuals, respectively <br> $d_I, d_V$ ≔ population's level of caution and safety factors, respectively <br><br> → <br> $\beta = \beta_0 f_I f_V$ |

| Paper | Title | Behavioural modelling solution |
|---|---|---|
| | | $\frac{dS}{dt} = -\beta SI - \alpha$<br>$\frac{dV}{dt} = -\beta(1-\eta)VI + \alpha$<br>$\frac{dI}{dt} = (\beta S + \beta(1-\eta)V - \mu\gamma - \gamma)I$<br>Where $S$ denotes the vaccine effectiveness, and $\eta$ is the vaccine effectiveness. |
| Wang et al. 2020 | Studying social awareness of physical distancing in mitigating COVID-19 transmission | $f(I) = \begin{cases} \alpha_1, & if\ 0 < I < I_1 \\ \alpha_2, & if\ I_1 < I < I_2 \\ \alpha_2, & if\ I_2 < I \end{cases}$<br>$f(I)$ ≔ reduction in the contact rate dependent on symptomatic isolated/hospitalized individuals $I$ (threshold levels are constants $I_1, I_2$)<br><br>→<br>$\frac{dS}{dt} = -\beta_1(1 - f(I))AS - \beta_2(1 - f(I))US$<br>$\beta_1, \beta_2$ ≔ transmission rates with asymptomatic $A$ and unreported $U$ cases, respectively[d] |
| Warne et al. 2020 | Hindsight is 2020 vision: a characterisation of the global response to the COVID-19 pandemic | $\mathcal{U}(A_t, R_t, D_t) = w_A A_t + w_R R_t + w_D D_t$<br>$g(A_t, R_t, D_t) = \alpha_0 + \alpha f(\mathcal{U}(A_t, R_t, D_t))$<br>$f(\mathcal{U}(\cdot)) = \frac{1}{1 + \mathcal{U}(\cdot)^n}$<br>$A_t, R_t, D_t$ ≔ confirmed active cases, case recoveries, and case fatalities, respectively. They are determined as $\dot{A}_t = \gamma I_t - (\beta + \delta)A_t, \dot{R}_t = \beta A_t, \dot{D}_t = \delta A_t$<br>$\mathcal{U}(A_t, R_t, D_t)$ ≔ reporting function<br>$g(A_t, R_t, D_t)$ ≔ nonlinear transmission rate<br>$f(\mathcal{U}(\cdot))$ ≔ response function<br><br>→<br>$\dot{S}_t = -g(A_t, R_t, D_t)S_t I_t / P$<br>$\dot{I}_t = -(\gamma + \eta\beta)I_t + g(A_t, R_t, D_t)S_t I_t / P$<br>Where $S, I, P$ denote the susceptible, infectious, and total population, respectively |
| Weitz et al. 2020 | Awareness-driven behavior changes can shift the shape of | Model A (short-term awareness of risk): $\dot{S} = -\frac{\beta SI}{[1+(\delta/\delta_c)^k]}, \delta = \dot{D} = f_D \gamma I$ |

[d] Authors assume that confirmed infected individuals cannot transmit the disease.

| Paper | Title | Behavioural modelling solution |
|---|---|---|
| | epidemics away from peaks and toward plateaus, shoulders, and oscillations | Where $S, I, D$ denote susceptible, infectious, and dead individuals, respectively<br><br>Model B (short-term awareness of risk, with lags): as model A, but changes $\dot{D}$ dynamics → $\dot{H} = f_D \gamma I - \gamma_H H, \dot{D} = \gamma_h H$<br>where $H$ denote hospitalization<br><br>Model C (short-and long-term awareness of risk): as model B, but changes $\dot{S}$ dynamics → $\dot{S} = -\frac{\beta SI}{[1+(\delta/\delta_c)^k + (D/D_c)^k]}$<br><br>Model D (adding fatigue): as model B, but changes $\dot{S}$ dynamics → $\dot{S} = -g(D)\beta SI,$ $\dot{\beta} = \frac{\epsilon}{2}\left[\frac{\left(\frac{\hat{\beta}}{1+(\delta/\delta_c)^k} - \beta\right)}{1+(D/D_c)^k} + (\hat{\beta} - \beta)\right]$<br><br>Model D.1: $g(D) = 1$; Model D.2: $g(D) = \frac{1}{1+(D/D_c)^k}$<br><br>$\delta_c, D_c$ ≔ half-saturation constants<br>$\hat{\beta}, \epsilon$ ≔ baseline behavior and time scale for behaviour change, respectively |
| Witbooi et al. 2023 | Control and elimination in an SEIR model for the disease dynamics of COVID-19 with vaccination | $\dot{S} = A_0 - \theta SE - (\mu + z)S$<br>$\dot{E} = A_1 + \theta SE - (\mu + \delta_1 + \alpha + \omega_1)E$<br>$\theta = \theta(I) = \frac{\beta}{1+bI}$<br>$\theta(I)$ ≔ awareness-dependent contact rate<br>$S, E, I$ ≔ susceptible, asymptomatic but infectious, and symptomatic infectious individuals, respectively<br><br>→<br>Awareness reduces the contact rate |
| Yousef 2022 | A fractional-order model of COVID-19 with a strong Allee effect considering the fear effect spread by social networks to the community and the existence of | $D^\alpha S(t) = \Lambda_1 + S(t)r\left(1 - \frac{S(t)}{K_1}\right)\frac{1}{1+\alpha_1 I(t)}(S(t) - \rho) - \beta_1 E(t)S(t) - \beta_2 I(t)S(t) - \eta S(t) + \delta_2 R(t)$<br>$D^\alpha Q(t) = \frac{Q(t)}{1+\alpha_2 D(t)} + \gamma_1 I(t) - \eta Q(t) - \mu Q(t) - \gamma_2 Q(t)$<br>$\frac{1}{1+\alpha_1 I}$ ≔ effect of the fear of getting infected from COVID-19, where $\alpha_1$ is the level of fear to get infected |

| Paper | Title | Behavioural modelling solution |
|---|---|---|
| | the silent spreaders during the pandemic stage | $\frac{1}{1+\alpha_2 D}$ := effect of the fear of dying from COVID-19, where $\alpha_2$ is the level of fear of the death from COVID-19<br>$S, I, Q, R$ := susceptible, exposed, infected, isolated/quarantined, and recovered individuals, respectively<br><br>→<br>Fear/awareness act on both the transmission rate and the quarantine rate |
| Zhou et al. 2023 | Transmission trends of the global COVID-19 pandemic with combined effects of adaptive behaviours and vaccination. | $\frac{dS(t)}{dt} = A - \lambda_1 f \frac{S(t)}{N(t)}\left(I_c(t) + \alpha I_b(t)\right) - (\vartheta + d)S(t)$<br>$f = \left(1 - \frac{I + \phi R}{N}\right)^k$<br>$\phi$ := long-term awareness of protection<br>$I_c, I_b$ := individuals with asymptomatic and clinical infection, respectively<br>$S, I, R$ := susceptible, infectious ($I_c + I_b$), and recovered individuals, respectively<br><br>→<br>Long-term awareness lowers the transmission rate |

## Supplementary Material S4.II: Modelling solutions adopted by papers in the "**Public perception and Government Response**" category

| Paper | Title | Behavioural modelling solution |
|---|---|---|
| Ajbar et al. 2020 | Bifurcation analysis of a SEIR epidemic system with governmental action and individual reaction | $\beta = \beta_0(1-\mu)\, exp\left(\frac{-\kappa P}{N}\right)$<br><br>$\beta, \beta_0$ ≔ perceived risk-varying and baseline transmission rate, respectively<br>$P$ ≔ public risk perception evolving according to infected individuals,<br>$\frac{dP}{dt} = e\gamma I - \lambda P$<br>$\mu$ ≔ governmental action strength<br>$\kappa$ ≔ strength of public response |
| Angulo et al. 2021 | A modified SEIR model to predict the behavior of the early stage in coronavirus and coronavirus-like outbreaks | $\beta_t = \beta_0(1-\alpha)[1-p(t)]^{\kappa}$<br><br>$\beta_t, \beta_0$ ≔ perceived risk-varying and baseline transmission rate, respectively<br>$p(t)$ ≔ public risk perception evolving according to individuals expected to die,<br>$\partial_t p(t) + \lambda p(t) - gd(t) = 0$<br>$\kappa$ ≔ strength of public response<br>$\alpha$ ≔ governmental action strength |
| Danane et al. 2023 | A fractional-order model of coronavirus disease 2019 (COVID-19) with governmental action and individual reaction | $\beta(t) = \beta_0(1-\rho)\, exp\left(1-\frac{D}{N}\right)^{\kappa}$<br><br>$\beta(t), \beta_0$ ≔ perceived risk-varying and baseline transmission rate, respectively<br>$D$ ≔ public risk awareness evolving according to<br>$D^{\alpha}D = d\gamma I - \lambda D$<br>$\kappa$ ≔ strength of public response<br>$\rho$ ≔ governmental action strength |
| Das et al. 2022 | A Fractional Order Model to Study the Effectiveness of Government Measures and Public Behaviours in COVID-19 Pandemic | $\Phi = (1-\alpha)[\beta_1^{\varepsilon} S I_s (1-Q)^{\kappa} + \beta_2^{\varepsilon} S I_a]$<br><br>$\Phi$ ≔ "infection function"<br>$Q$ ≔ social behavioral dynamics state variable, evolving according to symptomatic individuals,<br>${}_{t_0}^{C}D_t^{\varepsilon} Q(t) = d\rho_2^{\varepsilon} I_s - \lambda^{\varepsilon} Q$ |

| Paper | Title | Behavioural modelling solution |
| --- | --- | --- |
| | | $\kappa \coloneqq$ strength of public response<br>$\alpha \coloneqq$ governmental action strength<br>$I_a \coloneqq$ asymptomatic infected individuals<br>$\beta_1, \beta_2 \coloneqq$ rate of infection per unit of time by symptomatic and asymptomatic infected population, respectively |
| Furati et al. 2021 | Fractional model for the spread of COVID-19 subject to government intervention and public perception | $\beta(N, P) = \beta_0^q(1-\nu)\left(1-\frac{P(t)}{N(t)}\right)^\kappa$<br><br>$\beta, \beta_0 \coloneqq$ perceived risk-varying transmission and baseline transmission rate, respectively<br>$P \coloneqq$ public risk perception evolving according to<br>${}^cD^q P = \phi\gamma^q I - \omega^q P$<br>$\kappa \coloneqq$ strength of public response<br>$\nu \coloneqq$ governmental action strength<br>$I \coloneqq$infectious individuals<br>${}^cD^q \coloneqq$ Caputo fractional derivative of order $q$ |
| Kwuimy et al. 2020 | Nonlinear dynamic analysis of an epidemiological model for COVID 19 including public behavior and government action | $\Upsilon = (1-\alpha)[\beta_1 SI(1-D)^\kappa + \beta_2 SE]$<br><br>$\Upsilon \coloneqq$ infection function<br>$\beta_1, \beta_2 \coloneqq$ transmission rates for symptomatic (infectious) and asymptomatic (exposed) individuals, respectively<br>$D \coloneqq$ social behavioral dynamics evolving according to<br>$D' = d\gamma I - \lambda D$<br>$\kappa \coloneqq$ strength of public response<br>$\alpha \coloneqq$ governmental action strength<br>$S, I, E$:= susceptible, exposed, and infectious individuals, respectively |
| Lin et al. 2020 | A conceptual model for the coronavirus disease 2019 (COVID-19) outbreak in Wuhan, China with individual reaction and governmental action | $\beta(t) = \beta_0(1-\alpha)\left(1-\frac{D}{N}\right)^\kappa$<br><br>$\beta(t), \beta_0 \coloneqq$ perceived risk-varying transmission and baseline transmission rate, respectively<br>$D \coloneqq$ public risk awareness evolving according to<br>$\dot{D} = d\gamma I - \lambda D$<br>$\kappa \coloneqq$ strength of public response<br>$\alpha \coloneqq$ governmental action strength<br>$I \coloneqq$ infectious individuals |

| Paper | Title | Behavioural modelling solution |
|---|---|---|
| Mushanyu et al. 2021 | Modelling the impact of detection on COVID-19 transmission dynamics in Ghana | $\beta(t) = \beta_0(1-\alpha)\left(1-\frac{P}{N}\right)^{\kappa}$<br><br>$\beta(t), \beta_0$ ≔ perceived risk-varying transmission and baseline transmission rate, respectively<br>$P$ ≔ public perception evolving according to<br>$P' = \sigma_1 E + \sigma_2 I + \delta_2 Q - \omega P$<br>$\kappa$ ≔ strength of public response<br>$\alpha$ ≔ governmental action strength<br>$E, I, Q$ ≔ exposed, infected, and detected quarantined individuals, respectively |
| Mushayabasa et al. 2020 | On the role of governmental action and individual reaction on COVID-19 dynamics in South Africa: A mathematical modelling study. | $\beta(t) = \beta_0(1-\alpha)\left(1-\frac{P(t)}{N(t)}\right)^{\kappa}$<br><br>$\beta(t), \beta_0$ ≔ perceived risk-varying transmission and baseline transmission rate, respectively<br>$P$ ≔ public risk perception evolving according to<br>$P'(t) = p_1\gamma I_m(t) + p_2\delta I_s(t) + \omega A(t) + p_3\eta Q(t) - \lambda P(t)$<br>$\kappa$ ≔ strength of public response<br>$\alpha$ ≔ governmental action strength<br>$I_m, I_s, A, Q$ ≔ undetected symptomatic patients with mild symptoms, undetected symptomatic patients with severe symptoms, undetected asymptomatic, and detected patients, respectively |
| Oz 2022 | A Theoretical Model to Investigate the Influence of Temperature, Reactions of the Population and the Government on the COVID-19 Outbreak in Turkey | $\beta(t) = \beta_0 e^{-\xi T(t)}(1-\alpha(t))\left(1-\frac{P(t)}{N}\right)^{\kappa}$<br><br>$\beta(t), \beta_0$ ≔ perceived risk-varying transmission and baseline transmission rate, respectively<br>$P$ ≔ public risk perception evolving according to<br>$\frac{d}{dt}P(t) = d\gamma I(t) - \lambda P(t)$<br>$\kappa$ ≔ strength of public response<br>$\alpha$ ≔ governmental action strength<br>$I(t)$ ≔ infectious individuals<br>$T(t)$ ≔ temperature, whose effects are captured by the parameter $\xi$ |

# Supplementary Material S4.III: Modelling solutions adopted by papers in the “**Awareness**” category

*Awareness contagion process*

| Paper | Title | Behavioural modelling solution |
|---|---|---|
| Abbas et al. 2022 | Evolution and consequences of individual responses during the COVID-19 outbreak | Mod. II of Perra et al. 2011:<br>$\frac{dS}{dt} = \Lambda - \frac{\beta SI}{N} - \beta_f S(1 - e^{-\delta Q}) + \mu_f S_f \left(\frac{S+R}{N}\right) - \mu S$<br>$\frac{dS_f}{dt} = -\frac{r_\beta \beta S_f I}{N} + \beta_f S(1 - e^{-\delta Q}) - \mu_f S_f \left(\frac{S+R}{N}\right) - \mu S_f$<br>$S, S_f$ ≔ susceptible and behavior-changed susceptible individuals, respectively<br>$r_\beta$ ≔ reduction factor in the transmission rate due to behavioral change<br>$I, Q, R$ ≔ infected, quarantined, recovered individuals, respectively<br>$\beta_f$ ≔ “transmission rate of fear”<br><br>→<br>Probability of infection differs between adherent and non-adherent individuals (difference modulated by a constant $r_\beta$) |
| Althobaity et al. 2022 | Non-pharmaceutical interventions and their relevance in the COVID-19 vaccine rollout in Saudi Arabia and Arab Gulf countries | Mod. II of Perra et al. 2011:<br>$\frac{dS_m}{dt} = -\lambda_m S_m - (1 - e^{-g(\delta)})S_m - (1 - e^{-h(\gamma)})S_m^{NA}$<br>$\frac{dS_m^{NA}}{dt} = -r\lambda_m S_m^{NA} + (1 - e^{-g(\delta)})S_m - (1 - e^{h(\gamma)})S_m^{NA}$<br>$\frac{dV_m}{dt} = -(1 - VE_s)\lambda_m V_m - (1 - e^{-g(\delta)})V_m + (1 - e^{-h(\gamma)})V_m^{NA}$<br>$\frac{dV_m^{NA}}{dt} = -r(1 - VE_s)\lambda_m V_m^{NA} + (1 - e^{-g(\delta)})V_m - (1 - e^{-h(\gamma)})V_m^{NA}$<br>$\lambda_{X^{NA} \to X} = 1 - e^{-h(\gamma)} = 1 - e^{-\gamma d_{t-1}^0}$<br>$\lambda_{X \to X^{NA}} = 1 - e^{-g(\delta)}$<br><br>$d_{t-1}^0$ ≔ number of deaths per 100,000 recorded in the preceding time step<br>$\delta$ ≔ constant rate of transition from adherent to non-adherent<br>$S_m, S_m^{NA}$ ≔ susceptible of age group $m$ adhering and non-adhering to protective behaviours, respectively |

| Paper | Title | Behavioural modelling solution |
|---|---|---|
| | | $V_m, V_m^{NA}$ ≔ vaccinated of age group $m$ adhering and non-adhering to protective behaviours, respectively<br>$VE$ ≔ vaccine effectiveness<br><br>→<br>Probability of infection differs between adherent and non-adherent individuals (difference modulated by a constant $r$) |
| Gao et al. 2022 | Epidemic Spreading in Metapopulation Networks Coupled With Awareness Propagation | Mod. III of Perra et al. 2011:<br>$U + A \xrightarrow{\lambda} 2A, \quad A \xrightarrow{\delta} U$<br>$SIR: S + I \xrightarrow{\beta} 2I, \quad I \xrightarrow{\mu_1} S$<br>$SIS: S + I \xrightarrow{\beta} 2I, \quad I \xrightarrow{\mu_2} R$<br>$S, I, R, U, A$ ≔ susceptible, infected, recovered, unaware, and aware individuals, respectively<br><br>→<br>infected individuals become aware of the disease after infection;<br>being aware is related to a lower infection rate: $\gamma\beta^U = \beta^A, \gamma < 1$ |
| He et al. 2021 | Modeling the COVID-19 epidemic and awareness diffusion on multiplex networks | Mod. III of Perra et al. 2011:<br>$U + W \xrightarrow{\lambda} 2W, \quad W \xrightarrow{\delta} U$ [1]<br><br>→<br>infected individuals become aware of the disease after infection;<br>being aware is related to a lower infection rate: $\beta_S^W = \gamma\beta_S^U, \gamma \in [0,1]$ |
| Kim et al. 2020 | The impact of social distancing and public behavior changes on COVID-19 transmission dynamics in the Republic of Korea | Mod. II of Perra et al. 2011:<br>$\frac{dS}{dt} = -\beta\frac{I}{N}S - \beta_F(1 - e^{-\tau Q})S + \mu\frac{S+R}{N}S_F$<br>$\frac{dS_F}{dt} = \beta_F(1 - e^{-\tau Q})S - \frac{\delta\beta I}{N}S_F - \mu\frac{S+R}{N}S_F$<br>$S, S_F$ ≔ susceptible and behavior-changed susceptible, respectively<br>$Q, I, R$ ≔ quarantined, infected, recovered individuals, respectively<br>$\delta$ ≔ reduction factor in susceptibility for behavior-changed individuals |

[1] Such an expression is not available in the full-text as reported here, but just as plain text. We harmonized the notation with other papers included in the review for easier comparison.

| Paper | Title | Behavioural modelling solution |
| --- | --- | --- |
| | | $\beta_F, \mu$ ≔ rate of change from susceptible to behavior-changed susceptible and vice versa respectively<br><br>→<br>Probability of infection differs between susceptible and behaviour-changed individuals (difference modulated by a constant $\delta$) |
| Mendoza et al. 2023 | Managing bed capacity and timing of interventions: a COVID-19 model considering behavior and underreporting | Mod. II of Perra et al. 2011:<br>$\frac{dS}{dt} = -\beta S \frac{I + I_u}{\tilde{N}} + \mu S_F \frac{S + R}{\tilde{N}} - \beta_F S \frac{Q}{\tilde{N}}$<br>$\frac{dS_F}{dt} = -\delta\beta S_F \frac{I + I_u}{\tilde{N}} - \mu S_F \frac{S + R}{\tilde{N}} + \beta_F S \frac{Q}{\tilde{N}}$<br>$S, S_F$ ≔ susceptible and behavior-changed susceptible, respectively<br>$\delta$ ≔ reduction factor in transmission rate for behavior-changed susceptible,<br>$I, I_u, Q, R$ ≔ reported infectious, unreported infectious, isolated, and recovered individuals, respectively.<br>$\beta, \beta_F$ ≔ transmission rate of COVID-19 and of awareness or fear, respectively<br><br>→<br>Probability of infection differs between susceptible and susceptible-aware individuals (difference modulated by a constant $\delta$) |
| Teslya et al. 2020 | Impact of self-imposed prevention measures and short-term government-imposed social distancing on mitigating and delaying a COVID-19 epidemic: A modelling study | Mod. I of Perra et al. 2011:<br>$\frac{dS(t)}{dt} = -S(t)\lambda_{inf}(t) - kS(t)\lambda_{aware}(t) + \mu S^a(t)$<br>$\frac{dS^a(t)}{dt} = -S^a(t)\lambda^a_{inf}(t) + kS(t)\lambda_{aware}(t) - \mu S^a(t)$<br>$\lambda_{aware}(t) = \delta \cdot [I_D(t) + I^a_D(t)]$<br><br>$S, S^a$ = unaware and aware susceptible, respectively<br>$\lambda_{aware}(t)$ ≔ awareness acquisition rate, dependent on number of diagnosed unaware $I$ and aware $I^a$ individuals<br>$S, S^a$ ≔ susceptible unaware and aware, respectively<br><br>→<br>Difference in state of awareness (and disease severity) affects the transmission rates, which is captured by |

| Paper | Title | Behavioural modelling solution |
|---|---|---|
| | | $M(t) = \begin{array}{c} \\ unaware\ S \\ aware\ S \end{array} \begin{array}{cccc} unaware\ I_M & unaware\ I_S & aware\ I_M & aware\ I_S \\ M_{11}(t) & M_{12}(t) & M_{13}(t) & M_{14}(t) \\ M_{21}(t) & M_{22}(t) & M_{23}(t) & M_{24}(t) \end{array}$<br>The entries of the $M(t)$ matrix depend on the protective behaviour being modelled (mask-wearing, handwashing, self-imposed social distancing, short-term government-imposed social distancing). |
| Teslya et al. 2022 | The importance of sustained compliance with physical distancing during COVID-19 vaccination rollout. | Mod. I of Perra et al. 2011:<br>$\frac{dS(t)}{dt} = -\lambda_{\text{inf}}(t)S(t) - \Psi_C(t)S(t) + \mu(t)S^C(t) - \upsilon S(t)$<br>$\frac{dS^C(t)}{dt} = -\lambda_{\text{inf}}^C(t)S^C(t) + \Psi_C(t)S(t) - \mu(t)S^C(t) - \upsilon S^C(t)$<br>$\Psi_C(t) = \delta \cdot \alpha \cdot [E(t) + E^C(t) + E^V(t)]$<br>$\mu(t) = \mu_0 + \mu_1 \bar{V}(t)/N$<br>$S, S^C$ := susceptible and susceptible compliant with social distancing individuals, respectively<br>$\Psi_C(t)$ := rate of social distancing "compliance gain"<br>$\mu$ := average duration of compliance state<br>$E, E^C, E^V$ := exposed, exposed compliant, and exposed that vaccinated but did not gain immunity, respectively<br><br>→<br>Difference in state of awareness (but also disease severity and vaccination status) affects the transmission rates, which is captured by<br>$M = \frac{c\epsilon}{N(t)+r_1N^C(t)+r_2N^V(t)} \begin{bmatrix} 1 & r_1 & r_2 \\ r_1 & r_1^2 & r_1r_2 \\ r_2 & r_1r_2 & r_2^2 \end{bmatrix}$<br>Where $N, N^C, N^V$ denote the total non-compliant, total compliant, and total vaccinated but without immunity population. Matrix rows correspond to $S, S^C, S^V$, while matrix columns to $I, I^C, I^V$, respectively. $r_1, r_2$ are ratio of contact rates between classes (non-compliant vs compliant and vaccinated vs non-compliant, respectively), while $c, \epsilon$ are average contact rate of non-compliant individuals and probability of transmission per contact, respectively. |
| Zhao et al. 2021 | The impact of awareness diffusion on the spread of COVID-19 based on a two-layer SEIR/V-UA epidemic model | Mod. III of Perra et al. 2011:<br>$\frac{dS}{dt} = -\frac{\beta S}{N}I - p\frac{S}{N}A$<br>$\frac{dV}{dt} = \mu I + p\frac{S}{N}A$ |

| Paper | Title | Behavioural modelling solution |
|---|---|---|
| | | $\frac{dA}{dt} = \alpha \frac{U}{N} A + \gamma E - \mu I$<br>$\frac{dU}{dt} = -\alpha \frac{U}{N} A - \gamma E + \mu I$<br>$U, A \coloneqq$ unaware and aware individuals, respectively<br>$p \coloneqq$ probability of being vaccinated once individuals become aware<br>$S, E, I, V \coloneqq$ susceptible, exposed, infected, and vaccinated individuals, respectively<br><br>→<br>Awareness induces people to get vaccinated. Vaccinated individuals cannot get infected and are thus excluded from the transmission process |

*Awareness-inducing compartment ("semi-implicit")*

| Paper | Title | Behavioural modelling solution |
|---|---|---|
| Adeniyi et al. 2020 | Dynamic model of COVID-19 disease with exploratory data analysis | $\beta(H) = \beta_{\max} - H(\beta_{\max} - \beta_{\min})$<br>$H = H_0 + \frac{(1 - H_0)\omega_0 E}{1 + E}$<br>$\frac{dE}{dt} = \frac{\omega_0 I(t)}{\omega_1 + \omega_2 I(t)} - dE(t)$<br>$\lambda(H) = p_1 \beta(H) I$<br>$\lambda(H)$ ≔ force of infection<br>$\beta(H)$ ≔ transmission rate of the disease dependent on the sanitation level $H$, which in turn is dependent on educational campaigns $E$<br>$E$ ≔ educational campaigns compartment<br>$I$ ≔ infectious individuals<br><br>→<br>$\frac{d}{dt} S(t) = \Lambda - \lambda(H) S(t) - \mu S(t) + \alpha_2 Q(t) + \delta R(t)$<br>$\frac{d}{dt} Q(t) = \lambda(H) S(t) - (\alpha_1 + \alpha_2 + \mu) Q(t)$<br>Where $S, Q, R$ denote the susceptible, quarantined, recovered department, respectively |
| Agaba et al. 2020 | Analysing the spread of COVID-19 using delay epidemic model with awareness | $\frac{dS_n}{dt} = -\frac{\beta S_n I}{N} - \eta M(t - \tau) S_n + \lambda_k S_k$<br>$\frac{dS_k}{dt} = -\frac{\beta \phi S_k I}{N} + \eta M(t - \tau) S_n - \lambda_k S_k$<br>$\frac{dM}{dt} = \omega_0 + \alpha_0 I + \frac{\phi_0 S_k}{N} - \lambda_0 M$<br><br>$S_n, S_k$ ≔ unaware and aware susceptible population, respectively<br>$M$ ≔ awareness compartment, both from local ($\phi_0$: rate of awareness rising from aware individuals) and global sources<br>$I$ ≔ infective individuals (symptomatic + asymptomatic)<br><br>→<br>Probability of infection differs between unaware and aware individuals (difference modulated by a constant $\phi$) |

| Paper | Title | Behavioural modelling solution |
| --- | --- | --- |
| Baba et al. 2020 | Awareness as the most effective measure to mitigate the spread of COVID-19 in Nigeria | $\frac{dS}{dt} = \Lambda - \beta SI - \lambda SA - dS + \Upsilon I + \theta S_A$<br>$\frac{dS_A}{dt} = \lambda SA - dS_A - \theta S_A$<br>$\frac{dA}{dt} = \mu I - \phi A$<br>$A :=$ awareness programs compartment<br>$S, S_A :=$ susceptible without and with awareness, respectively<br>$I :=$ infective population<br><br>→<br>Probability of infection differs between susceptible with and without awareness (only individuals without awareness can be infected) |
| Chang et al. 2021 | Study on an SIHRS Model of COVID-19 Pandemic With Impulse and Time Delay Under Media Coverage | $\frac{dS(t)}{dt} = B - \beta cS(t)\left(I(t) + \theta H(t)\right) - \varphi_1 S(t)Y(t) + q_1 S_m(t) - dS(t)$<br>$\frac{dS_m(t)}{dt} = \varphi_1 S(t)Y(t) - q_1 S_m(t) + \gamma_1 R(t) - dS_m(t)$<br>$\frac{dI(t)}{dt} = \beta cS(t)\left(I(t) + \theta H(T)\right) - \varphi_2 I(t)Y(t) - \alpha I(t) + q_2 I_m(t) - dI(t)$<br>$\frac{dI_m(t)}{dt} = \varphi_2 I(t)Y(t) - \alpha I_m(t) - q_2 I_m(t)$<br>$Y(t) = \frac{M_1(t) - M_2(t) + M_3(t)}{K + M_1(t) + M_3(t)}$<br>$\frac{dM_1(t)}{dt} = e_1 + u_1 f\left(H(t) - \mu_0 M_1(t)\right)$<br>$\frac{dM_2(t)}{dt} = e_2 + u_2 f\left(H(t)\right) + u_3 f\left(w(t)\right) - u_5 M_3(t) - \mu_0 M_2(t)$<br>$\frac{dM_3(t)}{dt} = u_4 f(L) - \mu_0 M_3(t)$<br>$S, S_m :=$ susceptible without and with active isolation and protection awareness, respectively<br>$M_1, M_2, M_3 :=$ information coming from positive emotions, negative emotions, and policies/regulations (restraining negative emotions), respectively<br>$H, I, I_m, R, L :=$ hospitalized, infected without and infected with awareness, recovered individuals and "audiences who can be affected by policies and regulations information reports", respectively<br><br>→ |

| Paper | Title | Behavioural modelling solution |
| --- | --- | --- |
| | | Probability of infection differs between susceptible without and with awareness (only individuals without awareness can be infected), and probability of infect others differs between infected with and without awareness (only infected without awareness can infect) |
| Deng et al. 2021 | The impacts of anti-protective awareness and protective awareness programs on COVID-19 outbreaks | $\frac{dS_a}{dt} = (1-\lambda)\Lambda - \frac{\beta\varepsilon_1 S_a I}{N} + \delta_2 S_u M_2 - \delta_1 S_a M_1 - (p+d)S_a$<br>$\frac{dS_u}{dt} = \lambda\Lambda - \frac{\beta S_u I}{N} - \delta_2 S_u M_2 + \delta_1 S_a M_1 + pS_a - dS_u$<br>$\frac{dM_1}{dt} = \eta_1 S_u - \upsilon_1 M_1$<br>$\frac{dM_2}{dt} = \eta_2 I - \upsilon_2 M_2$<br>$S_a, S_u$ ≔ aware of the disease (with a certain protective consciousness) and unaware of the disease (without protective consciousness)<br>$M_1, M_2$ ≔ anti-protective and protective awareness programs induced by media<br>$I$ ≔ infected individuals<br><br>→<br>Probability of infection differs between unaware and aware individuals (difference modulated by a constant $\varepsilon_1$) |
| Guo et al. 2022 | Discrete epidemic modelling of COVID-19 transmission in Shaanxi Province with media reporting and imported cases | $S_{t+1} = S_t - \frac{\left(c(M_t)\beta + q(M_t)c(M_t)(1-\beta)\right)S_t I_t}{N_t} + \left(1-e^{-\lambda}\right)S_{q,t}$<br><br>$H_{t+1} = H_t + e^{-\alpha}\left(1-e^{-\delta_I}\right)I_t + (1-e^{-\sigma})E_{q,t} - \left(1-e^{-\delta_H}\right)H_t + \xi\bar{h}_t$<br>$M_{t+1} = M_t + \eta(H_t + H_{ct}) - (1-e^{-\mu_M})M_t$<br>$M_t$ ≔media reports<br>$H_t, H_{ct}$ ≔ hospitalized not confirmed and hospitalized confirmed, respectively<br>$S_t, S_{q,t}, I_t, E_t, E_{q,t}$ ≔ susceptible, quarantined susceptible, infected, exposed, and quarantined exposed, respectively<br><br>→<br>Media reports affect the transmission rate |
| Gutierrez-Jara et al. 2022 | Risk Perception Influence on Vaccination Program on COVID-19 in Chile: A Mathematical Model | $\beta_x = \beta_x^* \cdot \frac{P^*}{P},\ x \in \{a, i\}$<br>$\beta_x^v = \beta_x^* \cdot \frac{P_v^*}{P_v}$<br>$P' = -\Lambda_1(P - P^*) + \Lambda_2(I + I_v + T + T_v)$<br>$P_v' = -\Lambda_1^v(P_v - P_v^*) + \Lambda_2^v(I + I_v + T + T_v)$ |

| Paper | Title | Behavioural modelling solution |
| --- | --- | --- |
| | | $\beta_x, \beta_x^v$ ≔ transmission rate for unvaccinated and vaccinated individuals, respectively. $\{a, i\}$ denote transmission by asymptomatic and symptomatic individuals.<br>$P, P^*$ ≔ perception of risk and perception of average risk, respectively, for unvaccinated individuals<br>$P_v, P_v^*$ ≔ perception of risk and perception of average risk, respectively, for vaccinated individuals<br><br>→<br>$S' = bN - S[(\beta_a A + \beta_i I)/N_u + \delta(\beta_a A_v + \beta_i I_v)/N_v] + \psi R - (f + d)S + f_v S_v$<br>$S_v' = fS - \sigma S_v[(\beta_a^v A + \beta_i^v I)/N_u + \delta(\beta_a^v A_v + \beta_i^v I_v)/N_v] + \psi_v R_v - (f_v + d)S_v$<br>$E' = S[(\beta_a A + \beta_i I)/N_u + \delta(\beta_a A_v + \beta_i I_v)/N_u] - (a + d)E$<br>$E_v' = \sigma S_v[(\beta_a^v A + \beta_i^v I)/N_u + \delta(\beta_a^v A_v + \beta_i^v I_v)/N_v] - (a_v + d)E_v$<br>Where $S, S_v, I, I_v, N_v, N_u, E, E_v, A, A_v, R, R_v$ denote the susceptible, susceptible vaccinated, infected, infected vaccinated, total unvaccinated, total vaccinated, exposed, exposed vaccinated, asymptomatic, asymptomatic vaccinated, recovered, and recovered vaccinated populations. The subscripts $\{a, i\}$ denote transmission by asymptomatic and symptomatic individuals. |
| Johnston et al. 2020 | A dynamical framework for modeling fear of infection and frustration with social distancing in COVID-19 spread | $\frac{d\omega}{dt} = k_\omega(f(P_I, P_\omega) - \omega)$<br>$\frac{dP_I}{dt} = k_{P_I}(g(\lambda E) - P_I)$<br>$\frac{dP_\omega}{dt} = k_{P_\omega}(h(\omega) - P_\omega)$<br>$\omega$ ≔ social distancing behavior depending on the level of new daily infections<br>$P_I, P_\omega$ ≔ perceived fear of infection and perceived frustration with social distancing, respectively<br>$E$ ≔ exposed (non-infectious) individuals (s.t. $\lambda E$ is interpreted as the number of new active cases)<br><br>→<br>$\omega = \frac{(\lambda E)^q}{M^q + (1+\omega^*)(\lambda E)^q}$<br>$\frac{dS}{dt} = -\frac{\beta(1-\omega)}{N} SI$<br>$\frac{dE}{dt} = \frac{\beta(1-\omega)}{N} SI - \lambda E$<br>$\frac{dI}{dt} = \lambda E - \gamma I$ |
| Lacitignola et al. 2021 | Managing awareness can avoid hysteresis in disease spread: an | $\dot{S} = \Lambda - \mu S - \beta I\ (1 + aI)S - \gamma m S$<br>$\dot{m} = \rho I - \rho_0 m$ |

| Paper | Title | Behavioural modelling solution |
|---|---|---|
| | application to coronavirus Covid-19 | $m$ ≔ awareness programs<br>$I$ ≔ infective individuals<br><br>→<br>Aware individuals are removed from the infection process (they can't get infected) |
| Li et al. 2021 | Linking the disease transmission to information dissemination dynamics: An insight from a multi-scale model study | $\frac{dS_1}{dt} = \mu - \beta S_1 I_1 - \sigma_1 \beta S_1 I_2 - \mu S_1 - \alpha B S_1 + q S_2$<br>$\frac{dS_2}{dt} = -\sigma_S \beta S_2 I_1 - \sigma_1 \sigma_S \beta S_2 I_2 - \mu S_2 + \alpha B S_1 - q S_2$<br>$\frac{dI_1}{dt} = \beta S_1 I_1 + \sigma_1 \beta S_1 I_2 - \frac{\gamma I_1}{1 + h I_1} - \mu I_1 - \alpha B I_1 + q I_2$<br>$\frac{dI_2}{dt} = \sigma_S \beta S_2 I_1 + \sigma_S \sigma_I \beta S_2 I_2 - \frac{\gamma I_2}{1 + h I_2} - \mu + \alpha B I_1 - q I_2$<br>$\frac{dB}{dt} = \rho(I_1 + \theta I_2) - dB$<br>$S_1, S_2$ ≔ disease unaware and disease aware susceptible, respectively<br>$B$ ≔ average number of news items related to the outbreak<br>$\alpha$ ≔ rate at which the disease-unaware become disease-aware<br>$I_1, I_2$ ≔ disease unaware and disease aware infected, respectively<br><br>→<br>Transmission rate is reduced when infection occurs between disease-aware and disease-unaware populations (modulated by constants $\sigma_1$ and $\sigma_S$). |
| Li et al. 2022 | Complex dynamics of an epidemic model with saturated media coverage and recovery | $\frac{dS}{dt} = \Lambda - \beta e^{-\alpha M} SI - dS + \frac{\gamma I}{1 + h_1 I}$<br>$\frac{dE}{dt} = \beta e^{-\alpha M} SI - (d + \sigma_d)E$<br>$\frac{dM}{dt} = \frac{\delta}{1 + h_2 \sigma_d E} \sigma_d E - \mu_d M$<br>$M(t)$ ≔ average number of daily news items related to the outbreak<br>$S, E$ ≔ susceptible and exposed individuals, respectively (specifically, $\sigma_d E$: newly observed infectious individuals)<br>$d, h_2$ ≔ reporting rate and saturation factor, respectively<br><br>→<br>Media reports reduce the transmission factor |

| Paper | Title | Behavioural modelling solution |
|---|---|---|
| Liu et al. 2022 | Mathematical modeling and stability analysis of the time-delayed S AIM model for COVID-19 vaccination and media coverage | $\frac{dS(t)}{dt} = B - \beta S(t)I(t) - dS(t) - \lambda_0 M(t)S(t)$<br>$\frac{dA(t)}{dt} = \lambda_0 M(t)S(t) - dA(t) - \alpha A(t)I(t) + \nu I(t)$<br>$\frac{dM(t)}{dt} = \mu I(t-\tau) - \mu_0 M(t)$<br>$S, A :=$ non-antibody and antibody population; $S \rightarrow A$ is made through vaccination<br>$M :=$ Media coverage<br><br>→<br>Media coverage influences the willingness to get vaccinated. Vaccinated individuals have a lower chance to get infected (modulated by the parameter $\alpha$). |
| Nyabadza et al. 2023 | Modelling the Influence of Dynamic Social Processes on COVID-19 Infection Dynamics | $\dot{\rho} = \rho_0 + \alpha_1 \epsilon - \nu_1 \rho$<br>$\dot{\epsilon} = \epsilon_0 + \alpha_2 i_d - \nu_2 \epsilon$<br>$s(t) :=$ fraction of susceptible population<br>$\rho(t) :=$ social distancing, dependent on information dynamics $\epsilon(t)$ (in turn, dependent on the fraction of infected detected individuals $i_d$)<br><br>→<br>$\dot{s} = \mu - \beta(1-\rho)(i_d + \eta i_u)s - \mu s$<br>Where $s, i_u$ denote the fraction of susceptible and undetected infected individuals, respectively |
| Pal et al. 2024 | Examining the impact of incentives and vaccination on COVID-19 control in India: addressing environmental contamination and seasonal dynamics | $\frac{dS}{dt} = \Lambda - \left(\beta - \beta_1 \frac{\upsilon_1 M}{p + \upsilon_1 M}\right) SI - \psi_e \frac{E_c}{L + E_c} S - \phi\left(1 + \gamma \frac{\upsilon_2 M}{p + \upsilon_2 M}\right) S + \nu I + \delta V - dS$<br>$\frac{dI}{dt} = \left(\beta - \beta_1 \frac{\upsilon_1 M}{p + \upsilon_1 M}\right) SI + \psi_e \frac{E_c}{L + E_c} S - (\nu + \alpha + d)I$<br>$\frac{dE_c}{dt} = s_1 I - s_0 E_c - \eta \frac{\upsilon_3 M}{q + \upsilon_3 M} E_c$<br>$\frac{dM}{dt} = rM\left(1 - \frac{M}{K}\right) + \theta_1 IM + \theta_2 E_c M$<br>$M :=$ amount of incentives provided by the government (dollar)<br>$S, E_c, I, V :=$susceptible, concentration of environmental contamination, infected, and vaccinated individuals, respectively<br>$K, L :=$ “carrying capacity of the incentive provided by the government” and half-saturation constant, respectively<br><br>→ |

| Paper | Title | Behavioural modelling solution |
|---|---|---|
| | | Transmission rate is reduced by governmental incentives (modulated by a constant $\beta_1$) |
| Rai et al. 2022 | Impact of social media advertisements on the transmission dynamics of COVID-19 pandemic in India | $\frac{dS}{dt} = \Lambda - \beta_1 \left(1 - c_{m_s} \frac{M}{p+M}\right) SI_s - \beta_2 \left(1 - c_{m_a} \frac{M}{p+M}\right) SI_a - \lambda SM + \lambda_0 A + \delta R - dS$<br>$\frac{dE}{dt} = \beta_1 \left(1 - c_{m_s} \frac{M}{p+M}\right) SI_s + \beta_2 \left(1 - c_{m_a} \frac{M}{p+M}\right) SI_a - (\sigma + d)E$<br>$\frac{dA}{dt} = \lambda SM - (\lambda_0 + d)A$<br>$\frac{dM}{dt} = r\left(1 - \theta \frac{A}{\omega + A}\right) I_s - r_0(M - M_0)$<br>$A$ ≔ aware individuals<br>$M$ ≔ cumulative number of social media advertisements which includes internet information as well as TV, radio and print media<br>$M_0$ ≔ baseline number of social media advertisements<br>$S, I_s, I_a$ ≔ susceptible, infected symptomatic, and infected asymptomatic, respectively<br>$c_{m_s}, c_{m_a}$ ≔ efficacy of social media advertisements to reduce contacts with symptomatic and asymptomatic individuals, respectively (s.t. $c_{m_s} \gg c_{m_a}$)<br><br>→ awareness reduces contacts at risk of transmission for everyone and on top of that excludes some individuals from the transmission process entirely (the "aware" individuals) |
| Sardar et al. 2023 | Detection of multiple waves for COVID-19 and its optimal control through media awareness and vaccination: study based on some Indian states | $\frac{dS}{dt} = \Pi + \omega S_L - lS - \mu S - \frac{\beta S I_S}{N - \theta S_L - H} - \frac{\rho \beta S\, I_A}{N - \theta S_L - H} - \frac{\psi S M}{1 + M}$<br>$\frac{dS_L}{dt} = \frac{\psi S M}{1 + M} + lS - \mu S_L - \omega S_L - \frac{\beta(1-\theta) S_L I_S}{N - \theta S_L - H} - \frac{\rho\beta(1-\theta) S_L I_A}{N - \theta S_L - H} + \omega_r R$<br>$\frac{dE}{dt} = \frac{\beta S I_S}{N - \theta S_L - H} + \frac{\rho \beta S\, I_A}{N - \theta S_L - H} + \frac{\beta(1-\theta) S_L I_S}{N - \theta S_L - H} + \frac{\rho\beta(1-\theta) S_L I_A}{N - \theta S_L - H} - (\sigma + \mu)E$<br>$\frac{dH}{dt} = \tau I_S - (\mu + \delta)H - (1 + \xi)\gamma_2 H - \xi \gamma_2 \lambda H$<br>$\frac{dM}{dt} = \frac{aH}{1 + H} - dM$<br>$S, S_L$ ≔ susceptible and home-quarantined population, respectively<br>$M$ ≔ media and awareness compartment<br>$I_S, I_A, H, R$ ≔ number of symptomatic infected, asymptomatic infected, notified COVID-19 patients, and recovered individuals in the population, respectively<br><br>→ |

| Paper | Title | Behavioural modelling solution |
| --- | --- | --- |
| | | Part of the quarantined population ($\theta$) does not mix with the community as it "maintains proper social distancing" and thus is excluded from the transmission process. |
| Tahajuddin et al. 2022 | The impact of a power law-induced memory effect on the SARS-CoV-2 transmission | $\frac{dS(t)}{dt} = \Pi + \omega L(t) - \psi \frac{S(t)M(t)}{1+M(t)} - \rho\zeta S(t) - \frac{\beta_1}{N-\theta L} S(t)\Phi_1(t,0)\left[\int_0^t \frac{k(t-\hat{t})}{\Phi_1(\hat{t},0)} d\hat{t}\right]$ $- \frac{\beta_2\rho}{N-\theta L} S(t)\Phi_2(t,0)\left[\int_0^t \frac{k(t-\hat{t})H(\hat{t})}{\Phi_2(\hat{t},0)} d\hat{t}\right] - (\mu+l)S(t) + \omega_r\theta_r R(t)$<br>$\frac{dL(t)}{dt} = lS(t) + \psi \frac{S(t)M(t)}{1+M(t)} - \rho\zeta L(t) - \frac{\beta_1(1-\theta)}{N-\theta L} L(t)\Phi_1(t,0)\left[\int_0^t \frac{k(t-\hat{t})}{\Phi_1(\hat{t},0)} d\hat{t}\right] - (\mu+\omega)L(t)$ $+ \omega_r(1-\theta_r)R_t$<br>$\frac{dE(t)}{dt} = \frac{\beta_1}{N-\theta L} S(t)\Phi_1(t,0)\left[\int_0^t \frac{k(t-\hat{t})}{\Phi_1(\hat{t},0)} d\hat{t}\right] + \frac{\beta_2}{N-\theta L} S(t)\Phi_2(t,0)\left[\int_0^t \frac{k(t-\hat{t})}{\Phi_2(\hat{t},0)} d\hat{t}\right]$ $+ \frac{\beta_1(1-\theta)}{N-\theta L} L(t)\Phi_1(t,0)\left[\int_0^t \frac{k(t-\hat{t})}{\Phi_1(\hat{t},0)} d\hat{t}\right] - (\sigma+\mu)E(t)$<br><br>$\frac{dM(t)}{dt} = \frac{\eta H(t)}{1+H(t)} - dM(t)$<br>$M :=$ awareness density<br>$H :=$ number of confirmed COVID-19 cases in the population<br>$S, L, E, R, U :=$ susceptible, home quarantined, exposed, recovered, and undetected infected individuals, respectively<br><br>→<br>A fraction $\theta$ of the home-quarantined population maintains "proper social distancing" and thus is excluded from the transmission process |
| Tiwari et al. 2021 | Dynamics of coronavirus pandemic: effects of community awareness and global information campaigns | $\frac{dS}{dt} = \Lambda - \beta_1 S I_s - \beta_2 S I_a - \lambda S \frac{M}{p+M} - \rho S A_h + \lambda_{01} A_h + \lambda_{02} A_l + \delta R - dS$<br>$\frac{dA_h}{dt} = (1-b)\lambda S \frac{M}{p+M} + (1-c)\rho S A_h - \sigma_2 \beta_2 A_h I_a - (\lambda_{01} + d)A_h$<br>$\frac{dA_l}{dt} = b\lambda S \frac{M}{p+M} + c\rho S A_h - \sigma_3 \beta_3 A_l I_a - (\lambda_{02} + d)A_l$<br>$\frac{dE}{dt} = \beta_1 S I_s + \beta_2 S I_a + \sigma_2 \beta_2 A_h I_a + \sigma_3 \beta_3 A_l I_a - (\sigma_1 + d)E$<br>$\frac{dM}{dt} = r\left(1 - \theta \frac{A_h + A_l}{w + A_h + A_l}\right) I_s - r_0(M - M_0)$ |

| Paper | Title | Behavioural modelling solution |
|---|---|---|
| | | $M$ ≔ cumulative number of social media advertisements<br>$A_h, A_l$ ≔ number of highly active aware individuals (such as healthcare workers, careless people) and less active aware individuals (such as uneducated people, careless people), respectively<br>$S, E, I_a, I_s, R$ ≔susceptible, exposed, asymptomatic infected, symptomatic infected, and recovered individuals, respectively<br><br>→<br>Awareness reduces the transmission rates (as modulated by the constants $\sigma_2, \sigma_3$) |
| Tripathi et al. 2022 | A mathematical model to study the COVID-19 pandemic in India | $\frac{dX}{dt} = A - \beta\frac{XY}{N} - \lambda VX - \nu\frac{XM}{1+\gamma M} - dX + \lambda_{11}X_M + \beta_{11}Y_i + \alpha Y$<br>$\frac{dX_M}{dt} = \nu\frac{XM}{1+\gamma M} - dX_M - \lambda_{11}X_M$<br>$\frac{dY}{dt} = \beta\frac{XY}{N} + \lambda VX - dY - \alpha Y - \beta_1 Y$<br>$\frac{dM}{dt} = \mu Y - \sigma M$<br>$X(t), X_M(t)$ ≔ susceptible and aware susceptible, respectively<br>$M$ ≔ cumulative density representing media alert to make the population aware of the disease<br>$Y, Y_i, V$ ≔ asymptomatic infectives, symptomatic infectives (isolated for treatment/hospitalized), and cumulative density of the virus, respectively<br><br>→<br>Aware individuals are excluded from the transmission process |
| Wang et al. 2022 | Timely and effective media coverage's role in the spread of Corona Virus Disease 2019 | $\frac{dS}{dt} = -\beta c e^{-\eta M_1 + \varepsilon M_2} S(I + \theta E)$<br>$\frac{dM_1}{dt} = p\phi\sigma E - \tau M_1$<br>$\frac{dM_2}{dt} = q\phi\sigma E - \tau M_2$<br>$M_1, M_2$ ≔ positive and negative media coverage, respectively, $p + q = 1$<br>$S, E, I$ ≔ susceptible, exposed asymptomatic, infected symptomatic individuals, respectively<br>$p, q, \phi, \sigma$ ≔ ratio of positive coverage on COVID-19, ratio of negative coverage on COVID-19, media reporting rate of the figure for daily new COVID-19 cases, and conversion rate from $E$ to $I$, respectively<br><br>→ |

| Paper | Title | Behavioural modelling solution |
|---|---|---|
| | | Media attention and the nature of media attention impact the transmission rate through accordingly adapted behaviour |
| Zhou et al. 2020 | Effects of media reporting on mitigating spread of COVID-19 in the ear y phase of the outbreak | $S' = -\left(ce^{-pM}\beta + ce^{-pM}q(1-\beta)\right)S(I+\theta E) - \lambda_M SM + \lambda S_q$<br>$S_q' = ce^{-pM}(1-\beta)qS(I+\theta E) + \lambda_M SM - \lambda S_q$<br>$M' = \eta\left(\delta_I I + \delta_q E_q\right) - \mu_M M$<br>$S, S_q$ ≔ susceptible and quarantined susceptible, respectively<br>$E, E_q, I$ ≔ exposed, exposed quarantined, and infected symptomatic individuals, respectively<br>$M$ ≔ cumulative density of awareness programs driven by the media reports<br><br>→<br>Awareness reduces the contact rate and induces some people to enter into quarantine. Individuals in quarantine cannot get infected and are thus excluded from the transmission process. |
| Zuo et al. 2022 | Analyzing COVID-19 Vaccination Behavior Using an SEIRM/V Epidemic Model With Awareness Decay | $\frac{dS_u}{dt} = \frac{\beta S_u I}{N} - \eta M S_u + \delta S_a + \rho V$<br>$\frac{dS_a}{dt} = \frac{k\beta S_a I}{N} + \eta M S_u - \delta S_a - \varepsilon S_a$<br>$\frac{dV}{dt} = \varepsilon S_a - \rho V$<br>$\frac{dM}{dt} = \frac{\alpha S_a}{N} + \alpha_0 I - \lambda M$<br>$S_u, S_a$ ≔ unaware and aware susceptible, respectively<br>$V, I$ ≔ vaccinated and infected individuals, respectively<br>$M(t)$ ≔ accumulated density of awareness programs<br><br>→<br>Awareness induces people to get vaccinated. Vaccinated individuals cannot get infected and are thus excluded from the transmission process. |

## *Awareness – Other*

| Paper | Title | Behavioural modelling solution |
|---|---|---|
| Lux et al. 2021 | The social dynamics of COVID-19 | $x = \frac{n_{noSD}}{N - D_t}, \quad n_{SD} + n_{noSD} = N - D_t$<br>$l_{eff} = x \cdot \lambda$<br>$w(n_{SD} + 1, n_{noSD} - 1 \mid n_{SD}, n_{noSD}) = \upsilon_{SD} \exp(U_{SD})$<br>$w(n_{SD} - 1, n_{noSD} \mid n_{SD}, n_{noSD}) = \upsilon_{noSD} \exp(U_{noSD})$<br>$U_{SD,t} = \alpha_1 \frac{(\sum_{\tau=1}^{T} I_{t-\tau} - \sum_{\tau=T+1}^{2T} I_{t-\tau})}{N - D_t} + \alpha_2 \frac{(\sum_{\tau=1}^{T} D_{t-\tau} - \sum_{\tau=T+1}^{2T} D_{t-\tau})}{N - D_t}$<br>$U_{noSD} = -U_{SD}$<br>$\upsilon_{SD} = \max\left\{\alpha_3 \frac{(\sum_{\tau=1}^{T} I_{t-\tau} - \sum_{\tau=T+1}^{2T} I_{t-\tau})}{N - D_t} + \alpha_4 \frac{(\sum_{\tau=1}^{T} I_{t-\tau} - \sum_{\tau=T+1}^{2T} I_{t-\tau})}{N - D_t}, 0\right\}$<br>$\upsilon_{noSD} = \alpha_0$<br>$x$ ≔ index of social distancing in the population<br>$n_{SD}, n_{noSD}$ ≔ number of agents practicing and not practicing social distancing, respectively<br>$\lambda_{eff}$ ≔ effective reproduction rate<br>$w(\cdot)$ ≔ transition rates for agents to move between social distancing and no social distancing, based on the functions $U_{SD}, U_{noSD}$ evaluating available information and parameters $\upsilon_{SD}, \upsilon_{noSD}$ representing the ease of behavioural change<br>$\alpha_1, \alpha_2, \alpha_3, \alpha_4$ ≔parameters representing strength of agents' reaction<br><br>→<br>$w(I + 1, S - 1 \mid S, I, R, D) = \lambda_{eff} \left(\frac{I}{N}\right) S$ |
| Molla et al. 2023 | Pharmaceutical and non-pharmaceutical interventions for controlling the COVID-19 pandemic | $\mu^i = \mu^i_{\max} \left[\frac{[K_M - K_C]_+}{[K_M - K_C]_+ + K_0 - K_C}\right]\left[\frac{[M_A - M_C]_+}{M_A - M_C]_+ + M_0 - M_C}\right]$<br>$K_M = \frac{dM/dt}{M \ln(2)}$<br>$\mu^v = \mu^w = \frac{\mu^u}{2}$<br>Flow from social distancing class 1 to class 2 = $\frac{\mu^i}{2}$<br>$\nu^i = \nu^i_{\max} \left[\frac{[C - C_c]_+}{[C - C_C]_+ + C_0 - C_C}\right]$<br>$\dot{C} = \frac{N_{crit}}{N} \sum_{\xi \in \{u,v,w\}} \left(\left(S_2^\xi + E_2^\xi\right) + (1 - \delta)\left(S_1^\xi + E_1^\xi\right)\right)$ |

| Paper | Title | Behavioural modelling solution |
| --- | --- | --- |
| | | $\mu^i :=$ social distance transition function from class 0 to class 1 for individuals who are not vaccinated, whose maximal value is $\mu^i_{\max}$<br>$K_M :=$ doubling rate<br>$M :=$ newly reported cases<br>$M_C, K_C :=$ critical values for $M_A, K_A$ respectively to activate social distancing (where $M$ are the active cases)<br>$\mu^v, \mu^w :=$ social distance transition function from class 0 to class 1 for individuals who are vaccinated or have perceived immunity, respectively<br>$\nu^i :=$ decrease rate in social distancing, whose maximal value is $\nu^i_{\max}$, dependent on the cost $C$<br>$N_{crit} :=$ population above which hospital resources are strained<br>$S_1^\xi, S_2^\xi, E_1^\xi, E_2^\xi :=$susceptible with moderate and full social distancing, exposed with moderate and full social distancing, respectively. $\xi \in \{u, v, w\}$ denotes the vaccination status.<br><br>→<br>Each compartment can have one of three levels of social distancing, from no social distancing (0) to full isolation (2). Each social distancing level reduces the transmission: social distancing class 1 by a constant $\delta$, social distancing class 2 completely (zero contact probability) |
| Moyles et al. 2021 | Cost and social distancing dynamics in a mathematical model of COVID-19 with application to Ontario, Canada | $\mu = \mu_{\max}\left[\frac{[k_M - k_C]_+}{[k_M - k_C]_+ + k_0 - k_C}\right]\left[\frac{[M_A - M_C]_+}{M_A - M_C]_+ + M_0 - M_C}\right]$<br>$k_M = \frac{dM/dt}{M\ln(2)}$<br>Flow from social distancing class 1 to class 2 = $\frac{\mu}{2}$<br>$\nu = \nu_{\max}\left[\frac{[C - C_c]_+}{[C - C_C]_+ + C_0 - C_C}\right]$<br><br>$\dot{c} = kN_{crit}\big((S_2 + E_2) + (1 - \delta)(S_1 + E_1)\big)$<br>Flow from social distancing class 1 to class 0 = $\frac{\nu}{2}$<br>$\mu :=$ social distance transition function from class 0 with proportion $q_0$ and from class 1 to class 2 with proportion $1 - q_0$, whose maximal value is $\mu_{\max}$<br>$k_M :=$ doubling rate<br>$M :=$ total number of cases<br>$M_C, k_C :=$ critical values for $M_A, k_A$ respectively to activate social distancing (where $M$ are the active cases)<br>$\nu :=$ decrease rate in social distancing from class 2 to class 1 with proportion $q_2$ and to class 0 with proportion $1 - q_2$, proportional to the cost $C$ |

| Paper | Title | Behavioural modelling solution |
|---|---|---|
| | | $N_{crit}$ ≔ population above which hospital resources are strained<br>$S_1, S_2, E_1, E_2$ ≔susceptible with moderate and full social distancing, exposed with moderate and full social distancing, respectively.<br><br>→<br>Each compartment can have one of three levels of social distancing, from no social distancing (0) to full isolation (2). Each social distancing level reduces the transmission: social distancing class 1 by a constant $\delta$, social distancing class 2 completely (zero contact probability) |
| Pant et al. 2024 | Mathematical Assessment of the Role of Human Behavior Changes on SARS-CoV-2 Transmission Dynamics in the United States | $\psi_{12}^{m}(t) = \alpha_{12}^{m} q_{12}^{m} \left(1 - e^{-\zeta M(t)}\right)$<br>$\psi_{12}^{c}(t) = \alpha_{12}^{c} q_{12}^{c} \frac{N_2(t)}{N(t)}$<br>$\psi_{21}^{c}(t) = \alpha_{21}^{c} q_{21}^{c} \frac{N_1(t)}{N(t)}$<br>$\psi_{12}^{i}(t) = \alpha_{12}^{i} q_{12}^{i} \frac{I_1(t) + I_2(t)}{K + I_1(t) + I_2(t)}$<br>$\psi_{21}^{ns}(t) = \alpha_{21}^{ns} q_{21}^{ns} \frac{S_1(t) + S_2(t) + E_1(t) + E_2(t) + A_1(t) + A_2(t) + R_1(t) + R_2(t)}{N(t)}$<br>$\psi_{12}^{m}$ ≔ Rate at which non-adherent individuals become adherent due to daily disease-induced mortality $M(t)$<br>$\psi_{12}^{c}$ ≔ rate at which non-adherent individuals become adherent due to contact with adherents $N_2(t)$<br>$\psi_{12}^{i}$ ≔ rate at which non-adherent individuals become adherent due to the level of symptomatic transmission in the community $I_1(t), I_2(t)$<br>$\psi_{21}^{c}$ ≔ rate at which adherent individuals become non-adherent due to contact with non-adherents $N_1(t)$<br>$\psi_{21}^{ns}$ ≔ rate at which adherent individuals become non-adherent due to the proportion of non-symptomatic individuals in the community ($S_1(t), S_2(t), E_1(t), E_2(t), A_1(t), A_2(t), R_1(t), R_2(t)$ ≔ non-adherent and adherent susceptible, non-adherent and adherent exposed, non-adherent and adherent asymptomatic, non-adherent and adherent recovered individuals, respectively)<br><br>→<br>Adherent individuals have a lower chance to get infected |
| Retzlaff et al. 2022 | Fear, behavior and the COVID-19 pandemic: a city-scale agent-based model using socio- | $fear = 100 * 2.2 * (1 - \exp(-global\ cases * personal\ cases))$<br>$behavior = \frac{attitude + norm}{2} * \frac{fear}{100}$ |

| Paper | Title | Behavioural modelling solution |
|---|---|---|
| | demographic and spatial map data | →<br>Behavior decreases the risk of infecting others through adoption of protective measures and their decision to partake in social activities (by determining whether the individual adopts protective behaviours or not = different parameters, as if they were two different population groups); moreover, agents can choose to reject self-isolate and therefore risk infecting others. |
| Worby and Chang 2020 | Face mask use in the general population and optimal resource allocation during the COVID-19 pandemic | $\omega_A = \frac{1}{1+e^{-k_1(I_S-k_2)}}$<br>$\omega_A$ ≔ demand for masks<br>$I_S$: = number of symptomatic individuals<br>$k_1, k_2$ ≔ rate of demand increase and timing of demand, respectively<br><br>→<br>$\frac{dS^n}{dt} = -\frac{S^n}{N}\left(\beta_a(I_A^n + I_P^n + I_A^m r_t + I_P^m r_t) + \beta_S(I_S^n + I_S^m r_t)\right) + \mu S^m - \omega_A \frac{M}{N} S^n$<br>$\frac{dS^m}{dt} = -\frac{S^m r_s}{N}\left(\beta_A(I_A^n + I_P^n + I_A^m r_t + I_P^m r_t) + \beta_s(I_S^n + I_S^m r_t)\right) - \mu S^m + \omega_A \frac{M}{N} S^n$<br>$M/N$ ≔ mask supply<br>$S^n, S^m, I_A^n, I_A^m, I_P^n, I_P^m, I_S^n, I_S^m$ ≔ non-wearer and mask-wearer susceptible, non-wearer and mask-wearer asymptomatic,/mildly symptomatic, non-wearer and mask-wearer presymptomatic, non-wearer and mask-wearer symptomatic individuals, respectively<br>$r_t, r_s$ ≔ relative transmissibility and susceptibility with masks, respectively |
| Yao et al. 2022 | Modeling the spread of infectious diseases through influence maximization | $\sum_{j \in N_G(i)} x_{j(t-1)} \leq n_0 + M z_{it}, \quad \forall i \in V_2, t \in \{1, \dots, T\}$<br>$\sum_{j \in N_G(i)} x_{j(t-1)} \geq n_0\left(z_{it} - z_{i(t-1)}\right), \quad \forall i \in V_2, t \in \{1, \dots, T\}$<br>$V_1, V_2$ ≔ individuals who take and do not take precautions against the infectious disease, respectively<br>$z_{it}$ ≔ binary variable that indicates whether $i \in V_2$ takes precautions at time $t$<br><br>→<br>If the number of infectious neighbours of $i \in V_2$ exceed the threshold $n_0$, the individual will change their behaviour<br>$\sum_{k=1}^{t_0} \left\{\sum_{j \in N_{G_1}^-(i)} a x_{j(t-k)} + \sum_{j \in N_{G_2}^-(i)} \left[c x_{k(t-k)} + (a-c) x_{j(t-k)} z_{j(t-k)}\right]\right\} \geq s_i(x_{it} - x_{i(t-1)}, \quad \forall i \in V_1, t \in \{1, \dots, T\}$ |

| Paper | Title | Behavioural modelling solution |
|---|---|---|
| | | $\sum_{k=1}^{t_0}\left\{\sum_{j\in N_{G_1}^-(i)}\left[bx_{j(t-k)}+(a-b)x_{j(t-k)}z_{i(t-k)}\right]+z_{i(t-k)}\sum_{j\in N_{G_2}^-(i)}\left[cx_{j(t-k)}+(a-c)x_{j(t-k)}z_{j(t-k)}\right]+\left(1-z_{i(t-k)}\right)\sum_{j\in N_{G_2}^-(i)}\left[dx_{j(t-k)}+(b-d)x_{j(t-k)}z_{j(t-k)}\right]\right\}\geq s_i\left(x_{it}-x_{i(t-1)}\right),\ \forall i\in V_2, t\in\{1,\dots,T\}$<br>$x_{it}+y_{it}\leq 1,\ \forall i\in V, t\in\{1,\dots,T\}$ |
| Yedomonhan et al. 2023 | Modeling the effects of prophylactic behaviors on the spread of SARS-CoV-2 in West Africa | $O=\{-2,-1,1,2\}$<br>$\chi_{(2,1)}(t)=\frac{\omega_{-2}(t)S_{-2}(t)+\omega_{-1}S_{-1}(t)+(1-\phi)\omega_1(t)S_1(t)}{N(t)-Q(t)}$<br>$\chi_{(1,2)}(t)=\frac{\phi\omega_1(t)S_1(t)+\omega_2(t)S_2(t)}{N(t)-Q(t)}$<br>$\chi_{(1,-1)}(t)=\frac{\omega_{-2}(t)S_{-2}(t)+\omega_{-1}(t)S_{-1}(t)}{N(t)-Q(t)}$<br>$\chi_{(-1,1)}(t)=\frac{\omega_1(t)S_1(t)+\omega_2(t)S_{-1}(t)}{N(t)-Q(t)}$<br>$\chi_{(-1,-2)}(t)=\frac{\omega_{-2}(t)S_{-2}(t)+\phi\omega_{-1}(t)S_{-1}(t)}{N(t)-Q(t)}$<br>$\chi_{(-2,-1)}(t)=\frac{(1-\phi)\omega_{-1}(t)S_{-1}(t)+\omega_1(t)S_1(t)+\omega_2(t)S_2(t)}{N(t)-Q(t)}$<br><br>$O$ ≔ four levels of prophylactic behavior (use of the mask, physical distance, hand hygiene, coughing hygiene, avoidance of touching the face)<br>$S_i$ ≔ susceptible with attitude $i$ for any opinion $i\in\mathcal{O}$, s.t. $S_{-2}$ is the individual with the highest degree of prophylactic behavior (and so the least susceptible) and $S_2$ is the individual with the lowest degree of prophylactic behavior (and thus the most susceptible)<br>$\chi_{(i,j)}(t)$ ≔ opinion change rate $i\to j$<br>$\omega_i(I)$ ≔ influence functions, i.e. rate at which sensitive individual $S_i$ influences the rest of the susceptible population, depending on the infectious but undetected individuals $I$ [2]<br>$Q$ ≔ quarantined individuals<br><br>→<br>$\dot{S}_2(t)=\chi_{(1,2)}S_1(t)-\left(\chi_{(2,1)}+\lambda_2\right)S_2(t)$<br>$\dot{S}_1(t)=\chi_{(2,1)}S_2(t)-\left(\chi_{(1,2)}+\lambda_1\right)S_1(t)$<br>$\dot{S}_{-1}(t)=\chi_{(1,-1)}S_1(t)-\left(\chi_{(-1,1)}+\lambda_{-1}\right)S_{-1}(t)$<br>$\dot{S}_{-2}=\chi_{(-1,-2)}S_{-1}(t)-\left(\chi_{(-2,-1)}+\lambda_{-2}\right)S_{-2}(t)$ |

[2] Several functional forms of $\omega_i(I)$ are inspected in the paper, but they all depend on $I$. We will not report them here for conciseness.

| Paper | Title | Behavioural modelling solution |
| --- | --- | --- |
| | | $\dot{I}(t) = \lambda_2 S_2(t) + \lambda_1 S_1(t) + \lambda_{-1} S_{-1}(t) + \lambda_{-2} S_{-2}(t) - \alpha I(t) - \delta I(t)$<br>$\lambda_{i(t)} = \frac{\beta_i I(t)}{N(t) - Q(t)}$<br>Where $\beta_2 = \beta_0, \beta_1 = \frac{1}{a}\beta_0, \beta_{-1} = \frac{1}{a^2}\beta_0, \beta_3 = \frac{1}{a^3}\beta_0$ (i.e. prophylactic measures act on the transmission rate) |

## Supplementary Material S4.IV: Modelling solutions adopted by papers in the "**Evolutionary Game Theory**" category

| Paper | Title | Behavioural modelling solution |
|---|---|---|
| Agusto et al. 2022 | To isolate or not to isolate: the impact of changing behavior on COVID-19 transmission | $\frac{dx_s}{dt} = \kappa_s x_s(1-x_s)[I(t)+Q(t)+H(t)-\varepsilon_s L_s(t)]$<br>$\frac{dx_I}{dt} = \kappa_I x_I(1-x_I)[\delta_I I(t)+\delta_Q Q(t)+\delta_H H(t)-\varepsilon_I]$<br>$x_s$ ≔ proportion of people supporting closure<br>$x_I$ ≔ proportion of symptomatically infected individuals who elect to self-isolate<br>$L_s$ ≔ accumulated socio-economic losses<br>$I, Q, H$ ≔ infected, quarantined, hospitalized individuals<br>$\varepsilon_s, \varepsilon_I$ ≔ sensitivity to socio-economic losses relative to COVID-19 infection and to self-isolation relative to infecting others, respectively<br>$\kappa_s, \kappa_I$ ≔ support for closure social learning rate and self-isolation social learning rate, respectively<br>$\delta_.$ ≔ death rate ($\cdot \in \{I, Q, H\}$)<br><br>→<br>Recruitment in exposed compartment:<br>$\beta[1-C(t)][(1-x_I(t))I(t)+\eta_A A(t)+\eta_Q Q(t)+\eta_H H(t)]S(t)$, where<br>$C(t)$ ≔ social distancing measures, s.t.<br>$C(t) = \begin{cases} 0, & t < t_{close} \vee x_s < 1/2 \\ C_0, & t \geq t_{close} \wedge x_s \geq 1/2 \end{cases}$ |
| Amaral et al. 2021 | An epidemiological model with voluntary quarantine strategies governed by evolutionary game dynamics | $\Phi_S = S_Q(S_N+I_N)\Theta(\pi_Q,\pi_N) - S_N(S_Q+I_Q)\Theta(\pi_N,\pi_Q)$<br>$\Phi_I = I_Q(S_N+I_N)\Theta(\pi_Q,\pi_N) - I_N(S_Q+I_Q)\Theta(\pi_N,\pi_Q)$ where $\Theta(\pi_i,\pi_j) = \frac{1}{1+e^{-(\pi_i-\pi_j)/k}}$<br>$\Phi_S$ ≔ rate at which $S_Q$ agents (i.e., self-quarantining) convert to $S_N$ (i.e., acting normally)<br>$\Phi_I$ ≔ rate at which $I_Q$ agents (i.e., self-quarantining) convert to $I_N$ (i.e., acting normally)<br>$\pi_i, \pi_j$ ≔ payoffs<br><br>→<br>$\dot{S}_N = -S_N(\beta_N I_N + \beta_a I_Q) + \tau\Phi_S$ |

| Paper | Title | Behavioural modelling solution |
|---|---|---|
| | | $\dot{S}_Q = -S_Q(\beta_a I_N + \beta_Q I_Q) - \tau\Phi_S$<br>$\dot{I}_N = S_N(\beta_N I_N + \beta_a I_Q) - \gamma I_N + \tau\Phi_I$<br>$\dot{I}_Q = S_Q(\beta_a I_N + \beta_Q I_Q) - \gamma I_Q - \tau\Phi_I$<br>With $\beta_N > \beta_a > \beta_Q$ |
| Ge and Wang 2022 | Vaccination games in prevention of infectious diseases with application to COVID-19 | $\frac{dx}{dt} = \nu x(1-x)\left(\frac{1}{1+e^{-bM(t)}} - \frac{1}{2}\right)$<br>$M(t) = \varepsilon \cdot g(t) + \int_{t-\tau}^{t} g(\eta) \cdot \alpha e^{-\alpha(t-\eta)} d\eta$<br>$g(t) = m \cdot \underbrace{g_1(t)}_{=f_v - f_n} + (1-m) \cdot \underbrace{g_2(t)}_{=\beta Nsi}$<br>$f_v = -c_v - (1-\sigma)\beta i \cdot c_i - \sigma\phi \cdot \beta i \cdot c_i$<br>$f_n = -\beta i \cdot c_i$<br><br>$x$ ≔proportion of individuals who are voluntarily vaccinated<br>$f_v, f_n$ ≔ pay-off for vaccination and non-vaccination strategies, respectively<br>$m$ ≔ weight coefficient<br>$\alpha e^{-\alpha(t-\eta)}$ ≔ exponential fading memory function to account for fading of accumulated information<br>$s, i$ ≔ fraction of susceptible and infected individuals, respectively<br>$c_v, c_i$ ≔ costs of vaccination and infection, respectively<br><br>→<br>$\frac{dS}{dt} = \mu N - \sigma\rho[\theta S + (1-\theta)Sx] - \frac{\beta SI}{N} + \phi V + \kappa R - \mu S$<br>$\frac{dV}{dt} = \sigma\rho[\theta S + (1-\theta)Sx] - \phi V - \mu V$<br>Where $S, I, V, R$ denote the susceptible, infected, vaccinated, and recovered individuals, respectively. $\theta$ is the fraction of compulsorily vaccinated.<br>Vaccinated individuals cannot get infected. |
| He et al. 2024 | Modelling and analysis of the cross-impact of age heterogeneity and behavioural changes on the evolution of disease transmission | $p_i' = k_i p_i(1-p_i)\left(1 - m_i \sum_{j=1}^{l} (\varepsilon_i I_{jm}(\tau) + I_{js})\right)\frac{(S_i^v + S_i^{\bar{v}} + E_i^v + E_i^{\bar{v}})}{N} + \mu(1-2p_i)$<br>$p_i = \frac{S_{in}^v + S_{in}^{\bar{v}} + E_{in}^v + E_{in}^{\bar{v}}}{S_i^v + S_i^{\bar{v}} + E_i^v + E_i^{\bar{v}}}$<br><br>$p_i$ ≔ proportion of people who maintain normal behaviour among susceptible and exposed individuals in age group $i$, whose rate of change is dependent on the rate $\kappa_i$ of the speed of |

| Paper | Title | Behavioural modelling solution |
|---|---|---|
| | | behavioural changes in age group $i$, the risk threshold $m_i$ on which susceptible people make decisions, and the correction factors $\varepsilon_i$<br>$S_i^v, S_i^{\bar{v}}, E_i^v, E_i^{\bar{v}}, I_{im}, I_{is}$ ≔ susceptible vaccinated, susceptible unvaccinated, exposed vaccinated, exposed unvaccinated, infected with mild symptoms, and infected with severe symptoms in age group $i$, respectively<br><br>→<br>$S_i^{\bar{v}\prime} = -\sum_{j=1}^{l} \frac{(p_i\beta_{ijn}+(1-p_i)\beta_{ija})c_{ij}S_i^{\bar{v}}(I_{jm}+I_{js})}{N_j}$<br>$S_i^{v\prime} = -\sum_{j=1}^{l} \frac{\lambda(p_i\beta_{ijn}+(1-p_i)\beta_{ija})c_{ij}S_i^{v}(I_{jm}+I_{js})}{N_j}$<br>$E_i^{\bar{v}\prime} = \sum_{j=1}^{l} \frac{\lambda(p_i\beta_{ijn}+(1-p_i)\beta_{ija})c_{ij}S_i^{\bar{v}}(I_{jm}+I_{js})}{N_j} - \sigma E_i^{\bar{v}}$<br>$E_i^{v\prime} = \sum_{j=1}^{l} \frac{\lambda(p_i\beta_{ijn}+(1-p_i)\beta_{ija})c_{ij}S_i^{v}(I_{jm}+I_{js})}{N_j} - \sigma E_i^{v}$<br>Where $\lambda$ denotes the reduction in transmission probability due to vaccination, while $\beta_{ijn}$ and $\beta_{ija}$ are the transmission probability of contact for susceptibles that maintain normal behaviour and that adopt protective behaviours, respectively. |
| Jentsch et al. 2021 | Prioritising COVID-19 vaccination in changing social and epidemiological landscapes: a mathematical modelling study | $\frac{dx}{dt} = \kappa x(1-x)\left(\frac{\sum_{i=1}^{16}\alpha_i(I_{a_i}+I_{s_i})}{\sum_{i=1}^{16}N_i} - cx\right) + p_{ul}(1-2x)$<br>$x$ ≔ proportion of individuals practicing NPIs<br>$\kappa, c, \alpha_i, p_{ul}$ ≔ social learning rate, incentive not to practice NPIs, ascertainment rate ($I_a, I_{s_i}$: asymptomatic and symptomatic individuals in age group $i$, respectively) , and the effect of social heterogeneity and influence from external populations (preventing the system from remaining arbitrarily close to the extremes for too long)<br><br>→<br>$\frac{dS_i^1}{dt} = -r\rho_i\left[1+s\sin\left(\frac{2\pi}{365}(t-\phi)-\frac{\pi}{2}\right)\right]S_i^1\sum_{j=1}^{16}C_{ij}(t)\frac{(I_{s_j}+I_{a_j}+P_j)}{N_j} - \tau S_i^1$<br>$\frac{dS_i^2}{dt} = -r\rho_i\left[1+s\sin\left(\frac{2\pi}{365}(t-\phi)-\frac{\pi}{2}\right)\right]S_i^2\sum_{j=1}^{16}C_{ij}(t)\frac{(I_{s_j}+I_{a_j}+P_j)}{N_j} - \tau S_i^2$<br>$\frac{dE}{dt} = r_i\left[1+s\sin\left(\frac{2\pi}{365}(t-\phi)-\frac{\pi}{2}\right)\right](S_i^1+S_i^2)\sum_{j=1}^{16}C_{ij}(t)\frac{(I_{s_j}+I_{a_j}+P_j)}{N_j} - \sigma_0 E_i + \tau(S_i^1+S_i^2)$<br>$C_{ij}(t,x) = C_{ij}^W(t) + C_{ij}^S(t) + (1-\epsilon_P x)(\bar{C}_{ij}^O + \bar{C}_{ij}^H)$ |

| Paper | Title | Behavioural modelling solution |
|---|---|---|
| | | Where $S_i^1, S_i^2, E_i, I_{s_i}, I_{a_i}, P_i$ are the susceptible unvaccinated, susceptible vaccinated (but not immunized), exposed (but not yet infectious), symptomatic infectious, asymptomatic infectious, presymptomatic individuals of age group $i$, respectively.<br>$C_{ij}(t, x)$ :=number of contacts per day, stratified by setting ($W$: work, $S$: school, $H$: household, $O$: other locations). In $O$ and $H$, "[...] governed by social processes with imperfect ability of public health authorities to enforce mandates and hence are governed by voluntary population adherence to NPIs such as mask use and physical distancing as per the $\epsilon_P x(t)$ term", where $\epsilon_P$ is the behaviours' efficacy |
| Jia et al. 2022 | The effectiveness of human interventions against COVID -19 based on evolutionary game theory | $x(t) = x(1-x)\Delta\pi + \mu(1-x) - \mu x$<br>$\Delta\pi = \pi_D - \pi_C = c - (1-\sigma)Q$<br>$\mathcal{H} = \begin{cases} 1, & x \geq 0 \\ 0, & x < 0 \end{cases}$<br>$x(t)$ := fraction of individuals adopting the "positive strategy" $C$ (s.t. $1-x$ adopts the "negative strategy" $D$)<br>$\Delta\pi$ := net payoff between the cost of strategy $D$ and strategy $C$, dependent on the number of people in quarantine $Q$ (seen as a fraction of confirmed cases)<br>$\mathcal{H}$ := Heaviside function<br><br>→<br>Overall coupled model:<br>$\dot{S}(t) = -\lambda[\sigma S + xP(1-\sigma)]$<br>$\dot{E}(t) = \lambda[\sigma S + xP(1-\sigma)] - \theta E$<br>$\dot{I}(t) = \alpha E - \zeta I$<br>$\dot{Q}(t) = q_1 E + q_2 I - \gamma_2 Q$<br>$\dot{x}(t) = -x\lambda - x\frac{P'}{P} + \rho\mu\left(\frac{S}{P} - 2x\right) + \rho\Delta\pi[xzP + [S(1-z) - xP]\mathcal{H}(\Delta\pi)]$<br>$\dot{y}(t) = x\lambda - \theta y - y\frac{P'}{P} + \rho\mu\left(\frac{E}{P} - 2y\right) + \rho\Delta\pi[yzP + [E(1-z) - yP]\mathcal{H}(\Delta\pi)]$<br>$\dot{z}(t) = (1-z)\frac{P'}{P} + y(\alpha + q_1) - \rho[z(1-z)P\Delta\pi - \mu(1-2z)]$<br>Where $S, E, I, Q$ denote susceptible, exposed, infected, and quarantined individuals, respectively, $P(t)$ is the total number of players, and<br>$x = \frac{S_D(t)}{P(t)}, y = \frac{E_D(t)}{P(t)}, z = \frac{P_C(t)}{P(t)}$<br>s.t. |

| Paper | Title | Behavioural modelling solution |
|---|---|---|
| | | $S_D = xP, S_C = S - xP, E_D = yP, E_C = E - yP$ |
| Jing et al. 2023 | Vaccine hesitancy promotes emergence of new SARS-CoV-2 variants. | $\frac{dx}{dt} = kx(1-x)\Delta F$<br>$\Delta F = \underbrace{F_v}_{-r_v} - F_n(A_{WT}, A_{MT}, I_{WT}, I_{MT})$<br>$x$ :=proportion of people getting vaccinated<br>$\Delta F$ :=payoff from switching to a vaccination strategy, dependent on the perceived cost of being vaccinated $r_v$ and the cost associated with non-being vaccinated $F_n$ (i.e. the cost of infection)<br>$I_{WT}, I_{MT}, A_{WT}, A_{MT}$ :=symptomatic and asymptomatic infected individuals, respectively; $WT$ and $MT$: SARS-CoV-2 strains<br><br>→<br>$\frac{dS}{dt} = -\frac{\lambda_{WT}S}{N} - \frac{\lambda_{MT}S}{N} - pxS + \tau V$<br>$\frac{dV}{dt} = pxS - \frac{\lambda_{WT}\eta_{WT}V}{N} - \frac{\lambda_{MT}\eta_{MT}V}{N} - \tau V$<br>$\frac{dE_{WT}}{dt} = \frac{\lambda_{WT}[(1-u_S)S+(1-u_v)\eta_{WT}V]}{N} - \sigma_{WT}E_{WT}$<br>$\frac{dE_{MT}}{dt} = \frac{\lambda_{WT}(u_S S+u_v\eta_{WT}V)}{N} + \frac{\lambda_{MT}(S+\eta_{MT}V)}{N} - \sigma_{MT}E_{MT}$<br>where $S, V, E_{WT}, E_{MT}$ denote susceptible, vaccinated, exposed to strain $WT$ and to strain $MT$, respectively. Being vaccinated reduces the susceptibility of the individuals (governed by constants $\eta_{MT}, \eta_{WT}$) |
| Kabir et al. 2020 | Evolutionary game theory modelling to represent the behavioural dynamics of economic shutdowns and shield immunity in the COVID-19 pandemic | $\dot{\eta} = m \cdot \eta(t) \cdot [1-\eta(t)] \cdot [-C_Q \cdot Q(t) + C_\iota \cdot I^{tot}(t)]$<br>$\eta(t)$ :=proportion of quarantining people<br>$C_Q, C_\iota$ :=economic cost of stay-at-home and cost of infection, respectively<br>$Q, I^{tot}$ := perceived fraction of quarantined + non-infected individuals and total number of infected individuals, i.e. symptomatic + asymptomatic, respectively<br>$m$ := proportionality constant<br><br>→<br>$\dot{E} = \frac{\beta S(I^A+I^S q(C,t)Q+hH)}{1+\theta R} - \alpha E + \delta Q$<br>$\dot{Q} = \alpha\eta(t)E - \delta Q$<br>$\dot{I}^S = \alpha(1-\rho)(1-\eta(t))E - rI - \gamma_S I^S$<br>$\dot{I}^A = \alpha\rho(1-\eta(t))E - \gamma_A I^A$ |

| Paper | Title | Behavioural modelling solution |
|---|---|---|
| | | Where $E, I^S, I^A, Q, H, R$ denote the exposed, symptomatic infected, asymptomatic infected, quarantined, hospitalized, and recovered individuals, respectively. $q(C,t)$ is defined as *counter-compliance factor*, s.t. $q(C,t) = C \cdot [q - \exp(-q_0 t)]$ |
| Kabir et al. 2021 | An evolutionary game modeling to assess the effect of border enforcement measures and socio-economic cost: Export-importation epidemic dynamics | $\dot{\eta}_k(t) = m_k \cdot \eta_k(t) \cdot [1 - \eta_k(t)][-\alpha_k \cdot C_k \cdot Q_k(t) + I_k(t)]$<br>$\eta_k$ ≔ proportion of exposed individuals adopting the quarantine policy declared by the government<br>$C_k, C_i$ ≔ cost of staying at home and of infection, respectively<br>$m_k, \alpha_k$ ≔ proportionality constant and sensitivity along cost, respectively<br>$I_k$ ≔ total infected people<br><br>→<br>$\dot{Q}_k = a_k \eta_k E_k - \delta Q_k$<br>$\dot{I}_k = \alpha_k (1 - \eta_k) E_k - (\gamma_k + r_k + d_k) I_k$<br>$\dot{J}_k = r_k I_k - w_k J_k$<br>$\dot{R}_k = \delta_k Q_k + \gamma_k I_k + d_k I_k + w_k J_k$<br>Where $E_k, I_k, Q_k, J_k, R_k$ denote exposed, quarantined, infected individuals, hospitalized-isolation, and recovered individuals, respectively. $k$ refers to the number of countries/subpopulations considered. Only non-quarantined infected individuals can end in the hospitalized compartment. |
| Li et al. 2024 | Linking Spontaneous Behavioral Changes to Disease Transmission Dynamics: Behavior Change Includes Periodic Oscillation | $\dot{x}(t) = -(\mu + \gamma) x \frac{I}{S + E} + \phi x (1 - x)(S + E)(k - I)$<br>$x$ ≔ proportion of individuals adopting the "normal" behavior<br>$S, E, I$ ≔ susceptible, exposed, and infected individuals, respectively<br>$k, \phi$ ≔ inverse of the sensitivity of individuals to perceived infection and rate of speed of behaviour change, respectively<br><br>→<br>$\dot{S}(t) = \mu - [x + \eta(1 - x)]\beta S I - \mu S + \gamma I$<br>$\dot{E}(t) = [x + \eta(1 - x)]\beta S I - \delta E - \mu E$<br>Where $\eta$ denotes the reduction factor of the transmission rate for individuals with "altered" behaviour |
| Liu et al. 2022 | Herd Behaviors in Epidemics: A Dynamics-Coupled Evolutionary Games Approach | $\frac{1}{\tau} \dot{x}_i^p(t) = \sum_{j \in \mathcal{J}^p} x_j^p(t) R_{ji}^p(w(t), x(t)) - \sum_{j \in \mathcal{J}^p} x_i^p(t) R_{ij}^p(w(t), x(t)), \forall i \in \mathcal{J}^p, \forall p \in \mathcal{P}$<br>$F_i^p = s_i^p \underbrace{\mathcal{U}_{\text{ina}}}_{=r_{\text{ina}}} + (1 - s_i^p) \underbrace{\mathcal{U}_{act}^{p,i}}_{=-r_{\text{act}}\eta_i^p} = s_i^p \underbrace{r}_{=\frac{r_{\text{ina}}}{r_{\text{act}}}} - (1 - s_i^p)\eta_i^p$ |

| Paper | Title | Behavioural modelling solution |
|---|---|---|
| | | $x_i^p$ ≔ fraction of players with degree $p$ playing strategy $s_i^p$ (i.e., the social inactivity intensity level)<br>$R_{ji}^p$ ≔ "revision protocols" capturing individual perceptions of payoffs given current state and others' strategies<br>$F_i^p$ ≔ payoff of players with degree $p$ choosing strategy $s_i^p$, dependent on the cost of isolation $r_{\text{ina}}$ and the probability that a player in population $p$ playing strategy $s_i^p$ is infected $\eta_i^p = \mathcal{O}_i^p\left(\mathcal{I}_i^p(t)\right)$ ($\mathcal{O}_i^p$: function corresponding to the degree of observation of the infected fraction of players at the time of an information broadcast. In the paper, they consider perfect observation: $\eta_i^p = I_i^p(t)$)<br><br>→<br>$\dot{I}_i^p = -\gamma I_i^p(t) + \underbrace{\lambda_i^p}_{=\lambda(1-s_i^p)} \left(1 - I_i^p(t)\right) p\Theta(t),\ \Theta(t) = \frac{\sum_{p\in\mathcal{P}}(\sum_{i\mathcal{J}^p} p x_i^p I_i^p(t))}{\sum_{p\in\mathcal{P}} pm^p}$<br>Where $I_i^p$ denotes the fraction of infected players in population $p$ adopting strategy $s_i^p \in \mathcal{S}^p$ at time $t$. $\lambda_i^p$ is the activity-aware contagion rate of a player with degree $p$ and social inactivity level $(1 - s_i^p)$, and $\Theta(t)$ is the probability that a link is connected to an infected player. |
| Madeo et al. 2022 | Identification and Control of Game-Based Epidemic Models | $\beta(x) = \beta_0(1 - ex)$<br>$\dot{x} = x(1 - x)\left((\sigma_C^0 + \sigma_N^0)x - \sigma_N^0\right) \underbrace{G(I, D)}_{=\frac{I+D-a}{M}}$<br>$\beta(x), \beta_0$ ≔ cooperation-dependent and natural infection rate, respectively<br>$x$ ≔fraction of individuals cooperating, i.e. adhering to the imposed restrictions<br>$G(I, D)$ ≔ evaluation of the pandemic status by agents, dependent on detected infected individuals $I$ (quarantined or hospitalized) and dead individuals $D$.<br>$\sigma_C^0, \sigma_N^0$ ≔ net payoffs for cooperation and defection, respectively<br>$a$ ≔ tuning parameter (threshold)<br>$M$ ≔ total population<br><br>→<br>$\dot{S} = -\beta(x)\frac{SU}{M} + \alpha R$<br>$\dot{U} = \beta(x)\frac{SU}{M} - \gamma U - \tau U$<br>where $S, U, R$ denote susceptible, undetected infected, and recovered individuals, respectively. |
| Martcheva et al. 2021 | Effects of social-distancing on infectious disease dynamics: an | $x' = x(1 - x)(\delta_0 x - r_s + ci - \delta_0(1 - x))$<br>$x$ ≔ proportion of the population strictly maintaining social distancing |

| Paper | Title | Behavioural modelling solution |
|---|---|---|
| | evolutionary game theory and economic perspective | $\delta_0$ ≔ strength of social norms to follow health directives<br>$r_s$ ≔ costs for applying strict social distancing<br>$c$ ≔ proportionality constant<br><br>→<br>$i' = (1 - x)\beta i(1 - i) - (\mu + \gamma)i$<br>where $i$ denotes the infectious population and $\beta$ the transmission rate |
| Meng et al. 2023 | Coupled disease-vaccination behavior dynamic analysis and its application in COVID-19 pandemic | $\Delta_1 = S_N \cdot V_E P(NS \leftarrow VES) + S_N \cdot S_V P(NS \leftarrow VNS) + S_N \cdot I_V P(NS \leftarrow VNI)$<br>$\Delta_2 = S_V \cdot S_N P(VNS \leftarrow NS) + S_V \cdot I_N P(VNS \leftarrow NI)$<br>$P(A \leftarrow B)$ ≔ Fermi updating rule: probability an individual from A imitates a neighbour from B, here dependent on vaccination and infection cost (in turn, dependent on infection rate); VNS = vaccination-non-effective-strategy, NS = non-vaccination-susceptible-strategy, VES = vaccination-effective-strategy, VNI = vaccination-infection-strategy, NI = non-vaccination-infection-strategy<br><br>→<br>$\frac{dS_V}{dt} = -S_V\left[1 - (1 - \beta)^{\langle k\rangle\langle I_V + I_N\rangle}\right] + (1 - e)\Delta_1 - \Delta_2$<br>$\frac{dS_N}{dt} = -S_N\left[1 - (1 - \beta)^{\langle k\rangle\langle I_V + I_N\rangle}\right] - \Delta_1 + \Delta_2$<br>$\frac{dI_V}{dt} = S_V\left[1 - (1 - \beta)^{\langle k\rangle\langle I_V + I_N\rangle}\right] - \mu I_V$<br>$\frac{dI_N}{dt} = S_N\left[1 - (1 - \beta)^{\langle k\rangle\langle I_V + I_N\rangle}\right] - \mu I_N$<br>$\frac{dV_E}{dt} = e\Delta_1$<br>Where $S_V, S_N, I_V, I_N, V_E$ denote susceptible imperfectly vaccinated, susceptible non-vaccinated, infected imperfectly vaccinated, infected non-vaccinated, and perfectly vaccinated individuals, respectively. $e$ refers to the vaccination efficacy. Perfectly vaccinated individuals cannot get infected. |
| Ngonghala et al. 2021 | Human choice to self-isolate in the face of the COVID-19 pandemic: A game dynamic modelling approach | $\dot{x}_1 = \mu[x_1 x_2 \Delta G_{12} + x_1(1 - x_1 - x_2)\Delta G_{13}]$<br>$\dot{x}_2 = \mu[x_2(1 - x_1 - x_2)\Delta G_{23} - x_1 x_2 \Delta G_{12}]$<br>$x_1$ ≔ proportion of individuals who choose to self-quarantine<br>$x_2$ ≔ proportion of (non-quarantined) individuals who choose to social distance<br>$\Delta G_{ij} = P_i - P_j$, where<br>$P_1 = -\left[r_q + r_{sq}\psi_{sq}(m_a I_a + m_s I_s + m_i I_i)\right]$, |

| Paper | Title | Behavioural modelling solution |
| --- | --- | --- |
|  |  | $P_2 = -[r_d + r_{sd}\psi_{sd}(m_a I_a + m_s I_s + m_i I_i)]$,<br>$P_3 = -[r_n(m_a I_a + m_s I_s + m_i I_i) + r_\delta m_\delta(\delta_s I_s + \delta_i I_i)]$<br>Having $I_a, I_s, I_i$ denoting asymptomatic infectious, symptomatic infectious, and isolated infectious individuals, respectively. $r_q, r_d, r_n$ are the perceived cost of SQ, the perceived cost of SD, and the perceived risk of infection, respectively.<br><br>→<br>$\dot{S}_u = \xi_q S_q - (\lambda_u + \xi_u x_1)S_u$<br>$\dot{S}_q = \xi_u x_1 S_u - (\lambda_q + \xi_q)S_q$<br>$\dot{E}_u = (1-q)\lambda_u S_u - (\alpha_u + \sigma_u + \tau)E_u$<br>$\dot{E}_q = q\lambda_u S_u + \lambda_q S_q + \alpha_u E_u - \sigma_q E_q$<br>$\lambda_u = (1 - \epsilon x_2)(\beta_a I_a + \beta_s(I_s + \eta_i I_i))/(N - \omega_q(S_q + E_q))$<br>$\lambda_q = \phi_q(\beta_a I_a + \beta_s(I_s + \eta_i I_i))/(N - \omega_q(S_q + E_q))$<br>i.e., social distancing and self-quarantine reduce transmission in separate ways (modulated by constants $\epsilon$ and $\phi_q$. |
| Ovi et al. 2022 | Social distancing as a public-good dilemma for socio-economic cost: An evolutionary game approach | $\frac{dx_d}{dt} = m x_d(t)[1 - x_d(t)][(1-d)C - (I^U + I^D)]$<br>$x_d(t)$ ≔ proportion of people that adopt social distancing<br>$C$ ≔ social distancing cost<br>$I^U, I^D$ ≔ undetected infected and detected infected individuals, respectively<br><br>→<br>$\frac{dS}{dt} = -x_d \beta S(I^U + qI^D)$<br>$\frac{dE}{dt} = x_d \beta S(I^U + qI^D) - \alpha E$<br>Where $S, E$ denote the susceptible and exposed individuals, respectively |
| Pedro et al. 2021 | Conditions for a Second Wave of COVID-19 Due to Interactions Between Disease Dynamics and Social Processes | $\frac{dx}{dt} = \kappa x(1-x)(\omega I - \epsilon L)$<br>$x$ ≔fraction of population that supports school and workplace closure<br>$I$ ≔ infectious individuals<br>$L$ ≔ phenomenological representation of accumulated socio-economic losses<br><br>→<br>$\frac{dS}{dt} = -\beta(1 - C(t))SI$<br>$\frac{dE}{dt} = \beta(1 - C(t))SI - \sigma E$ |

| Paper | Title | Behavioural modelling solution |
|---|---|---|
| | | $C(t) \coloneqq$ social distancing measures, s.t.<br>$C(t) = \begin{cases} 0, & t < t_{close} \vee x_s < 50\% \\ C_0, & t \geq t_{close} \wedge x_s \geq 50\% \end{cases}$<br>$S, E, I \coloneqq$ susceptible, exposed, and infectious individuals, respectively |
| Steinegger et al. 2022 | Behavioural response to heterogeneous severity of COVID-19 explains temporal variation of cases among different age groups | $\Gamma_i^{P\to NP} = \Theta(P_i^{NP} - P_i^P)\frac{P_i^{NP} - P_i^P}{c_i + T_i}$<br>$\Gamma_i^{NP\to P} = \Theta(P_i^P - P_i^{NP})\frac{P_i^P - P_i^{NP}}{c_i + T_i}$<br>$P_i^{NP} = -T_i \sum_{j=1}^{G} f_j(I_j^{NP} + I_j^P)$<br>$P_i^P = -c_i - T_i(1-\gamma_i) \sum_{j=1}^{G} f_j(I_j^{NP} + I_j^P)$<br>$\Gamma_i^{P\to NP}, \Gamma_i^{NP\to P} \coloneqq$ transition probabilities<br>$P, NP \coloneqq$ people adopting and not adopting prophylactic measures, respectively<br>$I_i \coloneqq$ infected individuals in group $i$<br>$T_i \coloneqq$ infection cost for group $i$<br>$c_i \coloneqq$ adoption cost for group $i$<br>$f_i \coloneqq$ fraction of the population belonging to group $i$<br>$\gamma_i \coloneqq$ efficacy of prophylactic measures<br><br>→<br>$\dot{S}_i^{NP} = -\beta_i S_i^{NP} \sum_{j=1}^{G} C_{ij}\left(I_j^{NP} + (1-\gamma_j)I_j^P\right) + \alpha_i S_i^P \Gamma_i^{P\to NP} - \alpha_i S^{NP} \Gamma_i^{NP\to P}$<br>$\dot{S}_i^P = -\beta_i S_i^P (1-\gamma_i) \sum_{j=1}^{G} C_{ij}\left(I_j^{NP} + (1-\gamma_j)I_j^P\right) + \alpha_i S_i^{NP} \Gamma_i^{NP\to P} - \alpha_i S^P \Gamma_i^{P\to NP}$<br>$\dot{I}_i^{NP} = \beta_i S_i^{NP} \sum_{j=1}^{G} C_{ij}\left(I_j^{NP} + (1-\gamma_j)I_j^P\right) - \mu_i I_i^{NP}$<br>$\dot{I}_i^P = \beta_i S_i^P (1-\gamma_i) \sum_{j=1}^{G} C_{ij}\left(I_j^{NP} + \Gamma_i I_j^P\right) - \mu_i I_i^P$<br>Where $S^P, S^{NP}$ denote susceptible individuals adopting and not adopting prohyplactic measures, respectively. $\alpha_i$ is an adoption/selection rate.<br>Being in the compartments adopting prophylactic measures reduces the chances of transmission. |
| Tang et al. 2022 | Controlling Multiple COVID-19 Epidemic Waves: An Insight from a Multi-scale Model Linking the | $x' = \vartheta x(1-x)\phi\Delta P = \omega x(1-x)\Delta P$<br>$\Delta P = P_n - P_a$<br>$P_n(t) = -m_n M(t)$<br>$P_a(t) = -k - m_a M(t)$ |

| Paper | Title | Behavioural modelling solution |
|---|---|---|
| | Behaviour Change Dynamics to the Disease Transmission Dynamics | $M'(t) = \eta\delta_I(t)I - \nu M$<br>$x$ ≔ proportion of individuals adopting protective behaviours<br>$P_n, P_a$ ≔payoffs of normal and altered behaviour, respectively<br>$\eta$ ≔ response rate of awareness<br>$I$ ≔ number of infectives with symptoms<br>$M$ ≔ information variable<br><br>→<br>$S' = -\lambda[B + q(1-B)]S$<br>$E' = \lambda[B + q(1-B)]S - \sigma E$<br>$B' = \frac{\rho B(1-B)(1-mM)(S+E+A+R_A)}{N}$<br>Where $S, E, A, R_A$ denote total susceptible, total exposed, total asymptomatic infected, and asymptomatic recovered individuals, respectively. The imitation dynamics process is inglobated in the evolution of the ratio of individuals with normal behaviour over the total population $B = \frac{S_n+E_n+A_n+R_n}{S+R+A+R_A}$. $1 - mM$ represents the balance between payoffs associated with the two behaviours, with $1/m$ being the threshold determing the more beneficial behaviour and $\rho$ being the speed of behavioural change parameter.<br>The transmission rate for altered behaviour individuals is lower (as modulated by the constant $q$) |
| Vivekanandhan et al. 2022 | Investigation of vaccination game approach in spreading covid-19 epidemic model with considering the birth and death rates | $x(e + (1-e)\exp(-(1-\eta)R_0R(x,\infty)))$ ≔ fraction of healthy, vaccinated individuals at the end of the epidemic season<br>$(1-x)\exp(-R_0R(x,\infty))$ ≔ fraction of non-vaccinated, healthy individuals at the end of the epidemic season<br>$x(1-e)(1-\exp(-(1-\eta)R_0R(x,\infty)))$ ≔ fraction of vaccinated, infected individuals at the end of the epidemic season<br>$(1-x)(1-\exp(-R_0R(x,\infty)))$ ≔ fraction of non-vaccinated, infected individuals at the end of the epidemic season<br><br>→<br>At the beginning of each season[1], $V(x,0) = x, S(x,0) = 1 - x$<br>At the end of the epidemic season, the progression of the disease determines the new $x$ fraction<br>$V, S, R$ denote the vaccinated, susceptible, and recovered individuals, respectively |

[1] The authors inspect three different mechanisms for $x$ updating rule, which we do not report it for the sake of conciseness

| Paper | Title | Behavioural modelling solution |
| --- | --- | --- |
| Zhao et al. 2020 | Imitation dynamics in the mitigation of the novel coronavirus disease (COVID-19) outbreak in Wuhan, China from 2019 to 2020 | $p' = \kappa p(1-p)[(1-\alpha)I/N - r]$<br>$p$ ≔probability of an individual to take infection prevention actions<br>$I$ ≔ infectious population<br><br>→<br>$U' = -\beta I \frac{U}{N} + \xi[(1-p)M - pU]$<br>$M' = -\alpha\beta I \frac{M}{N} - \xi[(1-p)M - pU]$<br>$E' = \beta I \frac{U+\alpha M}{N} - \sigma E$<br>Where $U, M$ denote individuals not taking and taking infection prevention actions, respectively. $E$ denotes the exposed individuals. Awareness reduces the transmission rate (as modulated by the constant $\alpha$). |

## Supplementary Material S4.V: Modelling solutions adopted by papers in the "**Economic Epidemiology**" category

| Paper | Title | Behavioural modelling solution |
|---|---|---|
| Baril-Tremblay et al. 2021 | Self-isolation | $\bar{k}_I(t) = \frac{1}{i(t)} \int_{j \in I(t)} k_j(t) dj$<br>$\bar{k}_S(t) = \frac{1}{s(t)} \int_{j \in S(t)} k_j(t) dj$<br>$\dot{p}_i(t) = -p_i(t)\big(1 - p_i(t)\big) k_i(t) \beta \, \bar{k}_I(t), \;\; p_i(0) = 1 - \mu$<br>$v_I = -E\left[\int_0^{\min\{\tau_H, \tau_D\}} e^{-rt}(c_H + c_I)dt + \frac{c_D}{r} e^{-r\tau_D} \mathbf{1}_{\tau_D < \tau_H}\right] = -\frac{1}{r + \gamma + \nu}\left(c_H + c_I + \nu \frac{c_D}{r}\right)$<br>$v_i(t; k_i) = \int_t^T e^{-r(s-t)} e^{-\int_t^s p_i(u) k_i(u) \, \beta \bar{k}_I(u) i(u) du} (p_i(s) k_i(s) \beta \bar{k}_I\,(s) i(s) v_I - c_H(1 - k_i(s))) ds$<br>$\bigstar \begin{cases} \max_{k_i \in \mathcal{K}} v_i(0, k_i) \\ s.t. \;\; \dot{p}_i(t) = -p_i(t)\big(1 - p_i(t)\big) k_i(t) \beta \, \bar{k}_I(t), \forall t \\ p_i(0) = 1 - \mu \end{cases}$<br>$V_i(t) = V_i(t + dt)(1 - rdt) - c_H dt + \max_{k_i(t) \in [0,1]} k_i(t) \left(c_H - p_i(t) \beta \bar{k}_I(t)(V_i(t) - v_I)\right) dt$<br>$\bar{k}_I(t), \bar{k}_S(t)$ ≔ average fraction of time spent outside at $t$ by infected and susceptible individuals, respectively<br>$p_i(t)$ ≔ subjective belief of individual $i$ of being of type $\theta_s$ (i.e., of being the severe, as opposed to asymptomatic, type). Proportion of asymptomatic types in the population: $\mu \in (0,1)$<br>$c_H, c_I, c_D$ ≔ flow cost per unit of time spent at home, having symptoms, and being dead, respectively<br>$\tau_H, \tau_D$ ≔ random times of healing and death, respectively<br>$v_I$ ≔ expected continuation payoff to individual $i$ the first time she has symptoms<br>$v_i(t; k_i)$ ≔ expected payoff to individual $i$ at time $t \leq T$ if she plays strategy $k_i$<br>$r$ ≔ time discounting rate<br>$s(t), i(t)$ ≔ susceptible and infective fraction of the population, respectively<br>$\beta, \gamma, \nu$ ≔ transmission, recovery, and death rates, respectively<br>$\bigstar$≔ best-response problem faced by $i$, fixed a strategy $\boldsymbol{k}$<br>$V_i(t)$ ≔ Bellman equation that satisfies the best-response payoff at time $t$ |

| Paper | Title | Behavioural modelling solution |
|---|---|---|
| | | → <br> $\dot{s}(t) = -\beta \bar{k}_S(t) s(t) \bar{k}_I(t) i(t)$ <br> $i(t) = \beta \bar{k}_S(t) s(t) \bar{k}_I(t) i(t) - (\gamma + (1-\mu)\nu) i(t)$ |
| Chan et al. 2021 | Modeling COVID-19 Transmission Dynamics With Self-Learning Population Behavioral Change | $a_i = \omega_1 \theta\left(u_i^n(t), u_i^r(t)\right), b_i = \omega_1 \theta(u_i^r(t), u_i^n(t))$ <br> $u_i^n(t) := \max_{z_i^n} v_i^n(z_i^n, \boldsymbol{e}(t))$ <br> $u_i^r(t) := \max_{z^r} v_i^r(z^r, \boldsymbol{e}(t))$ <br> $v_i^n(t, \boldsymbol{e}) = \underbrace{\bar{v}\left(z_i^{n*}(t)\right)}_{\text{present value}} - \underbrace{c\omega_2 m(z^{n*}) \sum_{j=1}^{M} k_I I_j(t-\tau)}_{\text{Expected cost after infection}} - \underbrace{\left(1 - \omega_2 m(z^{n*}) \sum_{j=1}^{M} k_I I_j(t-\tau)\right)}_{\text{Expected cost of susceptible}}$ <br> $v_i^r(t, \boldsymbol{e}) = \underbrace{\bar{v}\left(z_i^{r*}(t)\right)}_{\text{present value}} - \underbrace{c\omega_2 m(z^{r*}) \sum_{j=1}^{M} k_I I_j(t-\tau)}_{\text{Expected cost after infection}} - \underbrace{\left(1 - \omega_2 m(z^{r*}) \sum_{j=1}^{M} k_I I_j(t-\tau)\right)}_{\text{Expected cost of susceptible}}$ <br><br> $z_i^{n*}, z_i^r$ := optimal response functions for normal activity and reduced activity type, respectively <br> $u_i^{n*}(t), u_i^r(t)$ := utility functions for normal activity and reduced activity type, respectively <br> $a_i, b_i$ := rate of change normal → reduced and reduced → normal type, respectively <br> $v_i^n, v_i^r$ := value functions for normal activity and reduced activity type, respectively <br> $M$ := groups in the population (e.g., age groups) <br> $I_i$ := infected population in the i-th group <br><br> → <br> $\dot{S}_i^r = a_i S_i^n - b_i S_i^r - \beta_i^r S_i^r$ <br> $\dot{S}_i^n = -a_i S_i^n + b_i S_i^r - \beta_i^n S_i^n$ <br> Where $S_i^n, S_i^r$ denote normal and reduced activity susceptible individuals, respectively |
| Dasaratha 2023 | Virus dynamics with behavioral responses | $q_S(t) = \text{argmax}_q \left\{ q - aq^2 - Cqq_S(t)\beta \cdot \frac{I(t)S(t)}{S(t)+I(t)} \right\}$ <br> $q_R(t) = \text{argmax}_q \{q - aq^2\}$ <br><br> $q_S(t)$ := level of activity of individuals who are not recovered yet (and does not know whether they are already infected or not) <br> $q_R(t)$ := level of activity of recovered individual |

| Paper | Title | Behavioural modelling solution |
| --- | --- | --- |
| | | $q$ ≔ equilibrium level of activity<br>$S, I$ ≔ susceptible and infected individuals, respectively<br>$C$ ≔ cost of behaviour change<br><br>→<br>$\dot{S}(t) = -q_S(t)^2 \beta S(t) I(t)$<br>$\dot{I}(t) = q_S(t)^2 \beta S(t) I(t) - \kappa I(t)$ |
| Di Guilmi et al. 2022 | A Behavioural SIR Model: Implications for Physical Distancing Decisions | $U_t^i = \alpha x_t + \epsilon_i$<br>$P_t^D = \frac{1}{1 + e^{-\alpha x_t}}$<br>$x_t = \begin{bmatrix} S_t - S_{t-1} \\ -c_n \end{bmatrix}, \alpha = [a_x \quad 1]$<br>$U_t^i$ ≔ utility of choosing $D$<br>$P_t^D$ ≔ probability of an agent choosing to social distance, $D$<br>$x_t, \epsilon_i$ ≔ vector of observable factors and unobservable individual characteristics, respectively ($\epsilon_i$ follows a logistic distribution)<br>$c_n$ ≔ social and economic costs of physical distancing<br>$\alpha_x$ ≔ relative measure of individuals' risk perception<br><br>→<br>$\beta_t = \beta(1 - P_t^D) = \frac{\beta \exp(\alpha_x(S_t - S_{t-1}) + c_n)}{1 + \exp(\alpha_x(S_t - S_{t-1}) + c_n)}$<br>$S_{t+1} = S_t - \frac{\beta_t S_t I_t}{N}$ |
| Espinoza et al. 2021 | Asymptomatic individuals can increase the final epidemic size under adaptive human behavior | $u_t = (bC_t^h - (C_t^h)^2)^\nu$<br>$V_t(S) = \max_{C_t^S}\{u(S, C_t^S) + \delta[(1 - P^I)V_{t+1}(S) + P^I V_{t+1}(E)]\}$<br>$V_t(E) = u(E, C_t^S) + \delta[(1 - P^E)V_{t+1}(E) + P^E(\sigma V_{t+1}(A) + (1 - \sigma)[(1 - l)V_{t+1}(I_S) + l V_{t+1}(I_C)])]$<br>$V_t(A) = U(A, C_t^S) + \delta[(1 - P^R)V_{t+1}(A) + P^R V_{t+1}(R)]$<br>$V_t(I_S) = U(I_S, C_t^{I_S}) + \delta[(1 - P^R)V_{t+1}(I_S) + P^R V_{t+1}(R)]$<br>$= \min_{C_t^{I_S}} \left\{ u(I_S, C_t^{I_S}) \sum_{j=1}^{\tau} (\delta(1 - P^R))^j \right\}$, subject to $\frac{V_t(I_S)}{\tau} > u_C$<br>$V_t(I_C) = U(I_C, C_t^{I_C}) + \delta[(1 - P^R)V_{t+1}(I_C) + P^R V_{t+1}(R)]$<br>$= u(I_C, C^{I_C *}) \sum_{j=1}^{\tau} \delta^j (1 - P^R)^j + u(R, C^{R*}) \sum_{j=1}^{\tau} \delta_j (1 - (1 - P^R)^j)$ |

| Paper | Title | Behavioural modelling solution |
|---|---|---|
| | | $V_t(R) = u(R, C_t^{R*}) + \delta V_{t+1}(R)$<br>$P^E = 1 - e^{-\kappa}$<br>$P^I = 1 - \exp\left(-\beta C_t^S S \frac{C_t^E \rho E + C_t^A \varepsilon A + C_t^{I_S} \eta I_S + C_t^{I_C} I_C}{C_t^S S + C_t^E E + C_t^A A + C_t^{I_S} I_S + C_t^{I_C} I_C + C_t^R R}\right)$<br>$P^R = 1 - e^{-\gamma}$<br><br>$C_t^h$ ≔ contact rate of a typical individual with health status $h$ (determined by maximization of expected utility for each health class)<br>$u(h, C_t^h)$ ≔ utility of individual of health class $h$ by making contacts $C_t$<br>$b, \nu$ ≔ maximum number of contacts possible and utility function shape parameter, respectively<br>$V_t(S), V_t(A), V_t(I_S), V_t(I_C), V_t(R)$ ≔ expected utility of susceptible individuals, asymptomatic, infected compliant, infected noncompliant, recovered at time $t$, respectively<br>$P^I, P^R$ ≔ probability of being infected and of recovery at time $t$, respectively<br>$P^E$ ≔ probability of moving from $E$ health class to either $A, I_S, I_C$<br><br>→<br>incidence term: $\beta C_t^S S \frac{C_t^E \rho E + C_t^A \varepsilon A + C_t^{I_S} \eta I_S + C_t^{I_C} I_C}{C_t^S S + C_t^E E + C_t^A A + C_t^{I_S} I_S + C_t^{I_C} I_C + C_t^R R}$ |
| Gostoli et al. 2023 | Self-Isolation and Testing Behaviour During the COVID-19 Pandemic: An Agent-Based Model | Self-isolation module:<br>$r_i = 1 - e^{-\beta I_i V_i}$<br>$V = \frac{C}{(1 - h\epsilon)^s}, \epsilon = \frac{1 - p^f}{p^f}, C = V_h + V_v + V_d$<br>$m_i = \frac{\sum_{j=1}^n r_j e^{\theta d_{ij}}}{\sum_{j=1}^n e^{\theta d_{ij}}}$<br>$r_i^n = r_i + \lambda(m_i - r_i)$<br>$q_i = \frac{\sum_{j=1}^n A_{ij} e^{-\phi d_{ij}}}{\sum_{j=1}^n e^{-\phi d_{ij}}}$<br>$b = \begin{cases} b_0 + (r_{\max} - b_0)(1 - e^{-\tau q_1}, & q_i > 0 \\ b_0 - b_0(1 - e^{-\tau q_1}), & q_i < 0 \end{cases}$<br>$s_i = r_i + (b - r_i)\gamma$<br>$r_i$ ≔ individual self-isolation propensity, where $I_i$ denotes the agent's income quintile and $V_i$ the value of the expected cost<br>$p^f$ ≔ probability of infection in case of no isolation |

| Paper | Title | Behavioural modelling solution |
|---|---|---|
| | | $C \coloneqq$ expected cost associated with infection, given by the sum of the expected cost associated with hospitalization, intensive care, and death<br>$m_i \coloneqq$ weighted average of the self-isolation propensities of the agents' friends<br>$r_i^n \coloneqq$socially-affected self-isolation propensity, where $\lambda$ is he agent sensitivity to social norms<br>$q_i \coloneqq$ help availability index, where $n$ is the number of households with a kin relationship with household $i$, $d_{ij}$ is the weight of household $j$ contribution, and $A_{ij}$ the difference in mean ages between $i, j$<br>$b_0, b \coloneqq$ household mean isolation propensity before and after help provision, respectively<br>($r_{\max} \coloneqq$ maximum individual self-isolation propensity within the household, i.e. unconstrained by income considerations). $\tau$ is the sensitivity of the household's mean self-isolation propensity<br>$s_i \coloneqq$ individual self-isolation rate (i.e., agent's self-isolation propensity adjusted by their household's mean propensity isolation)<br><br><u>Testing module</u>:<br>$p_{test} = 1 - e^{-\alpha^T \mu I w}$<br>$w = p_f V_i, \qquad p_f = \frac{(N+H)S}{2}$<br>$p_{test} \coloneqq$ probability of taking a test (dependent on social preferences $\mu$, agent's expected cost of infection $w$, agent's income quintile $I$, and a "parameter of the model" $\alpha^T$)<br>$p_f \coloneqq$ likelihood that the agent has been infected (dependent on probability of having been infected during a social interaction $N$, probability of having been infected thorugh the agent's interaction within household $H$, and a factor $S$ determined by the maximum value of agent's symptoms $s$ and the probability of being asymptomatic $p_a$)<br><br>→<br>The agent's isolation rate is used in the social interaction process to determine venue attendants (after being assigned to a pool, each agent s removed with a probability equal to the agent's self-isolation rate).<br>The agent's testing rate is used to represent the probability of taking a test. If the test is taken, the result can be positive if infected/false positive and negative if not infected/false negative. If positive, the agent self-isolates completely.<br>Thus, the two modules influence the entire transmission process. |

## Supplementary Material S4.VI: Modelling solutions adopted by papers in the "**Explicit - other**" category

| Paper | Title | Behavioural modelling solution |
|---|---|---|
| De Mooij et al. 2023 | A framework for modeling human behavior in large-scale agent-based epidemic simulations | $a(n) = \frac{1}{1 + e^{(-k\cdot(\bar{f}(n)-t))}}$<br>$f_3 = \bar{s}^l_{\Delta d}$<br><br>$a(n)$ ≔ attitude of the agent towards norm $n$, dependent on the average of the agent's values for the factors $\{f_1, \dots, f_n\}$, $\bar{f}(n)$<br>$f_3$ ≔ factor concerning agent's symptomatic neighbors (the more, the more likely to comply with norms), dependent on the average fraction of contacts at location $l$ in the past $d$ days $\bar{s}^l_{\Delta d}$<br><br>→ norm $n$ → attitude $a(n)$ → deliberation of norm-compliance (i.e. compliance with COIVD-19 regulations |
| Silk et al. 2021 | Improving pandemic mitigation policies across communities through coupled dynamics of risk perception and infection | • Social construction of concern: concern changes based on the proportion of connections in the communication layer that are adherent→ linear increase in concern on a logit scale, whose strength varies in size: $0, 0.1, 0.2, 0.5, 1$<br>• Awareness of incidence among communications contacts: concern changes based on the number of neighbors in the communication layer that are infected and show full symptoms in the previous day → linear increase in concern on a logit scale, whose strength varies in size: $0, 0.2, 0.4, 0.8, 1.6$<br>• Reassurance effect: concern changes based on the number of neighbors in the communication layer that do not show symptoms → decrease in concern, whose strength varies in size: drawn from a uniform distribution between $-0.2, -0.01$<br><br>→<br>adherence: depends on a Bernoulli draw based on concern construct; each time an individual becomes adherent, they cut their connections within the infection layer of the network while maintaining connectivity in the communication layer, thus reducing disease transmission |
| Zhai et al. 2024 | Optimizing the impact of virus transmission on society: | $\dot{x}_i(t) = -[\delta_{\min} + (\delta_i - \delta_{\min})(y_i(t) + 0.5)]x_i(t) + \big(1 - x_i(t)\big) \sum_{j\in T_i} [\beta_{ij} - u_{1i} - \big(\beta_{ij} - u_{1i} - \beta_{\min}\big)$ |

| Paper | Title | Behavioural modelling solution |
| --- | --- | --- |
| | decreasing infection rates and enhancing public awareness | $\dot{y}_i(t) = \left(x_i(t) - (y_i(t) + 0.5)\right) + u_{2i} + \sum_{j \in \overline{T}_i} \lvert \bar{a}_{ij} \rvert (sign(\bar{a}_{ij}) y_j(t) - y_i(t))$<br>$x_i \in [0,1]$ ≔ proportion of infected population<br>$y_i \in [-0.5,0.5]$ ≔ opinion of community $i$ on the severity of the epidemic |

# PRISMA 2020 Checklist

| Section and Topic | Item # | Checklist item | Location where item is reported |
|---|---|---|---|
| **TITLE** | | | |
| Title | 1 | Identify the report as a systematic review. | Title |
| **ABSTRACT** | | | |
| Abstract | 2 | See the PRISMA 2020 for Abstracts checklist. | n/a |
| **INTRODUCTION** | | | |
| Rationale | 3 | Describe the rationale for the review in the context of existing knowledge. | pp. 1-2 |
| Objectives | 4 | Provide an explicit statement of the objective(s) or question(s) the review addresses. | pp. 2-3 |
| **METHODS** | | | |
| Eligibility criteria | 5 | Specify the inclusion and exclusion criteria for the review and how studies were grouped for the syntheses. | pp. 14-15 + Protocol OSF |
| Information sources | 6 | Specify all databases, registers, websites, organisations, reference lists and other sources searched or consulted to identify studies. Specify the date when each source was last searched or consulted. | p. 14 + Protocol OSF |
| Search strategy | 7 | Present the full search strategies for all databases, registers and websites, including any filters and limits used. | pp. 14 + Protocol OSF |
| Selection process | 8 | Specify the methods used to decide whether a study met the inclusion criteria of the review, including how many reviewers screened each record and each report retrieved, whether they worked independently, and if applicable, details of automation tools used in the process. | pp. 15 + Protocol OSF |
| Data collection process | 9 | Specify the methods used to collect data from reports, including how many reviewers collected data from each report, whether they worked independently, any processes for obtaining or confirming data from study investigators, and if applicable, details of automation tools used in the process. | pp. 15-16 + Protocol OSF |
| Data items | 10a | List and define all outcomes for which data were sought. Specify whether all results that were compatible with each outcome domain in each study were sought (e.g. for all measures, time points, analyses), and if not, the methods used to decide which results to collect. | p. 15-16 + Protocol OSF |
| | 10b | List and define all other variables for which data were sought (e.g. participant and intervention characteristics, funding sources). Describe any assumptions made about any missing or unclear information. | p. 15-16 + Protocol OSF |
| Study risk of bias assessment | 11 | Specify the methods used to assess risk of bias in the included studies, including details of the tool(s) used, how many reviewers assessed each study and whether they worked independently, and if applicable, details of automation tools used in the process. | p. 16 + Protocol OSF |
| Effect measures | 12 | Specify for each outcome the effect measure(s) (e.g. risk ratio, mean difference) used in the synthesis or presentation of results. | n/a |
| Synthesis methods | 13a | Describe the processes used to decide which studies were eligible for each synthesis (e.g. tabulating the study intervention characteristics and comparing against the planned groups for each synthesis (item #5)). | Protocol OSF |
| | 13b | Describe any methods required to prepare the data for presentation or synthesis, such as handling of missing summary statistics, or data conversions. | n/a |
| | 13c | Describe any methods used to tabulate or visually display results of individual studies and syntheses. | Protocol OSF |
| | 13d | Describe any methods used to synthesize results and provide a rationale for the choice(s). If meta-analysis was performed, describe the model(s), method(s) to identify the presence and extent of statistical heterogeneity, and software package(s) used. | n/a |
| | 13e | Describe any methods used to explore possible causes of heterogeneity among study results (e.g. subgroup analysis, meta-regression). | n/a |
| | 13f | Describe any sensitivity analyses conducted to assess robustness of the synthesized results. | n/a |
| Reporting bias assessment | 14 | Describe any methods used to assess risk of bias due to missing results in a synthesis (arising from reporting biases). | n/a |

# PRISMA 2020 Checklist

| Section and Topic | Item # | Checklist item | Location where item is reported |
|---|---|---|---|
| Certainty assessment | 15 | Describe any methods used to assess certainty (or confidence) in the body of evidence for an outcome. | n/a |
| **RESULTS** | | | |
| Study selection | 16a | Describe the results of the search and selection process, from the number of records identified in the search to the number of studies included in the review, ideally using a flow diagram. | pp. 4, 15 + Supplementary Material S1 |
| | 16b | Cite studies that might appear to meet the inclusion criteria, but which were excluded, and explain why they were excluded. | Protocol OSF |
| Study characteristics | 17 | Cite each included study and present its characteristics. | pp. 4-9 |
| Risk of bias in studies | 18 | Present assessments of risk of bias for each included study. | Supplementary Material S2 |
| Results of individual studies | 19 | For all outcomes, present, for each study: (a) summary statistics for each group (where appropriate) and (b) an effect estimate and its precision (e.g. confidence/credible interval), ideally using structured tables or plots. | Supplementary Material S1 (where applicable) |
| Results of syntheses | 20a | For each synthesis, briefly summarise the characteristics and risk of bias among contributing studies. | Supplementary Material S1 + Supplementary Material S2 |
| | 20b | Present results of all statistical syntheses conducted. If meta-analysis was done, present for each the summary estimate and its precision (e.g. confidence/credible interval) and measures of statistical heterogeneity. If comparing groups, describe the direction of the effect. | n/a |
| | 20c | Present results of all investigations of possible causes of heterogeneity among study results. | n/a |
| | 20d | Present results of all sensitivity analyses conducted to assess the robustness of the synthesized results. | n/a |
| Reporting biases | 21 | Present assessments of risk of bias due to missing results (arising from reporting biases) for each synthesis assessed. | n/a |
| Certainty of evidence | 22 | Present assessments of certainty (or confidence) in the body of evidence for each outcome assessed. | n/a |
| **DISCUSSION** | | | |
| Discussion | 23a | Provide a general interpretation of the results in the context of other evidence. | pp. 9-10 |
| | 23b | Discuss any limitations of the evidence included in the review. | p. 13 |
| | 23c | Discuss any limitations of the review processes used. | p. 13 |
| | 23d | Discuss implications of the results for practice, policy, and future research. | pp. 9-13 |
| **OTHER INFORMATION** | | | |
| Registration and protocol | 24a | Provide registration information for the review, including register name and registration number, or state that the review was not registered. | pp. 3, 13-14 |
| | 24b | Indicate where the review protocol can be accessed, or state that a protocol was not prepared. | pp. 3, 13-14 |
| | 24c | Describe and explain any amendments to information provided at registration or in the protocol. | Protocol OSF |
| Support | 25 | Describe sources of financial or non-financial support for the review, and the role of the funders or sponsors in the review. | Protocol OSF |
| Competing | 26 | Declare any competing interests of review authors. | Protocol OSF |

# PRISMA 2020 Checklist

| Section and Topic | Item # | Checklist item | Location where item is reported |
|---|---|---|---|
| interests | | | |
| Availability of data, code and other materials | 27 | Report which of the following are publicly available and where they can be found: template data collection forms; data extracted from included studies; data used for all analyses; analytic code; any other materials used in the review. | Protocol OSF |